\pdfoutput = 1
\documentclass[12pt]{scrartcl}
\usepackage{hyphenat}
\usepackage[T1]{fontenc}         
\usepackage[utf8]{inputenc}
\usepackage[british]{babel}

\areaset{158mm}{238mm}
\usepackage{setspace}
\usepackage[all]{xypic}
\usepackage{scrlayer-scrpage}
\usepackage{tikz}
\usetikzlibrary{cd,positioning}

\usepackage{amsmath}
\usepackage{mathtools}
\usepackage{mathrsfs}
\usepackage{graphicx}

\usepackage{amssymb}
\usepackage[mathscr]{eucal}
\usepackage[unicode=true,linktoc=all]{hyperref}
\usepackage{cite}
\usepackage{bm}
\usepackage{slashed}

\newenvironment{tz}[1][]{%
			\begin{tikzpicture}[baseline={([yshift=-.8ex]current bounding 					box.center)},#1] %
				}{%
			\end{tikzpicture} %
			}
			
\pgfkeys{%
  /tikz/on layer/.code={
    \pgfonlayer{#1}\begingroup
    \aftergroup\endpgfonlayer
    \aftergroup\endgroup
  },
  /tikz/node on layer/.code={
     \gdef\node@@on@layer{%
      \setbox\tikz@tempbox=\hbox\bgroup\pgfonlayer{#1}\unhbox\tikz@tempbox\endpgfonlayer\egroup}
    \aftergroup\node@on@layer
  },
  /tikz/end node on layer/.code={
    \endpgfonlayer\endgroup\endgroup
  }
}

\def\node@on@layer{\aftergroup\node@@on@layer}

\tikzset{omor/.style={scale=0.6, color=black}}   

\usetikzlibrary{3d}
\def\dl{-135}
\def\dr{-45}
\def\ul{135}
\def\ur{45}

\pgfdeclarelayer{back}
\pgfdeclarelayer{front}
\pgfdeclarelayer{superfront}
\pgfdeclarelayer{superback}
\pgfdeclarelayer{fronta}
\pgfdeclarelayer{frontb}
\pgfdeclarelayer{frontc}
\pgfdeclarelayer{backa}
\pgfsetlayers{superback,back,backa,main,fronta,frontb,frontc,front,superfront}
\tikzset{xyplane/.style={canvas is yx plane at z=#1}}
\tikzset{xzplane/.style={canvas is yz plane at x=#1}}
\tikzset{yzplane/.style={canvas is xz plane at y=#1}}
\tikzset{td/.style={
					y={(0.4cm, 0.6cm)},
					x={(-1cm, 0cm)},					
                    z  = {(0cm,1cm)},
                    scale = 0.5}}
\tikzset{slice/.style={draw = gray!50, line width = 0.7pt}}
\tikzset{braid slice/.style={draw = white, double distance =0.7pt, line width =1.4pt, double = gray!50}}
\tikzset{braid slice red/.style={draw = white, double distance =0.7pt, line width =1.4pt, double = red}}
\tikzset{braid slice blue/.style={draw = white, double distance =0.7pt, line width =1.4pt, double = blue}}
\tikzset{wire/.style={black, line width=0.7pt}}
\tikzset{bluewire/.style={blue, line width=0.7pt}}
\tikzset{braid wire/.style={draw=white, double distance=0.7pt, line width=1.4pt, double=black}}
\tikzset{dot/.style={circle, scale=0.15, fill=black, thick, draw}}
\tikzset{obj/.style={scale=0.6, color=gray}}
\tikzset{omor/.style={scale=0.6, color=black}}   
\tikzset{tmor/.style={scale=0.6, color=black}}   
\def\lw{1.4pt}           
\tikzset{short/.style={shorten >=-0.26*\lw,shorten <=-0.25*\lw}}    
\def\h{3}
\tikzset{tinydash/.style={on layer=back,gray!50,line width = 0.4pt,densely dotted}}

\usepackage{tikz,pgf}
\usetikzlibrary{shapes}
\usetikzlibrary{calc}
\usetikzlibrary{decorations.pathmorphing}
\usetikzlibrary{decorations.pathreplacing,shapes.misc}
\usetikzlibrary{positioning}
\usetikzlibrary{decorations.markings}
\tikzset{->-/.style={decoration={
  markings,
  mark=at position .5 with {\arrow{>}}},postaction={decorate}}}
\tikzset{-<-/.style={decoration={
  markings,
  mark=at position .5 with {\arrow{<}}},postaction={decorate}}}

\usepackage{xcolor}
  \definecolor{rblue}{RGB}{81, 49, 193}
  \definecolor{rorange}{RGB}{255, 147, 40}
  \definecolor{rgreen}{RGB}{176, 233, 0}
  
  \usepackage{bbm}
\usepackage{lscape}

\newcommand{\ds}{\displaystyle}

\newcommand{\fP}{\mathfrak{P}}

\newcommand{\eg}{\textit{e.g.}}

\numberwithin{equation}{section}

\newcommand{\be}{\begin{equation}} \newcommand{\ee}{\end{equation}}
\newcommand{\bea}{\begin{equation} \begin{aligned}} 
\newcommand{\eea}{\end{aligned} \end{equation}}

\newcommand{\cA}{\mathcal{A}}

\newcommand{\cC}{\mathcal{C}}
\renewcommand{\cD}{\mathcal{D}}

\newcommand{\cK}{\mathcal{K}}
\renewcommand{\cL}{\mathcal{L}}
\newcommand{\cM}{\mathcal{M}}
\newcommand{\cN}{\mathcal{N}}
\newcommand{\cO}{\mathcal{O}}

\newcommand{\cS}{\mathcal{S}}
\newcommand{\cT}{\mathcal{T}}

\newcommand{\cV}{\mathcal{V}}
\newcommand{\cW}{\mathcal{W}}

\newcommand{\cZ}{\mathcal{Z}}

\newcommand{\bC}{\mathbb{C}}

\newcommand{\bQ}{\mathbb{Q}}
\newcommand{\bR}{\mathbb{R}}

\newcommand{\bZ}{\mathbb{Z}}

\newcommand{\fM}{\mathfrak{M}}

\newcommand{\unit}{\mathbbm{1}}

\def\repa{\raise4pt\hbox{$\square$}\mkern-14mu\raise-4pt\hbox{$\square$}}
\def\repab{\overline{\raise4pt\hbox{$\square$}\mkern-14mu\raise-4pt\hbox{$\square$}\mkern-1mu}}

\DeclareMathOperator{\End}{End}
\DeclareMathOperator{\Hom}{Hom}
\DeclareMathOperator{\id}{id}

\newcommand{\scat}[1]{\mathfrak{C}^{#1}}
\def\ob{\mathcal D} %
\def\sup{\Sigma} %
\def\mor{\boldsymbol{m}} %
\def\ssup{\widetilde{\Sigma}}%
\def\mmor{\boldsymbol{\eta}} %
\def\mmmor{\boldsymbol{\omega}} %
\def\id{\mathbf{1}} %
\def\cond{\mathcal C} %
\def\AF{\boldsymbol{F}}
\def\Uchiral{{\mathcal U^\chi}}
\def\Uonef{\mathfrak U^{(1)}}
\def\algA{\mathbb{A}} %
\def\salgA{\mathbb{B}} %
\def\ev{\boldsymbol{e}} %
\def\coev{\overline{\boldsymbol{e}}}

\newcommand\Dchiral[1]{\mathfrak D_{#1}}
\newcommand\scatchi[1]{\scat{#1}_\chi}

\newcommand{\lls}{[\![}
\newcommand{\rrs}{]\!]}

\usepackage[status=draft,inline,nomargin]{fixme}
\fxusetheme{color}
\FXRegisterAuthor{ko}{ako}{KO}

\title{\vspace{-2cm} Higher Structure of Chiral Symmetry}

\author{Christian Copetti$^{\ast \, \$}$, Michele Del Zotto$^{\dagger}$,\\ Kantaro Ohmori$^{\sharp}$, and Yifan Wang$^{\star}$
\\[1cm] 
        \footnotesize\slshape $^\ast$ Theoretical Particle Physics, SISSA %
 Via Bonomea, 34124 Trieste, Italy\\
  \footnotesize\slshape $^\$ $ INFN, Sezione di Trieste, %
 Via Valerio 2, 34127 Trieste, Italy\\
	\footnotesize\slshape $^\dagger$ Mathematics Institute, Uppsala University,  %
 Box 480, SE-75106 Uppsala, Sweden\\
	\footnotesize\slshape $^\dagger$ Department of Physics and Astronomy, Uppsala University, %
 Box 516, SE-75120 Uppsala, Sweden\\
	\footnotesize\slshape $^\sharp$ Department of Physics, University of Tokyo, %
 Hongo 7-3-1, Bunkyo, Tokyo, Japan\\
	\footnotesize\slshape $^\star$ Center for Cosmology and Particle Physics, New York University,  %
 New York, USA\\
	}
\date{}

\begin{document}
\thispagestyle{empty}

\maketitle

\vspace{-1cm}

\medskip

\begin{abstract}
\hspace{5cm}\paragraph{\large{Abstract}}

\noindent A recent development in our understanding of the theory of quantum fields is the fact that familiar gauge theories in spacetime dimensions greater than two can have non-invertible symmetries generated by topological defects. The hallmark of these non-invertible symmetries is that the fusion rule deviates from the usual group-like structure, and in particular the fusion coefficients take values in  topological field theories (TFTs) rather than in mere numbers. In this paper we begin an exploration of the associativity structure of non-invertible symmetries in higher dimensions. %
The first layer of associativity is captured by F-symbols,
which we find to assume values in TFTs that have one dimension lower than that of the defect. We undertake an explicit analysis of the F-symbols for the non-invertible chiral symmetry that is preserved by the massless QED and explore their physical implications. In particular, we show the F-symbol TFTs can be detected by probing the correlators of topological defects with 't Hooft lines. Furthermore, we derive the Ward-Takahashi identity that arises from the chiral symmetry on a large class of four-dimensional manifolds with non-trivial topologies directly from the topological data of the symmetry defects, without referring to a Lagrangian formulation of the theory.  %

\end{abstract}

\vfill{}
--------------------------

May, 2023

\newpage
\tableofcontents

\vspace{2cm}

\section{Introduction}

Recent advances in our understanding of symmetries of quantum field theories (QFTs) indicate that they are captured by topological subsectors in the spectrum of extended operators with fusion rules that widely generalize the notion of groups \cite{Gaiotto:2014kfa}. The study of various generalized symmetries and their interrelations has led to many powerful results for QFTs in two dimensions \cite{Bhardwaj:2017xup,Chang:2018iay,Komargodski:2020mxz,Lin:2019kpn,Thorngren:2019iar,Thorngren:2021yso,Jacobsen:2023isq} (see also e.g.\ \cite{Verlinde:1988sn,Petkova:2000ip,Fuchs:2002cm} for earlier references) and in higher dimensions as well (see e.g.\ \cite{Aharony:2013hda,Gaiotto:2017yup,Gaiotto:2017tne,Choi:2021kmx,Choi:2022zal,Kaidi:2022uux}).\footnote{\ Especially during the last 2-3 years there has been an explosion of activity regarding the study of generalized (categorical) symmetries in higher dimensions. See e.g. \cite{Gaiotto:2010be,Kapustin:2013qsa,Kapustin:2013uxa,Aharony:2013hda,DelZotto:2015isa,Sharpe:2015mja,Tachikawa:2017gyf,Cordova:2018cvg,Hsin:2018vcg,Wan:2018bns,GarciaEtxebarria:2019caf,Eckhard:2019jgg,Wan:2019soo,Bergman:2020ifi,Morrison:2020ool,Albertini:2020mdx,Hsin:2020nts,Bah:2020uev,DelZotto:2020esg,Hason:2020yqf,Aasen:2020jwb,Bhardwaj:2020phs,Apruzzi:2020zot,Cordova:2020tij,Thorngren:2020aph,DelZotto:2020sop,BenettiGenolini:2020doj,Yu:2020twi,Bhardwaj:2020ymp,DeWolfe:2020uzb,Gukov:2020btk,Iqbal:2020lrt,Hidaka:2020izy,Brennan:2020ehu,Komargodski:2020mxz,Closset:2020afy,Thorngren:2020yht,Closset:2020scj,Bhardwaj:2021pfz,Nguyen:2021naa,Heidenreich:2021xpr,Apruzzi:2021phx,Apruzzi:2021vcu,Hosseini:2021ged,Cvetic:2021sxm,Buican:2021xhs,Bhardwaj:2021zrt,Iqbal:2021rkn,Braun:2021sex,Cvetic:2021maf,Closset:2021lhd,Thorngren:2021yso,Sharpe:2021srf,Bhardwaj:2021wif,Hidaka:2021mml,Lee:2021obi,Lee:2021crt,Hidaka:2021kkf,Koide:2021zxj,Apruzzi:2021mlh,Kaidi:2021xfk,Choi:2021kmx,Bah:2021brs,Gukov:2021swm,Closset:2021lwy,Yu:2021zmu,Apruzzi:2021nmk,Beratto:2021xmn,Bhardwaj:2021mzl,Wang:2021vki,Cvetic:2022uuu,DelZotto:2022fnw,Cvetic:2022imb,DelZotto:2022joo,DelZotto:2022ras,Bhardwaj:2022yxj,Hayashi:2022fkw,Kaidi:2022uux,Roumpedakis:2022aik,Choi:2022jqy, Choi:2022zal,Arias-Tamargo:2022nlf,Cordova:2022ieu,Bhardwaj:2022dyt,Benedetti:2022zbb, DelZotto:2022ohj,Bhardwaj:2022scy,Antinucci:2022eat,Carta:2022spy, Apruzzi:2022dlm, Heckman:2022suy, Choi:2022rfe, Bhardwaj:2022lsg, Lin:2022xod, Bartsch:2022mpm,Apruzzi:2022rei,GarciaEtxebarria:2022vzq,Cherman:2022eml,Heckman:2022muc, Lu:2022ver, Niro:2022ctq, Kaidi:2022cpf,Mekareeya:2022spm,vanBeest:2022fss,Giaccari:2022xgs,Bashmakov:2022uek,Cordova:2022fhg, GarciaEtxebarria:2022jky, Choi:2022fgx, Robbins:2022wlr, Bhardwaj:2022kot, Bhardwaj:2022maz, Bartsch:2022ytj, Gaiotto:2020iye,Robbins:2021ibx, Robbins:2021xce,Huang:2021zvu,Inamura:2021szw,Cherman:2021nox,Sharpe:2022ene,Bashmakov:2022jtl, Lee:2022swr, Inamura:2022lun, Damia:2022bcd, Lin:2022dhv,Burbano:2021loy, Damia:2022rxw, Apte:2022xtu, Chen:2022cyw, Nawata:2023rdx, Bhardwaj:2023zix, Kaidi:2023maf, Etheredge:2023ler, Lin:2023uvm, Amariti:2023hev, Bhardwaj:2023wzd, Bartsch:2023pzl, Carta:2023bqn, Zhang:2023wlu, Cao:2023doz, Putrov:2023jqi, Acharya:2023bth,Inamura:2023qzl,Dierigl:2023jdp,Antinucci:2023uzq,Cvetic:2023plv}. }

The aim of this paper is to expose some of the higher structures that underpin these generalized symmetries, which correspond to the intricate relationships that topological operators of varying dimensions can possess. 
An example of higher structure is given by the Postnikov class in the context of higher-group symmetry, which encodes a mixture of invertible symmetries across multiple codimensions \cite{Barkeshli:2014cna,Benini:2018reh,Cordova:2018cvg,Delmastro:2022pfo,Brennan:2022tyl}. 
Several related works have analyzed the symmetry structure under generalized gauging procedures \cite{Fuchs:2002cm,Bhardwaj:2017xup,Tachikawa:2017gyf,Bhardwaj:2022yxj,Bhardwaj:2022lsg,Bhardwaj:2022maz},
while higher groups and their representations have been studied from a more formal perspective (see e.g.\   \cite{Bartsch:2022ytj,Bartsch:2023pzl,Bhardwaj:2023wzd} and references therein).

Our main result is that nontrivial higher structures of non-invertible symmetries are crucial ingredients to ensure the consistency of the topological networks formed by symmetry defects of various codimensions.
These higher structures are also
physical, as they can be detected by suitable correlators with extended (non-topological) operator insertions. Moreover, they lead to nontrivial Ward identities on spacetimes with non-trivial topologies, 
as well as potentially encoding appropriate generalizations of the 't Hooft anomalies. For concreteness, we focus on the non-invertible chiral symmetry in the massless QED to illustrate these ideas, but our consideration can be generalized to more general non-invertible symmetries.

The precise mathematical framework capturing a general finite higher and non-invertible symmetry of a $D$-dimensional QFT is that of a fusion $(D-1)$-category \cite{douglas2018fusion,Johnson-Freyd:2020usu}, which is denoted by $\scat{}$ in this paper. 
To roughly describe this concept and to give a physical intuition about this idea, it is useful to consider the two-dimensional example where the symmetry is a fusion 1-category (see e.g. \cite{Bhardwaj:2017xup,Chang:2018iay}). In two-dimensions we can view each topological defect line as an infra-red (IR) fixed point of a worldline quantum mechanics that is coupled to the 2d bulk. The  objects $\ob_1,\ob_2 \in \scat{}$ symbolize these topological defect lines. Here, $\Hom(\ob_1,\ob_2)$ is seen as the space of topological interfaces (i.e. the vector space of line-changing operators) between the two corresponding worldline quantum mechanics theories, while $\Hom(\ob_1,\ob_1)$ becomes the vector space of symmetry/topological operators in the same worldline quantum mechanics. %
Categories whose hom-sets are vector spaces are called linear categories, and fusion 1-categories are special cases. Let us proceed by generalizing this idea to the $D$-dimensional case. In $D$-dimensions we can regard a topological codimension-one defect as a $(D-1)$-dimensional topological field theory (TFT) coupled to the bulk non-topological QFT. An example of this  is given by the topological operator encoding the non-invertible chiral symmetry in the $D=4$ massless Quantum Electrodynamics (QED) \cite{Cordova:2022ieu,Choi:2022jqy}. The latter is constructed as 
\begin{equation}
    \Dchiral{p/N} = \Uchiral_{p/N} \otimes \mathcal{A}^{N, \, p}, \ \ \ \ \gcd(N,p)=1 \, ,
\end{equation} 
where $\mathcal{A}^{N,\,p}$ \cite{Hsin:2018vcg} is the minimal 3d TFT coupled to the 4d bulk $U(1)$ gauge field. An object $\ob$ of a fusion $(D-1)$-category $\scat{}$ corresponds to such a $(D-1)$-dimensional topological surface. Then, the endomorphism set $\End_{\scat{}}(\ob) = \Hom_{\scat{}}(\ob,\ob)$ represents a symmetry of the coupled TFT, which must be a (multi)fusion $(D-2)$-category by induction. %
In a more formal language, an $N$-category is defined as a category \textit{enriched} over the category of $(N-1)$-categories. This in particular implies that the hom-set between the objects are $(N-1)$-categories themselves. This definition resonates with physical intuition above, portraying a network of topological defects as a system of TFTs and interfaces coupled across various dimensions.

Moving further, let us delve into the structure encoded in a fusion $(D-1)$-category. We will expand the following descriptions in Section \ref{sec:category}. Considering the $D=2$ case, two objects can be fused, which can be thought of the operator-product-expansion (OPE) of the topological defects,
\begin{equation}
(\ob_i,\ob_j)\rightsquigarrow \ob_i\otimes \ob_j = \sum_{\ob_k: \,\text{simples}} N_{ij}^k \, \ob_k\,.
\end{equation}
Usually, it is understood that the fusion coefficient is a non-negative integer $N_{ij}^k \in \mathbb{Z}_{\geq 0}$. However, a more appropriate perspective is to view it as a decoupled topological quantum mechanics, that is, a vector space: the QM Hilbert space of vacua that account for the multiple fusion channels (via topological junctions among the defects).\footnote{ \
For a symmetry in a fermionic system, $N_{ij}^k$ is a super vector space graded by fermion parity.
}
When $\ob_i,\ob_j$ are seen as quantum mechanics coupled to the bulk, it is natural to expect that such a decoupled factor might emerge upon fusion. The fusion product is associative, but up to the associator morphisms
\cite{Moore:1988qv}:
\begin{equation}
F_{\ob_1,\ob_2,\ob_3}:(\ob_1\otimes \ob_2) \otimes \ob_3 \stackrel{\cong}{\to} \ob_1\otimes (\ob_2 \otimes \ob_3)
\end{equation}
This symbol represents a linear map that operates on the line-defect quantum mechanics. By fixing the basis of the vector spaces, the associator can be represented as a matrix whose elements are commonly known as F-symbols. A crucial property of the associators is their compliance with the pentagon identity, which is expressed as the commutativity of the following diagram:
\bea
\xymatrix{
  & ((\ob_1 \otimes \ob_2) \otimes \ob_3 ) \otimes \ob_4  \ar@{..>}[dr]^{F_{\ob_1, \ob_2, \ob_3}} \ar@{..>}[dl]_{F_{\ob_1 \otimes \ob_2, \ob_3, \ob_4}} &  \\
 (\ob_1 \otimes \ob_2) \otimes (\ob_3 \otimes \ob_4) \ar@{..>}[d]_{F_{\ob_1, \ob_2, \ob_3 \otimes \ob_4}}  &  & (\ob_1 \otimes (\ob_2 \otimes \ob_3)) \otimes \ob_4 \ar@{..>}[d]^{F_{\ob_1, \ob_2 \otimes \ob_3, \ob_4}} \\
 \ob_1 \otimes ( \ob_2 \otimes (\ob_3 \otimes \ob_4))  & & \ob_1 \otimes ((\ob_2 \otimes \ob_3) \otimes \ob_4) \ar@{..>}[ll]^{F_{\ob_2, \ob_3, \ob_4}}
}
\eea
In the case of $D>2$, there is already an interesting structure contained in a single simple object.
Here, an object/topological surface is simple when it cannot be written as a direct-sum of the others (see Section~\ref{sec:category} for details). 
As we emphasized above, given a simple object $\ob\in \scat{}$, the endomorphism fusion $(D-2)$-category $\End_{\scat{}}(\ob,\ob)$ is not necessarily trivial, as it captures the symmetries of $\ob$.
Furthermore, a gaugeable part, i.e. an algebra object $\algA$ (with properties dependent on $\ob$), might exist in $\End_{\scat{}}(\ob,\ob)$.
In this scenario, a new object $\ob/\algA$ can be obtained by gauging/condensing $\algA$ on $\ob$.\footnote{ \
The existence of $\ob/\algA$ in $\scat{}$ given $\algA$ satisfying an appropriate axioms is called Karoubi-completeness\cite{Gaiotto:2019xmp,douglas2018fusion}, which we assume among the properties of symmetry categories.} 
In addition, we can gauge $A$ on a half of the worldvolume of $\ob$, resulting in a non-trivial hom-category $\Hom_{\scat{}}(\ob,\ob/\algA)$.

This highlights that for $N$-fusion categories, when $N>1$, we expect that  nontrivial 1-morphism between two simple objects may exist.
Moreover, we can also stack a $(D-1)$-dimensional TFT decoupled from the bulk onto a $(D-1)$-dimensional surface defect, and then gauge/condense a part of the symmetry of the stacked defect. 
We let $\lls \ob \rrs$ to denote the equivalence class of $\ob$ up to this procedure, and let $\scat{}/\mathord{\sim}$ be the set of equivalence classes.\footnote{ \
This is a slight abuse of the notation because it is implicit that we consider only simple objects before taking the equivalence classes.}

With the above caveat in mind, we can consider the fusion of two objects: 
\begin{equation}
(\ob_i,\ob_j)\rightsquigarrow \ob_i\otimes \ob_j = \sum_{\ob_k:\, \text{simples}} \mathcal{C}_{ij}^k \, \ob_k
\end{equation}
Here, as $\ob_i,\ob_j$ are $(D-1)$-dimensional TFTs coupled to the bulk, the fusion of these can emit a decoupled $(D-1)$-dimensional TFT, which corresponds to $\mathcal{C}_{ij}^k$. As we emphasized above the fusion channel isn't unique; it is dictated by gauging or condensation data on the symmetry defect.  This fusion process typically is not closed with a finite number of simple objects even when $\scat{}/\mathord{\sim}$ is finite.
\begin{equation}
\lls\ob_i\rrs\otimes \lls\ob_j\rrs = \sum_{\lls\ob_k\rrs} N_{ij}^k \, \lls\ob_k\rrs,
\end{equation}
In this case, the fusion coefficient $N_{ij}^k$ is a non-negative integer as in the $D=2$ case for the fusion 1-category.
In known cases in $D>2$,
this fusion product on $\scat{}/\mathord{\sim}$ is group-like \cite{Kaidi:2021xfk,Choi:2021kmx,Choi:2022zal,Kaidi:2022uux} even when $\scat{}$ is non-invertible.\footnote{\ In most cases the group $\scat{}/\mathord{\sim}$ is abelian. However there are examples in theories of class $\cS$ where such group can also be non-abelian \cite{Bashmakov:2022uek,Antinucci:2022cdi}. For symmetries obtained by gauging a non-normal subgroup \cite{Bhardwaj:2022yxj,Antinucci:2022eat}, there are non-invertible topological lines  which correspond to higher morphisms in the category.}
This implies that for a non-invertible defect $\ob$, its absolute square $\ob \otimes \overline{\ob}$ ($\overline{\ob}$ is the orientation reversal) is a condensation of some lower-dimensional operator which is in the connected component of the trivial operator $\unit$.
Specifically, for $\scat{}$ being the chiral symmetry in QED, we find $\scat{}/\mathord{\sim}= \mathbb{Q}/\mathbb{Z}$.\footnote{ \
We refer to \cite{Putrov:2022pua} for related study of the $\bQ/\bZ$ grading of the non-invertible chiral symmetry and some of its implications.
}

We also define associators in the same way as in the fusion 1-category case:
\begin{equation}
\AF_{\ob_1,\ob_2,\ob_3}:(\ob_1\otimes \ob_2) \otimes \ob_3 \stackrel{\cong}{\to} \ob_1\otimes (\ob_2 \otimes \ob_3)
\end{equation}
The F-symbol now is given by a $(D-2)$-dimensional TFT, which specifies the associator interface between the $(D-1)$-dimensional defects. To be precise, to extract a TFT, we have to  specify the representatives from the equivalence classes and the gauging data at each junction. Also, it can contain $(D-2)$-dimensional topological operators that exists in the theory. This will be the one of the points emphasised in Section \ref{sec:Fsymbols}.

The next level introduces the pentagonator morphism. When $D>2$, F-symbols are TFTs rather than numbers, thus potentially possessing nontrivial self-interfaces on their worldvolume. Consequently, consideration of the pentagon diagram induces an additional data, called the ``pentagonator'' in \cite{douglas2018fusion}, instead of a constraint on the structures at lower levels. The structure continues to higher associators. If we name the usual associators 1-associators and the pentagonators 2-associators, a $(D-1)$-category will have up to $(D-1)$-associators, followed by identities among $(D-1)$-associators. When $\scat{}$ is invertible, the $(D-1)$-associators are nothing but the familiar 't Hooft anomalies.

An immediate physical implication of the higher structure we study explicitly here is that the associator interfaces (morphisms) can be detected by consideration of extended charged (non-topological) operators probing a network of topological defects. 
Another physical consequence of the associators are generalized Ward identities on spacetimes with non-trivial topologies, which allow us to extract this higher structure from topological manipulations of the defect network in a given correlator.

In the ensuing parts of this paper, we spell out the above structure in greater detail. In particular this paper is organized as follows. In Section \ref{sec:category} we spell out the structure and physical meaning of the higher categories briefly introduced above, reviewing some known examples using our notation. In Section \ref{sec:associati} we explicitly describe some layers of the aforementioned structure in the case of the non-invertible chiral symmetry. In particular we spell out the structure of one-morphisms and the F-symbols. Our goal is to uncover a detailed picture of the higher structures in the chiral symmetry and their interconnections.
In Section \ref{sec: Ward} we discuss the derivation of Ward identities on general four-manifolds. Our perspective relies on the surgery presentation of four-dimensional manifolds and should allow to streamline the derivation of such identities for more general (categorical) symmetries.

Our results give a first hint on how to concretely describe the higher structure of general non-invertible symmetries in higher dimensions, without resorting to an underlying group structure. It would be desirable to generalize our formalism further. A natural direction is to extend our analysis to more examples of non-invertible symmetries both in four and in other dimensions. It is likely that the non-invertible symmetry structure of axion electrodynamics for example might have a richer higher structure \cite{Choi:2022fgx,Yokokura:2022alv}. Similarly, it would be interesting to explore non-invertible symmetries in five and six dimensions from this perspective \cite{Damia:2022bcd}. Another natural future objective is to clarify and spell out the structure of 't Hooft anomalies for these symmetries and how to efficiently extract them from the higher structure. Finally, 
it would also be very intriguing to recast these data in the language of the Symmetry TFT \cite{Freed:2012bs,Gaiotto:2020iye,Apruzzi:2021nmk,Freed:2022qnc,Kaidi:2022cpf}.

\section{Categorical Aspects of Symmetries}
\label{sec:category}

Generalized global symmetries in QFT are associated with  topological operators (defects).  
The main purpose of this section is to discuss general structure of topological operators in QFT from the categorical point of view. Here we set up the notation for the subsequent case study in Section~\ref{sec:associati}, and along the way
 we extend and complement several aspects in the current literature  on the subject.

 In what follows, we take $\mathcal T$ to be a $D$-dimensional unitary Poincar\'e invariant QFT.
We denote its symmetry (sub)category by $\scat{}$, which encodes all axiomatic properties of a generalized global symmetry, and therefore we will often refer to $\scat{}$ simply as a symmetry of $\cT$. We denote a topological operator of codimension $q+1$ by $\cD^{(q)}$ which generates a (potentially non-invertible) $q$-form global symmetry.
Unless otherwise specified, we will work with the Euclidean QFT $\mathcal T$ on a closed oriented $D$-dimensional spacetime manifold $X$ without torsion.\footnote{\ Dropping these assumptions on $X$ is possible, but it requires some extra technicalities which we want to suppress in what follows.}

\subsection{Topological Defects}

We start with a quick recap of the topological defects and their salient properties in QFT.
Slightly abusing the notation, we will use $\scat{}$ to also denote the set of topological defects, which is naturally graded by their codimensions as below,
\be\label{eq:scat}
\scat{} = 
\left(\scat{(0)} , \scat{(1)} ,\dots  ,  \scat{(D-1)}\right)\,, 
\quad \cD^{(q)}\in \scat{(q)}\,.
\ee

\paragraph{Ward identities from isotopy invariance} The Ward identities for the generalized symmetry are encoded in the topological property of $\cD^{(q)}$. Given a general correlation function in $\mathcal T$ with the defect $\ob^{(k_i)} \in \scat{(k_i)}$ inserted along an oriented connected $(D-k_i-1)$-dimensional submanifold $\sup_i \subseteq X$, which we denote as 
\be
\langle \ob^{(k_i)}(\sup_i) \cdots \rangle\,,
\label{topset}
\ee
where the dots stand for possible additional operators (which can be non-topological in general) inserted away from $\Sigma_i$, the dependence on the support $\Sigma_i$ is \textit{quasi-topological}, meaning that it depends only on the homotopy type of $\Sigma_i \subset X$ (i.e. isotopy) up to contact terms between the topological defect and charged operator insertions, ultimately responsible for the action of the symmetry (and Ward identities). 
We will review more about the relation between topological operators and Ward identities in Section \ref{sec: Ward}.

\paragraph{Additive structure}
Topological defects at a given codimension have a natural additive structure. Given $\ob_1^{(q)},\ob_2^{(q)}\in \scat{(q)}$, their direct sum $\ob_1^{(q)} \oplus \ob_2^{(q)} \in \scat{(q)} $ is defined such that a general correlation function splits into a sum as follows,
\be
\langle (\ob_1^{(q)} \oplus \ob_2^{(q)} )(\sup) \cdots \rangle = \langle \ob_1^{(q)}(\sup) \cdots \rangle + \langle\ob_2^{(q)}(\sup) \cdots \rangle\,.
\ee

\paragraph{Simple topological defects} A topological defect $\cD^{(q)} \in \scat{(q)}$ is simple if it cannot be written as a direct sum of other topological defects in $\scat{(q)}$. Equivalently, a simple topological defect does not host point-like topological operators on its support.\footnote{\ The presence of a non-trivial point-like topological operator $\cO_\ob$ on the worldvolume of the topological defect $\cD$ implies that the defect Hilbert space has multiple vacua on any closed spatial manifold, which can be constructed using Euclidean path integral on a semi-infinite cylinder with insertions of $(\cO_\ob)^n$, thus leading to multiple superselection sectors in the limit of infinite spatial volume. With a suitable basis for the topological local operators, this 
amounts to the decomposition of a generic topological defect into its simple components.} A semisimple topological defect is a direct sum of the simple ones.

\paragraph{Trivial defect} 
There is a trivial defect $\id$ in each codimension, such that
\be
\langle \id^{(q)}(\Sigma)\cdots\rangle
=\langle \cdots\rangle
\ee
for all correlators and all $(D-q-1)$-dimensional submanifolds $\Sigma$.

\paragraph{Dual} 
For each topological defect $\cD^{(q)}\in \scat{(q)}$, we define its dual defect $\bar\cD^{(q)}$ as its orientation reversal. Explicitly, this means correlation functions that involve the dual $\bar\cD$ inserted on $\Sigma$ is equivalent to that with the defect $\cD$ inserted on the orientation-reversed support $\bar\Sigma$,
\be
\langle  \bar\cD(\Sigma) \cdots\rangle  = \langle  \cD(\bar\Sigma)  \cdots\rangle\,,
\ee
up to potential orientation-reversal anomalies (see for example \cite{Chang:2018iay,Cordova:2019wpi}).

\paragraph{Twisted sectors}
The topological defects not only act on charged operators, they also give rise to twisted sectors. For example, the $\ob^{(q)}$-twisted sector contains $(D-q-2)$-dimensional operators on which the topological defect $\ob^{(q)}$ can end. For $D=2$ and $q=0$, these are the usual twist fields (vertex operators). More generally the twisted sector on a spacetime submanifold $\Sigma$ which we denote as $T_{\cD_1,\cD_2,\dots,\cD_n}(\Sigma)$ is specified by a collection of co-oriented topological defects $\cD_i$ that intersect at $\Sigma$.
The twisted sector operators are generally non-topological defect operators. 

\paragraph{Topological junctions} The topological defects can form topological junctions, which can be thought of as the subset of topological operators $V_{\cD_1,\cD_2,\dots,\cD_n}(\Sigma)\subset T_{\cD_1,\cD_2,\dots,\cD_n}(\Sigma)$ in the corresponding twisted sector. In particular, this includes the topological interface between a pair of topological defects of the same codimension. In dimension $D=2$, the topological junctions between topological defect lines are topological point operators. In general $D>2$, there can be further topological junctions among the junctions of lower codimensions.
These topological junctions 
give rise to a network of defects that is isotopy invariant and describes the most general symmetry background associated with $\scat{}$ for the QFT $\cT$.

\paragraph{Genuine vs non-genuine topological defects}
For later convenience, we introduce the notion of genuine and non-genuine topological defects as in \cite{Bhardwaj:2022yxj}. The difference is the latter are only defined on the worldvolume of other topological defects of lower codimensions. More precisely, non-genuine topological defects are topological junctions that involve non-identity topological defects.

\paragraph{General symmetry enriched observable}
The most general observable in the QFT $\cT$ enriched by the symmetry $\scat{}$ is the correlation function that involves a topological network of   defects $\ob^{(q)}$ from $\scat{}$ that ends on operators in the twisted sector, in the presence of additional operators from the untwisted sector.\footnote{\
For a TFT, the axiomatic treatment of such general correlators are exposed in e.g.\ \cite{Carqueville:2018sld}.
}

\paragraph{Locality} The general symmetry enriched observables are subject to stringent constraints from locality, namely the consistency under cutting and gluing. In particular, the defect extended in time gives rise to a modified Hilbert space, known as the defect Hilbert space.

\paragraph{Parallel fusion}
For each fixed codimension, given two topological defects $\ob_1^{(q)}$ and $\ob_2^{(q)}$ in $\scat{(q)}$, we can form a third one $\ob_1^{(q)}\otimes \ob_2^{(q)}$ in $\scat{(k)}$ with the same support by parallel fusion:
\be
\begin{gathered}
\begin{tz}[td,scale=3]
\begin{scope}[yzplane=-0.5, on layer=superfront]
\draw[slice] (-1,0) to (1,0) to (1,\h) to (-1,\h) to (-1,0);
\node[left] at (-1,0.3) {$\ob_2^{(k)}(\sup)$};
\end{scope}
\begin{scope}[yzplane=0.5]
\draw[slice] (-1,0) to (1,0) to (1,\h) to (-1,\h) to (-1,0);
\node[left] at (-1,0.9*\h) {$\ob_1^{(k)}(\sup)$};
\end{scope}
\end{tz}
\end{gathered} = \begin{gathered}
\begin{tz}[td,scale=3]
\begin{scope}[yzplane=0, on layer=superfront]
\draw[slice] (-1,0) to (1,0) to (1,\h) to (-1,\h) to (-1,0);
\node[left] at (-1.2,\h/2) {$\ob_1^{(k)} \otimes \ob_2^{(k)}(\sup)\quad$};
\end{scope}
\end{tz}
\end{gathered} 
\ee
which holds as an operator equation in any correlation function. 
The trivial defect $\id^{(k)}$ acts as an identity for this fusion operation.
Note that the resulting defect from fusion is in general a direct sum of simple defects and we refer to the summands as the fusion channels,
\be 
\ob_1^{(q)}\otimes \ob_2^{(q)} (\sup)=\sum_{i} 
 {\cal C}^{(q)}_i(\sup) \ob_i^{(q)}(\sup)\,,
 \label{parallelfusion}
\ee
and the ``coefficient'' in each fusion channel (${\cal C}^{(q)}_i(\sup)$ above) is a decoupled $D-q-1$-dimensional TFT. The fusion product is associative but in general non-commutative for $q=0$, though in this work we will only consider topological defects that obey commutative fusion rules. We emphasize that although it is physically expected, the way associativity manifest itself in the structures of generalized symmetry in $D>2$ is rather nontrivial and a key point of this work is to elucidate this property in the concrete example of non-invertible chiral symmetry in $D=4$.

\paragraph{Partial fusion} Instead of the parallel fusion described above, thanks to their topological property, we can consider the local or partial fusion of a pair of topological defects $\cD_1^{(q)},\cD_2^{(q)}\in \scat{(q)}$ on an open submanifold $\Xi$ of codimension $q+1$, creating a $D-q-2$-dimensional topological junction located at $\partial \Xi$ among $\cD_1^{(q)},\cD_2^{(q)}$ and  the defect $\cD_i^{(q)}$ that appears in the parallel fusion \eqref{parallelfusion}. 
This generalizes in an obvious way to the partial fusion of more than two topological defects.

\paragraph{Symmetry fractionalization} 
The intersection with a topological defect $\ob^{(k)}$ can induce a topological interface between topological defects in $\scat{(q)}$, giving rise to an action of $\scat{(k)}$ on $\scat{(q)}$. 
One way to see this is to consider the configuration of $\ob^{(k)}$ wrapping another topological defect $\ob_1^{(q)}\in \scat{(q)}$ extended on $\Sigma$ with $q>k$. 
If we collapse the wrapping configuration everywhere onto $\Sigma$, we produce another topological defect $\ob_2^{(q)}\in \scat{(q)}$. 
Instead, if we only collapse the wrapping configuration over an open region $\Sigma'\subset \Sigma$ in the support of $\ob_1^{(q)}$, we create a topological interface induced by $\ob^{(k)}$ located at $\partial \Sigma'\subset \Sigma$ that separates  $\ob_1^{(q)}$ and $\ob_2^{(q)}$. See Figure~\ref{fig:fractionalization} for an example.

\begin{figure}[!htb]
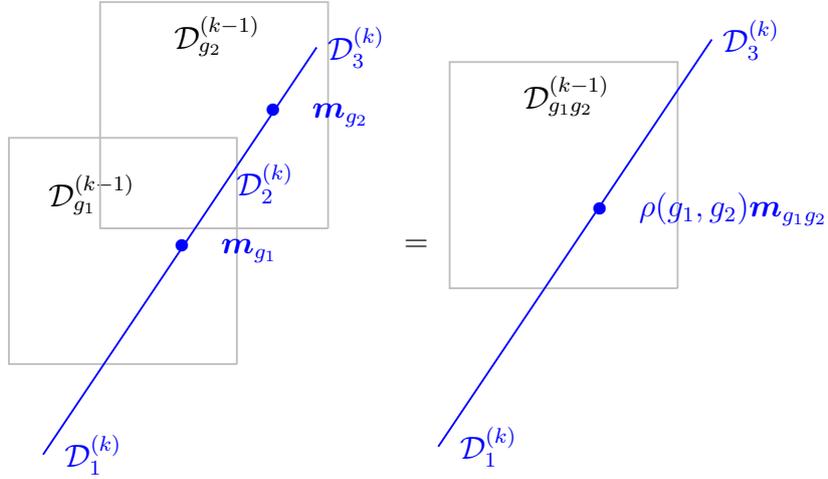

\begin{center}
$$
\begin{gathered}
\begin{tz}[td,scale=3]
\begin{scope}[yzplane=2]
\draw[slice] (0.5,-1) to (2.5,-1) to (2.5,1) to (0.5,1) to (0.5,-1) ;
\node[right] at (1.15,0) {\color{blue}{$\bullet \quad \mor_{g_1}$}};
\node[left] at (1.3,0.5) {$\ob_{g_1}^{(k-1)}$};
\end{scope}
\begin{scope}[yzplane=4]
\draw[slice] (0.5,-1) to (2.5,-1) to (2.5,1) to (0.5,1) to (0.5,-1) ;
\node[right] at (1.15,0) {\color{blue}{$\bullet \quad \mor_{g_2}$}};
\node[left] at (1,0.7) {$\ob_{g_2}^{(k-1)}$};
\end{scope}
\begin{scope}[xyplane=0]
\draw[bluewire] (-1,1) to (5,1);
\node[right] at (-1,1) {\color{blue}{$\,\,\ob_1^{(k)}$}};
\node[right] at (3,1) {\color{blue}{$\ob_2^{(k)}$}};
\node[right] at (5,1) {\color{blue}{$\ob_3^{(k)}$}};
\end{scope}
\end{tz}
\end{gathered} = \begin{gathered}\begin{gathered}
\begin{tz}[td,scale=3]
\begin{scope}[yzplane=3]
\draw[slice] (0.5,-1) to (2.5,-1) to (2.5,1) to (0.5,1) to (0.5,-1) ;
\node[right] at (1.35,-0.3) {\color{blue}{$\bullet \quad \rho(g_1,g_2) \mor_{g_1g_2}$}};
\node[left] at (1,0.7) {$\ob^{(k-1)}_{g_1 g_2}$};
\end{scope}
\begin{scope}[xyplane=0]
\draw[bluewire] (-1,1) to (5,1);
\node[right] at (-1,1) {\color{blue}{$\,\,\ob_1^{(k)}$}};
\node[right] at (5,1) {\color{blue}{$\ob_3^{(k)}$}};
\end{scope}
\end{tz}
\end{gathered}
\end{gathered}
$$
\caption{Illustration of mixed 't Hooft anomaly: in this example we have a group symmetry $\scat{(k-1)} = G^{(k-1)}$ that induces morphism on $\scat{(k)} = A^{(k)}$: the latter are fully specified by additional data. In this example, a class $\rho$ in $H^2(BG, \widehat{A})$, where $\widehat{A}$ is the Pontryagin dual group of $A$}\label{fig:mixedthooft}.
\end{center}
\end{figure}

That this induced interface (or action of $\scat{(k)}$ on $\scat{(q)}$)  is part of the definition of the symmetry was remarked recently \cite{Yu:2020twi,Delmastro:2022pfo}. It is interesting to note that the representation of symmetry defects in $\scat{(k)}$ along the higher codimensional defects in $\scat{(q)}$ can be projective (fractionalized) in general, and hence needs further additional data to be fully characterized. Such data are known as symmetry fractionalization classes and have been studied in \cite{PhysRevB.62.7850,Essin:2013rca,Barkeshli:2014cna,Chen_2015,Tarantino_2016,Yu:2020twi,Bulmash:2021hmb,Delmastro:2022pfo}. 
See Figure \ref{fig:mixedthooft} for an example where a mixed 't Hooft anomaly is responsible for inducing this action.
\begin{figure}[!htb]
\begin{center}
$$
\begin{gathered}
\begin{tz}[td,scale=3]
\begin{scope}[yzplane=\h/2]
\draw[slice] (0.5,-1) to (2.5,-1) to (2.5,1) to (0.5,1) to (0.5,-1) ;
\node[right] at (1.55,0) {\color{blue}{$\bullet \quad \mor$}};
\node[left] at (1,0.7) {$\ob^{(k)}$};
\end{scope}
\begin{scope}[xyplane=0]
\draw[bluewire] (-1,2) to (3,1);
\node[right] at (-1,2) {\color{blue}{$\,\,\ob_1^{(q)}$}};
\node[right] at (3,1) {\color{blue}{$\ob_2^{(q)}$}};
\end{scope}
\end{tz}
\end{gathered}
$$
\caption{Illustration of an induced topological interface ${\bm m}$ from a topological defect in $\scat{(q)}$ on $\scat{(\ell)}$.}\label{fig:fractionalization}
\end{center}
\end{figure}

\subsection{Decoupled TFTs and  Equivalent Classes of Topological Defects}\label{sec:witteq}

Although in $D=2$ QFTs, where in the absence of continuous symmetries, there are often a finite number of simple topological defect lines (i.e. $\ob^{(q)}$ $q=0$), the simplicity constraint is rather weak in higher dimensions. This is due to the large zoo of TFTs (with no topological point-like operator) in dimension $d>1$ that can be stacked on topological defects in the same dimension. In particular, all TFTs of dimension $d\leq D-1$ define topological defects in the $D$-dimensional QFT by stacking on the trivial defect ${\bm 1}^{(q)}$ with $d=D-q-1$. Nonetheless these topological defects are physically uninteresting as they do not interact with the rest of the QFT. More precisely, their structure cannot be detected by other observables in theory, and  are thus \textit{decoupled}.
 
This motivates us to define the 
equivalence classes $\lls \ob^{(q)}\rrs$ of topological defects $\ob^{(q)}$ where the equivalence relation comes from topological operations that involve stacking decoupled TFTs and 
discrete gauging of non-anomalous 
decoupled symmetries on their worldvolume. 

To be concrete, let us focus on the case of codimension-one topological defects (i.e. $q=0$) in $D=4$. This includes  the example of non-invertible chiral symmetry in $D=4$ which we study in detail in the next section, where each codimension-one topological defect has an explicit definition that involves a coupled $d=3$ TFT $\cA$ on its worldvolume. However the choice of $\cA$ is not unique, which reflects the above-mentioned equivalence relation. In particular, the equivalence relation from discrete gauging leads to equivalent classes that generalizes the Witt classes of 3d TFTs (i.e. MTCs), which we refer to as the \textit{twisted} Witt classes $[\cA,\mathbb{B}]$, where the twist (abstractly denoted by $\mathbb{B}$) keeps track of the coupling of this TFT to the bulk. For mathematical treatment of untwisted Witt class, see \cite{davydov2013witt}. In Section~\ref{sec:associati}, we discuss concrete examples of the twisted Witt classes relevant for the non-invertible chiral symmetry in $D=4$.

Physically, the conventional Witt equivalence relation $\cA_1 \sim \cA_2$ corresponds to a topological interface between the two 3d TFTs $\cA_1$ and $\cA_2$. Here the twisted Witt equivalence relation is required to also preserve the coupling to other topological defects in the bulk. Equivalently, this means the interface implementing this equivalence relation is neutral under bulk symmetries. As is with the conventional Witt classes, the twisted Witt classes form an abelian group, where the product structure comes from stacking the TFTs and taking the diagonal coupling to the bulk, and the inverse  comes from orientation reversal. We denote the conventional Witt group by $\cW$ and the Witt group twisted by coupling to the bulk as $\cW_{\mathbb{B}}$. The former is a subgroup  $\cW\subset \cW_{\mathbb{B}}$ corresponding to decoupled TFTs on the defect worldvolume.
Then the equivalence classes $\lls \ob^{(0)}\rrs$ of topological defects in this context are given by elements in the quotient $\cW_{\mathbb{B}}/\cW$.

While in general there is no preferred choice for a representative for each equivalent class of topological defects, as we review in the case of the chiral symmetry, there is a natural
``gauge'' choice, where the representatives are given by worldvolume TFTs $\cA$ that are ``minimal'' (i.e. having the minimal total quantum dimension among all representatives).\footnote{\ This is not a complete ``gauge-fixing'' since it is insensitive to stacking with SPTs. We thank Pavel Putrov for related discussions.}

However, the fusion product between topological defects does not preserve this ``gauge'' in general, and to restore it leads to the TFT-valued fusion coefficients which captures the change in the representative for the equivalence class that appears in a fusion channel. In a similar way, this leads to TFT-valued higher structure ``constants'' for the symmetry, as we explain in Section~\ref{sec:associati}.

\subsection{Condensation Defects}

As pointed out in \cite{Kapustin:2010if, Gaiotto:2019xmp,Kong:2020cie,Roumpedakis:2022aik}, in a given QFT, a special family of topological defects in $\scat{(q)}$ can be identified as composites of topological defects in $\scat{(p)}$ of higher codimensions $p\geq q$, and are known as condensation defects. We denote them as $\cC^{(q)}$ to differentiate from more general topological defects in $\scat{(q)}$. We discuss their special properties below.

From a physics perspective,  inserting a condensation defect $\cC^{(q)}$ in a general QFT observable is equivalent to gauging a generalized symmetry $\algA^{(q)}$ on a codimension-$q+1$ submanifold, also known as higher-gauging \cite{Roumpedakis:2022aik}.

More explicitly,
 the condensation defects $\cC^{(q)}$ can be defined by choosing a triangulation $\Delta$ over the submanifold $\Sigma^{(q)}$ and placing a sufficiently fine mesh of higher codimension defects, which we denote as $\algA^{(q)}$, on $\Sigma^{(q)}$. Suppose that $\ell \geq q$ determines the largest codimension (i.e. $\ell+1$) of the topological defects that constitute $\algA^{(q)}$,   we start by tiling these topological defects on the $(\ell-q)$-faces of $\Delta$, and then decorate the $( \ell-q - 1)$-faces by topological junctions and continue this way until we reach the vertices. We require consistency conditions in the definition of $\algA^{(q)}$ ensuring that the defect is well-defined: it does not depend on local changes in the mesh nor on substituting it for a finer one.
When $\cond^{(q)}$ is built from topological defect lines only, the conditions on $\algA^{(q)}$ can be spelled out explicitly and define a symmetric Frobenius algebra \cite{Fuchs:2002cm}. We will give a precise characterization of these condensation defects relevant for the non-invertible chiral symmetry in Section~\ref{sec:associati}.

Owing to their composite nature, condensation defect cannot in general detect other higher codimensional insertions (topological or non-topological). Explicitly, let $\ell+1\geq q+1$ be the highest codimension of the constituent topological defects for the condensation defect $\cC^{(q)}$  represented by a topological mesh $\algA^{(q)}$ as above. Any insertion with codimension $>D-\ell$ (or, dimension $< \ell $)  can freely pass through $\algA^{(q)}$ by local topological deformations of the mesh. 
The condensation defect $\cond^{(q)}$ here thus defines a $\ell$-porous defect.

Relatedly, a topological defect $\ob^{(q)}$ is simple if the topological interfaces between $\ob^{(q)}$ and itself consist of $q+1$-porous condensation defects $\cond^{(q+1)}$.

Despite the porous feature,
it should be stressed that such condensation defects contain nontrivial physical information. For example they are known to generate all of the zero-form symmetries in 3d TFTs \cite{Roumpedakis:2022aik}.
Moreover, it has been recently appreciated and we will elaborate further in subsequent sections that, the condensation defects play an important role in the higher structures of the generalized symmetry.

In contrast to general topological defects, an important property of condensation defects $\cC^{(q)}$ is the existence of a universal topological boundary condition on which it can terminate. In terms of the higher gauging picture, this defines a Dirichlet-type boundary condition on the topological mesh (network) which corresponds to a topological twist defect $\boldsymbol{\tau}\in \scat{(q+1)}$ (i.e. a topological junction between $\cC^{(q)}$ and the trivial defect ${\bm 1}^{(q)}$).\footnote{\ Such construction has been recently used to describe self-duality defects in $d>2$ \cite{Kaidi:2022cpf,Antinucci:2022vyk,Bashmakov:2022uek,Antinucci:2022cdi}.}
 
The fact that condensation defects can end topologically implies that they are in the ``trivial'' equivalence class of topological defects introduced in the previous section. In particular, fusion with a condensation defect preserves the equivalence class. 

Condensation defects are particularly well understood in the context of 3d G-crossed Modular Tensor Categories\cite{Barkeshli:2014cna}. In this setup any symmetry defect $U_g$ for the $G$ zero-form symmetry can be described as a condensation defect for a symmetric Frobenius algebra $\salgA_g$\cite{Fuchs:2002cm,Kapustin:2010if,Roumpedakis:2022aik}. 
The statement is a consequence of the absence of local charged operators in the theory. This allows to open up holes in the symmetry defect at no cost. When local charged operators are present instead the twisted sector of the symmetry defect must be non-topological and the construction fails.
\bea
\begin{tz}
\filldraw[color=white!90!blue] (0,0) -- (0,2) -- (2,2) -- (2,0) -- cycle ;
\node[right] at (2.5,1) {\large $\overset{\text{open holes}}{\simeq}$} ;
\begin{scope}[shift={(5,0)}]
\filldraw[color=white!90!blue] (0,0) -- (0,2) -- (2,2) -- (2,0) -- cycle ;  
 \filldraw[color=white, fill=white] (0,0) -- (0.5,0)  arc (0:90:0.5) -- (0,0);
 \draw[color=red] (0.5,0)  arc (0:90:0.5);
 \filldraw[color=white] (2,0) -- (1.5,0)  arc (180:90:0.5 and 0.5) -- cycle;
 \draw[color=blue] (1.5,0) arc (180:90:0.5 and 0.5);
 \filldraw[color=white] (0,2) -- (0.5,2)  arc (0:-90:0.5 and 0.5) -- cycle;
 \draw[color=blue] (0.5,2)  arc (0:-90:0.5 and 0.5) ;
   \filldraw[color=white] (2,2) -- (1.5,2)  arc (180:270:0.5 and 0.5) -- cycle;
   \draw[color=blue] (1.5,2)  arc (180:270:0.5 and 0.5);
\node[right] at (2.5,1) {\large $\overset{\text{expand holes}}{\simeq}$} ;
\end{scope}
\begin{scope}[shift={(10,0)}]
\draw[slice, color=blue] (1,0) node[below] {$\salgA_g$} to[out=up, in=\dr] (0.75,0.75) node[dot] {} to (1.25,1.25) node[dot] {}; 
\draw[slice, color=blue] (1.25,1.25) to[out=\ul, in=down] (1,2);
\draw[slice,color=blue] (0,1) to[out=right, in =\ul] (0.75,0.75);  \draw[slice,color=blue] (2,1) to[out=left, in =\dr] (1.25,1.25);  
\end{scope}
\end{tz}
\eea
Through this manipulation, the symmetry defect can be reduced to a (consistent) 2d topological network of lines, which defines the condensation of an algebra $\salgA_g$.

\subsection{Objects and Higher Morphisms in the Higher Category} 

Having laid the physical foundations for the generalized symmetries and their higher structures in terms of the topological defects and their properties, here we discuss how these topological data is formalized in terms of the symmetry category. As we will see in Section~\ref{sec:associati}, the categorical perspective will provide us with the natural mathematical language to elucidate the higher structures of these symmetries.

We start by reviewing the most basic building blocks of the symmetry category $\mathfrak{C}$, namely the objects and the (higher) morphisms, which provide a different way to organize the topological defects, that naturally comes from the recursive definition of the higher category and allows us to analyze the higher structures step by step. As we will see, one upshot is that general topological defects are equivalent to (higher) morphisms and the corresponding coefficient rings are generated by TFTs.  

Focusing on the genuine topological defects in \eqref{topset}, we define $\ell+1$ as the lowest codimension of the defects.
The symmetry category $\mathfrak{C}$ is a fusion $N$-category with $N=D-\ell-1$. For $D=2$, the topological defect lines (i.e. $\ell=1$) generate a symmetry 1-category which is an ordinary fusion category whereas for $D>2$ the symmetry category is a higher category in general. For the chiral symmetry category which we discuss at length in Section~\ref{sec:associati}, $D=4$ and $\ell= 0$, hence the corresponding $\scat{}$ is a 3-category. 
The fusion $N$-category has a layered structure that echoes the structure of topological defects of different codimensions which form complex topological networks as discussed in the previous sections.

\paragraph{Objects} There are $N+1$ layers in the $N$-category, starting from the objects at the first (bottom) level, also known as the $0$-morphisms. They correspond to  topological defects $\cD^{(\ell)}$ which have the highest dimensions. The objects have an additive structure coming from the direct sum of the topological defects, with simple objects corresponding to simple defects $\cD^{(\ell)}$ and semi-simple objects from direct sums of finitely many simple objects. 

The parallel fusion of topological defects induces a tensor product structure on the objects, which we denote by $\otimes$. 

\begin{figure}
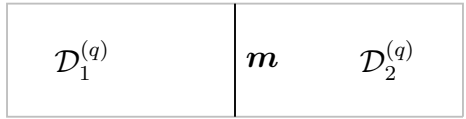

\begin{center}
$$
\begin{gathered}
\begin{tz}[td,scale=3]
\begin{scope}[yzplane=0, on layer=superfront]
\draw[slice] (-2,0) to (2,0) to (2,1) to (-2,1) to (-2,0);
\draw[wire] (0,0) to (0,1);
\node[right] at (0,0.5) {$\mor$};
\node[left] at (1,0.5) {$\ob_1^{(q)}$};
\node[right] at (-1,0.5) {$\ob_2^{(q)}$};
\end{scope}
\end{tz}
\end{gathered} 
$$
\caption{Morphism between two topological defects in $\scat{(q)}$.}\label{fig:morph}
\end{center}
\end{figure}

\paragraph{Morphisms}
At the second level of the $N$-category,
we have the $1$-morphisms between the objects, which corresponds to topological interfaces between a pair of topological defects $\ob_1^{(\ell)},\ob_2^{(\ell)}$ of the same codimension and thus topological defects in $\scat{(\ell+1)}$ (see Figure \ref{fig:morph}). For trivial $\ob_1^{(\ell)}=\ob_2^{(\ell)}={\bm 1}^{(\ell)}$, the 1-morphisms (endomorphisms of the identity defect) are genuine topological defects in $\scat{(\ell+1)}$. The direct sum for the topological defects induces an additive structure on $\scat{(\ell+1)}$.

  More explicitly, consider the two topological defects $\ob_1^{(\ell)}(\Sigma_1),\ob_2^{(\ell)}(\Sigma_2)$ inserted on two $(D-\ell-1)$-dimensional oriented submanifolds $\Sigma_1$ and $\Sigma_2$ of $X$ respectively which share a common $(D-\ell-2)$-dimensional boundary $\ssup$,
  \be
  \partial\sup_1 = \ssup\,,\quad \partial\sup_2 = \overline{\ssup}\,,
  \ee
  up to an orientation reversal. This topological configuration is specified by a choice of topological interface on $\ssup$, which corresponds to a 1-morphism
\be
\mor \in \text{Hom}(\ob_1^{(\ell)},\ob_2^{(\ell)})\,.
\ee
It give rise to correlators as the following form,
\be
\langle \ob_1^{(\ell)}(\sup_1)||\mor(\ssup)||  \ob_2^{(\ell)}(\sup_2) \cdots \rangle\,,
\ee
that depend only on the topology of the configuration up to homotopy and contact terms with charged operators, which now can also include further contact terms of the operators of $\mathcal T$ with the topological defect representing the morphism $\mor(\ssup)$. 

Topological junctions that connect multiple topological defects $\cD_i^{(\ell)}(\Sigma_i)$ at their common $(D-\ell-2)$-dimensional intersection $\ssup$ also correspond to 1-morphisms.
For example, in the case of three intersecting topological defects, each topological junction is specified by a 1-morphism,
\be
\mor \in \text{Hom}(\ob_1^{(\ell)} \otimes \ob_2^{(\ell)} ,\ob_3^{(\ell)})\,.
\ee

\paragraph{Higher morphisms} In the higher $N$-category $\scat{}$, the higher $n$-morphisms are defined recursively as morphisms between $(n-1)$-morphisms, and physically correspond to topological interfaces between a pair of topological defects in $\scat{(\ell+n-1)}$.
This produces further nested configurations of topological defects that decorate correlation functions in $\cT$. For instance, below is an example where a 2-morphism arises as an interface between 1-morphisms ${\bm m}_1$ and ${\bm m}_2$:
\be
\begin{gathered}
\begin{tz}[td,scale=3]
\begin{scope}[yzplane=0, on layer=superfront]
\draw[slice] (-2,0) to (2,0) to (2,\h) to (-2,\h) to (-2,0);
\draw[wire] (0,0) to (0,\h);
\node[right] at (0,0.8*\h) {$\mor_2$};
\node[right] at (0,0.2*\h) {$\mor_1$};
\node at (0,\h/2) {$\bullet$};
\node[right] at (0,\h/2) {$\mmor$};
\node[left] at (-1.5,\h/2) {$\ob_1^{(\ell)}$};
\node[right] at (1.5,\h/2) {$\ob_2^{(\ell)}$};
\end{scope}
\end{tz}
\end{gathered} 
\ee
Above we present a two-dimensional section of the full spacetime manifold $X$, while the remaining $D-\ell-3$ dimensions shared by all constituent defects are suppressed.

\begin{figure}[!htb]
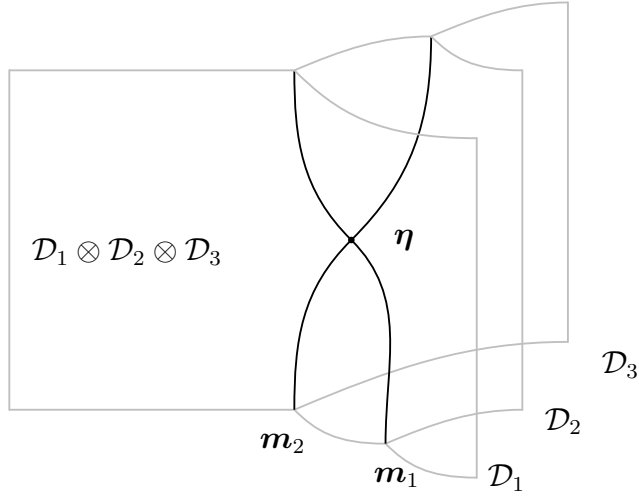

\begin{center}
\begin{tz}[td,scale=3]
\begin{scope}[xyplane=0]
\draw[slice] (0,0) to [out=up, in=\dl] (0.5,1) to [out=up, in=\dl] (1,2) to (1,4.5);
\draw[slice] (1,0) to [out=up, in=\dr] (0.5,1);
\draw[slice] (2,0) to [out=up, in=\dr] (1,2);
\end{scope}
\begin{scope}[xyplane=\h, on layer=superfront]
\draw[slice] (0,0) to [out=up, in=\dl] (1,2) to (1,4.5);
\draw[slice] (1,0) to [out=up, in=\dl] (1.5,1) to [out=up, in=\dr] (1,2);
\draw[slice] (2,0) to [out=up, in=\dr] (1.5,1);
\end{scope}
\begin{scope}[xzplane=0]
\draw[slice,short] (1,0) to (1,\h);
\draw[slice,short] (0,0) to (0,\h);
\draw[slice,short] (2,0) to (2,\h);
\end{scope}
\begin{scope}[xzplane=4.5]
\draw[slice,short] (1,0) to (1,\h);
\end{scope}
\coordinate (A) at (1.5,1,0.5*\h);
\draw[wire] (1,0.5,0) to [out=up, in=\dr] (A) to [out=\ur, in=down] (1,1.5,\h);
\draw[wire] (2,1,0) to [out=up, in=\dl] (A) to [out=\ul, in=down] (2,1,\h);
\node[dot] at (A){};
\node[right] at (A) {$\quad \mmor$};
\node[right] at (4,0,2) {$\ob_1 \otimes \ob_2 \otimes \ob_3$};
\node[right] at (-1,0,1) {$\ob_3$};
\node[right] at (-0.5,0,0.5) {$\ob_2$};
\node[right] at (0,0,0) {$\ob_1$};
\node[right] at (2,0,0.3) {$\mor_2$};
\node[right] at (1,0,0) {$\mor_1$};
\end{tz}
\end{center}
\caption{Associator 2-morphism $\mmor$ arising from a topological junction of 1-morphisms. This 2-morphism corresponds to an associator/F-symbol (see Section~\ref{sec:Fsymbols}).}\label{fig:eta}
\end{figure}

Similar remarks as in the case of 1-morphisms extend to higher morphisms. On the one hand, $p$-morphisms between topological defects in $\scat{(q)}$ give rise to special topological defects in $\scat{(k+q)}$. On the other hand, nested configurations of topological defects can give rise to $p$-morphisms that connect different partial fusion channels of the objects (see Figure \ref{fig:eta} for an example). In this paper we will encounter several examples of these, hence we will be brief here and refer the readers to Section~\ref{sec:associati} below.

\paragraph{Existence of the dual} All objects and morphisms in the symmetry category $\scat{}$ have a dual from the orientation reversal of the corresponding topological defect.
In $D=2$ --- that is, for 1-categories --- this is equivalent to the existence of evaluation $\ev$ and coevaluation $\coev$ maps  \cite{Bhardwaj:2017xup}
\be
\ev \; \in \Hom(\ob \otimes \overline{\ob}, \unit) \, , \ \ \ \coev \; \in \Hom(\unit, \ob \otimes \overline{\ob}) \,.
\label{evcoevmaps}
\ee 
For a self-dual object $\cN$, the composition of the evaluation and co-evaluation maps,
\be
\ev_\cN \circ \coev_\cN = \epsilon_\cN \; \text{id}_\cN \, ,
\ee
defines a ``gauge invariant'' datum $\epsilon_\cN = \pm 1$ of the defect $\cN$, which is called the Frobenius-Schur indicator \cite{Kitaev:2005hzj}.

For higher-categories in $D>2$, this notion of dual, while intuitively correct, needs to be slightly refined. Consider an object $\ob\in\scat{(k)}$ and denote its opposite $\overline{\ob}$. Roughly speaking the dual of a symmetry defect is obtained by folding $\ob$ on itself, reversing its orientation. 

This gives rise to a fusion channel that contains the identity. More precisely we require that for all $\ob^{(q)} \in \scat{(q)}$ there is a $\overline{\ob}^{(q)} \in \scat{(q)}$ such that
\be
\overline{\ob}^{(q)}(\sup) \otimes \ob^{(q)}(\sup)\cdots  =  \id^{(q)}(\sup) \otimes \cond_q(\sup)\cdots \oplus \dots\,,
\ee
where $\cond_q$ is a $q$-porous condensation defect. This requires the existence of evaluation and co-evaluation 1-morphisms, which we write as $\ev_{(0)}$ and $\coev_{(0)}$ and they correspond to the (co)evaluation maps \eqref{evcoevmaps} for $D=2$. This structure can be further extended to higher evaluation and co-evaluation $q+1$-morphisms  which we denote as $\ev_{(q)}$ and  $\coev_{(q)}$, for the $q$-morphisms. The physical intuition being that the orientation of these topological defects (topological junctions) can be reversed as an intrinsic operation over their worldvolume (e.g. by folding or bending the defect). In the same way, for self-dual topological defects in $\scat{(q)}$,
there should be a generalization of the Frobenius-Schur indicator for 1-categories to the higher categories that defines ``gauge invariants'' out of the higher (co)evaluation $q+1$-morphisms, in the form of a $D-q-2$-dimensional TFT.\footnote{\ One potential interesting case that arise in the $D=4$ non-invertible chiral symmetry is the time-reversal invariant topological defect $\Dchiral{1/2}\in \scat{(0)}$.}

\subsection{Associators (F-symbols)}\label{sec:Fsymbols}
In order to fully specify a symmetry category as a fusion $N$-category, we need to also include isomorphisms implementing associativity for the fusion of multiple objects in $\scat{}$.

The most basic example involves three objects $\cD_1,\cD_2,\cD_3$ (topological defects in $\scat{(\ell)}$) in two different fusion channels,
\be
\AF_{\ob_1,\ob_2,\ob_3} \in \Hom((\ob_1\otimes \ob_2)\otimes \ob_3, \ob_1\otimes(\ob_2\otimes \ob_3))
\label{eq: F-symbol morphism}
\ee
which, following \cite{Bhardwaj:2017xup}, we will refer to as the associator (morphism) for the tensor product. As a 1-morphism, $\AF_{\ob_1,\ob_2,\ob_3}$ can be determined in the following way.\footnote{\ In \eqref{eq: F-symbol morphism}, we regard the two objects $(\ob_1\otimes \ob_2)\otimes\ob_3$ and $\ob_1\otimes(\ob_2\otimes \ob_3)$ as different but equivalent objects. In the following we fix an equivalence between them and regard them the same objects, i.e. taking the skeleton of $\scat{}$.}

We start with the following diagram that describe a topological configuration obtained by partial fusion of the defects $\cD_1,\cD_2,\cD_3$, with fixed fusion channels (specified by the topological junctions represented by the solid vertical lines below)
\be
\begin{gathered}
\begin{tz}[td,scale=3]
\begin{scope}[xyplane=0, on layer=superfront]
\draw[slice] (0,0) to [out=up, in=\dl] (1,2) to (1,4);
\draw[slice] (1,0) to [out=up, in=\dl] (1.5,1) to [out=up, in=\dr] (1,2);
\draw[slice] (2,0) to [out=up, in=\dr] (1.5,1);
\draw[slice] (0,0) to [out=down, in=\ul] (0.5,-1) to [out=down, in=\ul] (1,-2) to (1,-4);
\draw[slice] (1,0) to [out=down, in=\ur] (0.5,-1);
\draw[slice] (2,0) to [out=down, in=\ur] (1,-2);
\end{scope}
\begin{scope}[xzplane=2]
\draw[wire,short] (1,0) to (1,\h);
\end{scope}
\begin{scope}[xzplane=1]
\draw[wire,short] (1.5,0) to (1.5,\h);
\end{scope}
\begin{scope}[xyplane=0.3, on layer=superfront]
\node[right] at (1,4.5) {$(\ob_1 \otimes \ob_2 \otimes \ob_3)$};
\node[right] at (1.3,2.25)  {$(\ob_2 \otimes \ob_3)$} ;
\node[right] at (0,0.75)  {$\ob_1$};
\node[right] at (1.9,0.25)  {$\ob_3$};
\node[right] at (-0.25,-1.5) {$(\ob_1 \otimes \ob_2)$} ;
\node[right] at (1,-2.4) {$(\ob_1 \otimes \ob_2 \otimes \ob_3)$};
\node[right] at (1,0.4) {$\ob_2$};
\end{scope}
\begin{scope}[xyplane=\h, on layer=superfront]
\draw[slice] (0,0) to [out=up, in=\dl] (1,2) to (1,4);
\draw[slice] (1,0) to [out=up, in=\dl] (1.5,1) to [out=up, in=\dr] (1,2);
\draw[slice] (2,0) to [out=up, in=\dr] (1.5,1);
\draw[slice] (0,0) to [out=down, in=\ul] (0.5,-1) to [out=down, in=\ul] (1,-2) to (1,-4);
\draw[slice] (1,0) to [out=down, in=\ur] (0.5,-1);
\draw[slice] (2,0) to [out=down, in=\ur] (1,-2);
\end{scope}
\begin{scope}[xzplane=-1]
\draw[wire,short] (0.5,0) to (0.5,\h);
\end{scope}
\begin{scope}[xzplane=-2]
\draw[wire,short] (1,0) to (1,\h);
\end{scope}
\end{tz}
\end{gathered}
\label{eq:FsymbolD1D2D3}
\ee
where $(\ob_1 \otimes \ob_2)$ denotes a fixed constituent object in the fusion product $\ob_1 \otimes \ob_2$ and so on.
Shrinking the cylindrical configuration in the middle defines a 1-morphism:
\be\label{eq:associator0}
\begin{gathered}
\begin{tz}[td,scale=3]
\begin{scope}[yzplane=0]
\draw[slice] (-3.3,0) to (-3.3,\h/2) to (3.3,\h/2) to (3.3,0) to (-3.3,0) ;
\end{scope}
\draw[wire] (0,0,0) to (0,0,\h/2);
\node[right] at (0,0,0.3*\h) {$\AF_{\ob_1,\ob_2,\ob_3}$};
\node[left] at (1.25,0,\h/8) {$(\ob_1 \otimes \ob_2 \otimes \ob_3)$};
\node[right] at (-1.25,0,\h/8) {$(\ob_1 \otimes \ob_2 \otimes \ob_3)$};
\end{tz}
\end{gathered} 
\ee
In $D=2$, this produces the standard $F$-symbol \cite{Moore:1988qv}. Therefore, we will also refer to the associators in higher $D>2$ as F-symbols. 

In higher categories, the two decompositions of $\ob_1 \otimes \ob_2 \otimes \ob_3$ into $(\ob_1 \otimes \ob_2 \otimes \ob_3)$ via different fusion channels are related by a 2-morphism:
\be
\xymatrix{
&\ob_1 \otimes \ob_2 \otimes \ob_3\ar@{..>}[dl]_{J_{12}}\ar@{..>}[dr]^{J_{23}}&\\
(\ob_1 \otimes \ob_2) \otimes \ob_3 \ar@{..>}[dr]_{J_{(12)3}}\ar@{->}[rr]^{\mmor_{\ob_1,\ob_2,\ob_3}} && \ob_1 \otimes (\ob_2 \otimes \ob_3)\ar@{..>}[dl]^{J_{1(23)}}\\
& (\ob_1 \otimes \ob_2 \otimes \ob_3)
}
\ee
where the dashed arrows represent fusions, while the solid arrow represents the 2-morphism:
\be
J_{(12)3} \circ J_{12} \xrightarrow{\quad \mmor_{123}\quad} \, J_{1(23)} \circ J_{23} \,,
\ee

The relation between $\AF_{\ob_1, \ob_2, \ob_3}$ and $\mmor_{\ob_1, \ob_2, \ob_3}$ follows from applying $\check{\mmor}_{\ob_1, \ob_2, \ob_3} \equiv \mmor_{\ob_1, \ob_2, \ob_3} \circ e_{\ob_3} \circ e_{(\ob_1 \otimes \ob_2 \otimes \ob_3)}$ on top of the diagram \eqref{eq:FsymbolD1D2D3} and shrinking the middle part of the diagram. On the top part we get the identity morphism $(\ob_1 \otimes \ob_2 \otimes \ob_3)  \xrightarrow{\unit} (\ob_1 \otimes \ob_2 \otimes \ob_3) $, while on the bottom we have the morphism from the associator $(\ob_1 \otimes \ob_2 \otimes \ob_3) \xrightarrow{\AF_{\ob_1, \ob_2, \ob_3}} (\ob_1 \otimes \ob_2 \otimes \ob_3)$. The two are connected by the two-morphism $\check{\mmor}_{\ob_1, \ob_2, \ob_3}$,
\be\label{eq:associator}
\begin{gathered}
\begin{tz}[td,scale=3]
\begin{scope}[yzplane=0]
\draw[slice] (-3.3,0) to (-3.3,\h/2) to (3.3,\h/2) to (3.3,0) to (-3.3,0) ;
\end{scope}
\draw[wire] (0,0,0) to (0,0,\h/4);
\node[dot] at (0,0,\h/4){0.1};
\node[above] at (0,0,\h/4) {$\check{\mmor}_{\ob_1, \ob_2, \ob_3}$};
\node[right] at (0,0,\h/8) {$\AF_{\ob_1,\ob_2,\ob_3}$};
\node[left] at (2,0,3*\h/8) {$(\ob_1 \otimes \ob_2 \otimes \ob_3)$};
\node[right] at (-2,0,3*\h/8) {$(\ob_1 \otimes \ob_2 \otimes \ob_3)$};
\end{tz}
\end{gathered} 
\ee

\medskip
In physical terms we should think of $\AF_{\ob_1, \ob_2, \ob_3}$ as an topological interface on the topological defect $(\ob_1 \otimes \ob_2 \otimes \ob_3)$, while $\check{\mmor}_{\ob_1, \ob_2, \ob_3}$ is a Dirichlet-type (topological) boundary condition for $\AF_{\ob_1, \ob_2, \ob_3}$.

It is interesting to remark following Douglas and Reutter \cite{douglas2018fusion} that we have a representation of the associator as a 2-morphism (see Figure \ref{fig:eta}). In the configuration in equation \eqref{eq:associator} above we are computing a 1-morphism which is capturing the image of $\mmor$.

\subsection{Higher Associators and Anomalies} 

Anticipating a generalization in higher categories,
we refer to the associator 1-morphism discussed in the last section as the 1-associator. 
For a higher $N$-category we can proceed and construct further higher associators as follows. These are obtained by considering the consecutive junction of morphisms implementing the associativity for a higher number of objects. It is interesting to remark that the consistency of these higher diagrams gives rise to stringent conditions. Let us consider for instance the case of the associativity condition on the fusion of four objects. In that case, one obtains the following cubic diagram (the so-called pentagonator):\footnote{\ For simplicity of this illustration, we focus on the case where the fusion among the equivalence classes of topological defect is group-like. The more general case is similar but one has to keep track of the multiple fusion channels which complicates the diagram.}
\be
\begin{gathered}
\xymatrix{
&\ob_1 \otimes \ob_2 \otimes \ob_3\otimes \ob_4\ar@{..>}[ddl]\ar@{..>}[dd]\ar@{..>}[ddr]&\\
&&\\
(\ob_1 \otimes \ob_2) \otimes \ob_3 \otimes \ob_4 \ar@{..>}[dd]\ar@{..>}[ddr] \ar@[red][r]^{\textcolor{red}{\mmor_{123}}}& \ob_1 \otimes(\ob_2 \otimes \ob_3) \otimes \ob_4 \ar@{..>}[ddr]\ar@{..>}[ddl] \ar@[red][r]^{\textcolor{red}{\mmor_{234}}}& \ob_1 \otimes \ob_2 \otimes (\ob_3 \otimes \ob_4)\ar@{..>}[dd]\ar@{..>}[ddl]\\
&&\\
(\ob_1 \otimes \ob_2 \otimes \ob_3) \otimes \ob_4\ar@{..>}[ddr] \ar@[red][r]^{\textcolor{red}{\mmor_{(12)34}}}\ar@[red]@/^2.0pc/[rr]^{\textcolor{red}{\mmor_{1(23)4}}}& (\ob_1 \otimes \ob_2) \otimes (\ob_3 \otimes \ob_4)\ar@{..>}[dd]\ar@[red][r]^{\textcolor{red}{\mmor_{12(34)}}} & \ob_1 \otimes (\ob_2 \otimes \ob_3 \otimes \ob_4)\ar@{..>}[ddl]\\
&&\\
&(\ob_1\otimes\ob_2\otimes \ob_3\otimes \ob_4)&\\
}
\end{gathered}
\ee
Each face of the cube but the top one corresponds to a non-trivial 1-associator. The top face of the cube is associated to the trivial identity
\be
\xymatrix{
&\ob_1 \otimes \ob_2 \otimes \ob_3 \otimes \ob_4\ar@{..>}[dl]\ar@{..>}[dr]&\\
(\ob_1 \otimes \ob_2) \otimes \ob_3 \otimes \ob_4 \ar@{..>}[dr]\ar@{->}[rr]^{\equiv} && \ob_1 \otimes \ob_2 \otimes (\ob_3\otimes \ob_4) \ar@{..>}[dl]\\
&(\ob_1 \otimes \ob_2) \otimes (\ob_3 \otimes \ob_4)
}
\ee
The possible compositions of dashed paths from the top to the bottom of this figure correspond to inequivalent ways of ordering the junction morphisms, and we can construct several identities among them which corresponds to the various 1-associators:
\be\label{eq:3way}
\begin{aligned}
J_{(123)4} \circ J_{(12)3} \circ J_{12} &\textcolor{red}{\xrightarrow{\quad\mmor_{123}\quad}}\, J_{(123)4} \circ J_{1(23)} \circ J_{23}\\
&\textcolor{red}{\xrightarrow{\quad\mmor_{1(23)4}\quad}}\, J_{1(234)} \circ J_{(23)4} \circ J_{23}\\
&\textcolor{red}{\xrightarrow{\quad\mmor_{234}\quad}}\, J_{1(234)} \circ J_{2(34)} \circ J_{34}\\
\end{aligned}
\ee
and
\be\label{eq:2way}
\begin{aligned}
J_{(123)4} \circ J_{(12)3} \circ J_{12} &\textcolor{red}{\xrightarrow{\quad\mmor_{(12)34}\quad}}\, J_{(12)(34)} \circ J_{34} \circ J_{12}\\
&\textcolor{red}{\xrightarrow{\quad\equiv\quad}}\, J_{(12)(34)}  \circ J_{12}\circ J_{34}\\
&\textcolor{red}{\xrightarrow{\quad\mmor_{12(34)}\quad}}\, J_{1(234)}  \circ J_{2(34)}\circ J_{34}\,.\\
\end{aligned}
\ee
Comparing \eqref{eq:3way} with \eqref{eq:2way} one is lead to conclude that a three-morphism $\mmmor_{1234}$ must exists such that
\be
\mmor_{234} \circ \mmor_{1(23)4} \circ \mmor_{123} \xrightarrow{\mmmor_{1234}} \mmor_{12(34)} \circ \mmor_{(12)34}\,.
\ee
Which is usually called the pentagonator identity. It is a generalization of the familiar pentagon equation which appears in the study of modular tensor categories. The existence of the three-morphism $\mmmor_{1234}$ is a necessary datum which specifies the higher associativity.

\medskip

This is an extremely powerful remark: \textit{associativity conditions for the fusion of more and more objects correspond to the requirement of further structure on the symmetry category in the form of higher morphisms}. Indeed, the associativity condition for 3 objects leads to a 2-morphism, similarly the associativity condition for 4 objects gives a 3-morphism,\footnote{ \ If the objects $\ob_i$ are graded by an abelian group the 3-morphism is described by an hypercube on whose faces the 2-morphisms $\mmor$ are performed. This extends to higher associators by iteration.} and so on: the associativity for $(N+1)$-objects gives an $N$-morphism. Notice however that higher and higher morphisms correspond to configuration of higher and higher codimension within a given symmetry defect worldvolume (see e.g. Figure \ref{fig:eta}): a one-morphism is codimension one, a 2-morphism is codimension 2, and so on: an $N$-morphism corresponds to a configuration of codimension $N$ within the defect worldvolume. This leads to the following remark: as long as $N$ does not exceed the dimensionality of the defect, the associativity conditions are just extra data which enter in the definition of the symmetry (following the same pattern of symmetry fractionalization discussed above). When instead we reach an associativity condition that leads to a morphism that corresponds to a configuration of codimension higher than the defect worldvolume itself, that corresponds to the appropriate generalization of a 't Hooft anomaly, meaning that if it does not vanish, it causes an obstruction to gauging.

Similar observations have been used to describe anomalies of invertible symmetries \eg \cite{Gaiotto:2017zba,Delmastro:2021xox}

\medskip

\subsubsection{Example 1: Quantum Mechanics}
In order to better understand the remark made above let us consider the case $d=0+1$, which is the familiar the textbook example of symmetries in quantum mechanics. According to the Wigner's theorem, a quantum mechanical system with Hilbert space $\mathcal H$ has a symmetry $G$ provided for all $g \in G$ there is an (anti)unitary transformation $U_g : \mathcal H \to \mathcal H$ for which $\mathcal H$ is in a projective representation.
The projectivity of the representation means that group multiplication is defined only up to a phase, namely
\be\label{eq:grouplaw}
U_g U_h = e^{i \alpha(g,h)} U_{gh}
\ee
where $\alpha : G \times G \to \mathbb R / 2\pi \mathbb R$. In this context, we have that the only constraining feature is the associativity (as the operators are already supported at fixed point in time, the associativity is itself a constraint which is one dimension too high)
\be
U_g (U_h U_k) = (U_g U_h) U_k
\ee
which is satisfied provided
\be
\alpha(g,hk) + \alpha(h,k) - \alpha(gh,k) - \alpha(g,h) = 0.
\ee
This is called the 2-cocycle condition, namely the statement that $\delta \alpha = 0$ in group co-homology. In the definition of a symmetry in quantum mechanics we are also allowed to freely rescale all operators by phases, e.g. 
\be
U_g \to e^{i\phi(g)} U_g 
\ee
where $\phi: G \to \mathbb R / 2 \pi \bZ$. We see that the group multiplication law in \eqref{eq:grouplaw} is still satisfied provided upon such a shift we transform $\alpha$ as follows:
\be
\alpha(g,h) \to \alpha(g,h) + \phi(gh) - \phi(g) - \phi(h)\,,
\ee
which means we are free to shift $\alpha(g,h)$ by a co-boundary in group co-homology. Therefore, to completely specify $U_g$ we need also to prescribe a class in $H^2(BG,U(1))$, which becomes part of the definition of a $G$-symmetry for a quantum system. This class is indeed the anomaly of the symmetry $G$ in quantum mechanics, which was expected from the general discussion above.

\subsubsection{Example 2: F-symbol in 2D theories}
The next step is to consider a fusion 1-category $\cC$  describing two dimensional topological line defects. 
In this case objects $\ob$ are lines and the morphisms $\mor_{1 2}^3 \in \Hom(\ob_1 \otimes \ob_2, \ob_3)$ ($N_{i,j}^k:=\dim\Hom(\ob_1 \otimes \ob_2, \ob_3)$) are local line-changing operators, 
which forms a (finite dimensional) vector spaces over $\bC$. For explicitness, it is convenient to choose a basis $(v_{i,j}^k)_{a=1,\cdots N_{i,j}^k}$ of the two-to-one junction space $\Hom(\ob_i\otimes \ob_j,\ob_k)$ for each triple $(\ob_i,\ob_j,\ob_k)$ of simple objects in $\cC$.

For the three-to-one junction spaces $\Hom(\ob_i\otimes \ob_j\otimes \ob_k,\ob_\ell)$, there are two possible basis made of the above basis $(v_{i,j}^k)_a$, i.e.\ 
\begin{equation}
    \{(v_{i,m}^\ell)_b \circ (\id_{\ob_i}\otimes (v_{j,k}^m)_a)\}_{\ob_m\in \cC,a,b}
    \text{, and}\quad
    \{(v_{n,k}^\ell)_d \circ ( (v_{i,j}^n)_c\otimes \id_{\ob_k})\}_{\ob_n\in \cC,c,d}.
\end{equation}
The F-symbol is the linear relationship between the two bases:
\begin{equation}
    \sum_{n;c,d}\AF\begin{bmatrix}
    a & b \\ c & d\end{bmatrix}^{\ell}_{i,j,k}(v_{n,k}^\ell)_d \circ ( (v_{i,j}^n)_c\otimes \id_{\ob_k})
    =
    (v_{i,m}^\ell)_b \circ (\id_{\ob_i}\otimes (v_{j,k}^m)_a).
\end{equation}

The pentagonator now has to be trivial, i.e.\ we have the pentagon \text{identity}, because the F-symbol maps a vector to a vector which has no nontrivial auto-equivalence (as opposed to a TFT). 
If $\cC$ is invertible, i.e. $\cC \sim \text{Vec}_G^\omega$, the F-symbols is nothing but the standard bosonic anomaly:
\be
\AF_{g , h , k} = \omega(g,h,k) \in H^3(G, U(1)) \, .
\ee

\subsubsection{Example 3: 2-Groups}
The final example concerns discrete two groups $\Gamma$ \cite{Benini:2018reh} (see also \cite{Cui:2016bmd,Bartsch:2022ytj,Bartsch:2023pzl} for a related treatment using categorical language). We focus on $d=3$ for concreteness.
The two group is specified by an action $\rho: G \to \text{Aut}(A)$ and by a Postnikov class:
\be
e \, \in \, H^3_\rho(BG, \, A) \, .
\ee
For simplicity let us consider a trivial $\rho$ action. As a category, objects of the 2-Group are zero-form symmetry defects $U_g$, $g \in G$. One morphisms between them are all invertible and can be decorated by one-form symmetry surfaces $L_{a}$, $a \in A$, thus:
\be
\Hom(U_g \otimes U_h, \, U_k) = \boldsymbol{\delta}_{gh, \, k} \, ( i_{gh} \, \otimes \, A^{(1)} ) \, .
\ee
The Postnikov class specifies a 1-associator
\be
\boldsymbol{e}(g,h,k) : (i_{(gh) k} \otimes b) \circ (i_{gh} \otimes a) \xrightarrow{\quad} (i_{g(hk)} \otimes d) \circ (i_{hk} \otimes c) \, ,
\ee
between these one-morphisms. More explicitly\footnote{ \ We use the representation $e(g,h,k)$ as a 3-cochain in group cohomology $H^3(G, A)$.}
\bea
\begin{tz}[td,scale=2]\begin{scope}[xyplane=0]
\node at (1.15*\h,10) {$\ds \xymatrix{
&U_g \otimes U_h \otimes U_k\ar@{..>}[dl]_{i_{gh} \otimes \, a}\ar@{..>}[dr]^{i_{hk} \otimes \, c}&\\
U_{gh} \otimes U_k \ar@{..>}[dr]_{i_{(gh)k} \otimes \, b}\ar@{->}[rr]^{\boldsymbol{e}} && U_g \otimes U_{hk}\ar@{..>}[dl]^{i_{g(hk)} \otimes \, d}\\
& U_{ghk}
}$};
\draw[slice] (0,0) to [out=up, in=\dl] (0.5,1) to [out=up, in=\dl] (1,2) to (1,4.5);
\draw[slice] (1,0) to [out=up, in=\dr] (0.5,1);
\draw[slice] (2,0) to [out=up, in=\dr] (1,2);
\end{scope}
\begin{scope}[xyplane=\h, on layer=superfront]
\draw[slice] (0,0) to [out=up, in=\dl] (1,2) to (1,4.5);
\draw[slice] (1,0) to [out=up, in=\dl] (1.5,1) to [out=up, in=\dr] (1,2);
\draw[slice] (2,0) to [out=up, in=\dr] (1.5,1);
\end{scope}
\begin{scope}[xzplane=0]
\draw[slice,short] (1,0) to (1,\h);
\draw[slice,short] (0,0) to (0,\h);
\draw[slice,short] (2,0) to (2,\h);
\end{scope}
\begin{scope}[xzplane=4.5]
\draw[slice,short] (1,0) to (1,\h);
\end{scope}
\coordinate (A) at (1.5,1,0.5*\h);
\draw[wire,blue] (1,0.5,0) to [out=up, in=\dr] (A) to [out=\ur, in=down] (1,1.5,\h);
\draw[wire,blue] (2,1,0) to [out=up, in=\dl] (A) to [out=\ul, in=down] (2,1,\h);
\node[dot] at (A){};
\node[right] at (A) {$\quad \boldsymbol{e}$};
\node[right] at (4,0,2) {$U_{ghk}$};
\node[right] at (-1,0,1) {$U_k$};
\node[right] at (-0.5,0,0.5) {$U_h$};
\node[right] at (0,0,0) {$U_g$};
\node[right] at (2.25,0,0.3) {$i_{(gh)k}$};
\node[right] at (1,0,0) {$i_{gh}$};
\node[above] at (1,1.5,\h) {$i_{hk}$};
\node[above] at (2,1,\h) {$i_{g(hk)}$};
\node[color=blue] at (1,1,0) {$a$}; 
\node[color=blue] at (2,1.5,0) {$b$}; 
\node[color=blue] at (1,1,\h) {$c$};
\node[color=blue,fill=white,left] at (1.75,0.25,\h) {$a b c^{-1} \, e(g,h,k)$};
\end{tz}
\eea
With the two-morphism explicitly given by the vector space
\be
\boldsymbol{e}(g,h,k)\begin{bmatrix}
    \ \ a & b \ \ \\ \ \ c & d \ \
\end{bmatrix} = \boldsymbol{\delta}_{d, \, ab c^{-1} \, e(g,h,k)} \ \bC^{\times} \, .
\ee
Redefining the one-form symmetry dressing by a closed 1-chain amounts to a shift of $\boldsymbol{e}$ by an exact 2-cochain: $e(g,h,k) \to e(g,h,k) + d\mu(g,h,k)$. 
The pentagonator requires:
\be
e(g,h,k) \, e^{-1}(gh,k,l) \, e(g,hk,l) \, e^{-1}(g,h,kl) \, e(h,k,l) = de(g,h,k,l) = 1\, ,
\ee
also that $\boldsymbol{e}$ defines a class $e \in H^3(BG, A)$.
The pentagonator furthermore defines an isomorphism $\omega$ between the one-dimensional vector spaces:
\be
\omega(g,h,k,l) : \boldsymbol{e}(gh,k,l) \otimes \boldsymbol{e}(g,h,kl) \xrightarrow{\quad} \boldsymbol{e}(g,h,k) \otimes \boldsymbol{e}(g,hk,l) \otimes \boldsymbol{e}(h,k,l)  \, ,
\ee
where we have suppressed the dependence on the $A^{(1)}$ dressing for simplicity. The 2-associator gives a constraint on $\omega$ by examining the action of the pentagonator on the vector space
\be
\boldsymbol{e}(g,h,klm) \otimes \boldsymbol{e}(gh,k,lm) \otimes \boldsymbol{e}(ghk,l,m)
\ee
of five-fold junctions.
If the 2-Group splits (that is, $\boldsymbol{e}$ is trivial), the four co-chain $\omega$ is constrained by the 2-associator to satisfy:
\be
d\omega(g,h,k,l,m) = 0 \, ,
\ee
defining a class $[\omega] \in H^4(BG, U(1))$ encoding the pure zero-form symmetry anomaly. To construct the full anomaly of a discrete 2-Group we must introduce further structure, such as the braiding 2-morphisms:
\be
\boldsymbol{b} : a \otimes b \xrightarrow{\quad} b \otimes a \, , \ \ \ \ 
\boldsymbol{\tilde{b}} : a \otimes i_{g h} \xrightarrow{\quad} i_{g h} \otimes a \, .
\ee
The 2-morphism $\boldsymbol{b}$ satisfies the same pentagon and hexagon identity as those of an MTC, encoding the pure one-form symmetry anomaly. This is described by a quadratic refinement $q \in \widehat{\Gamma(A^{(1)})}$, with $\Gamma(A)$ the universal quadratic group of $A$. Physically $q(a)$ determines the spin of the $a$ line $\theta_a = e^{2 \pi i q(a)}$. Similarly $\tilde{\boldsymbol{b}}$ must be consistent with the 1-associator for $g,h,k,l$. This leads to a mixed 0-form/ 1-form anomaly determined by a two co-chain $\lambda \in C^2(BG, \widehat{A})$ satisfying:
\be
d \lambda(g,h,k) = \langle e(g,h,k), \, \cdot \rangle_q \, ,
\ee
with $\langle \, , \, \rangle_q$ the bilinear pairing induced by $q$. Clearly $\lambda$ is a torsor over $H^2(BG, \, \widehat{A})$. Finally the 2-associator also makes $\omega$ into a torsor, see \cite{Benini:2018reh} for details.

\section{Higher Structure of Chiral Symmetry}\label{sec:associati}
In this section, we discuss the example of a theory with non-invertible chiral symmetry \cite{Cordova:2022ieu, Choi:2022jqy} in four dimensions to illustrate several features of the more abstract symmetry category we introduced above. This example will help to highlight the description of the categorical structure through TFTs.
The discrete version of this symmetry category (in which one only considers the coupling to a $\bZ_N^{(1)}$ one-form symmetry) corresponds to the category  $3\text{Rep}(\Gamma)$ \cite{Bartsch:2022ytj} of 3-representations for the 3-Group $\Gamma$ described by the following sequence:
\be
1 \xrightarrow{\quad} \bZ_{2N}^{(2)} \xrightarrow{\quad} \Gamma \xrightarrow{\quad} \bZ_{N}^{(1)} \xrightarrow{\quad} 1 \, ,
\ee
with Postnikov class $e \in H^4(B^2\bZ_N, \bZ_{2N})$ given by (a multiple of) the Pontryagin square operation.

\medskip

Let us give a brief summary of the results derived in this Section. The non-invertible chiral symmetry defects are described by $\bQ/\bZ$-graded objects $\Dchiral{p/N}$ obtained by stacking the naive chiral symmetry $\Uchiral_{p/N} \in \bQ/\bZ$ with a coupled minimal TFT $\cA^{N, \, p}$. This choice, although minimal, is not unique. It should rather be thought of as a choice of representative for the class $\lls \Dchiral{p/N} \rrs$.
 The categorical structure respects the grading is described explicitly below.

\paragraph{1-morphisms} $\mor_L$: $\lls \Dchiral{p/N} \rrs \otimes \lls \Dchiral{p'/N'} \rrs \, \xrightarrow{\quad \mor_L \quad} \, \lls \Dchiral{p/N +p'/N' \mod 1} \rrs$ are obtained by gauging of a $\bZ_L$ one-form symmetry $\algA$, the relation between $L$ and $p/N , \, p'/N'$ can is explained below \eqref{eq: 1morphisms}. The topological interface is described by the category of $\algA$-modules $\text{Mod}_\algA$.
 \bea
\begin{tz}
 \draw[slice] (0,0) -- (6,0) -- (6,2) -- (0,2) -- cycle;
 \draw[wire] (3,0) node[below] {$\mor_L$} to (3,2) node[above] {$\text{Mod}_{\algA}$};
 \node[fill=white] at (0.5,1) {$\lls \Dchiral{p/N} \otimes \Dchiral{p'/N'} \rrs$};
 \node[fill=white] at (5.5,1) {$\lls \Dchiral{p/N +p'/N'} \rrs$};
\end{tz}
\eea
Repeating such construction and gauging sequentially is sufficient to construct all one morphisms in arbitrary tensor products (up to composition with 1-endomorphisms). Notice that the composition law for equivalence classes $\lls \, . \, \rrs$ is exactly the group $\bQ/\bZ$.
\paragraph{2-morphisms} $\mmor_{\algA^+ \, \algA^-}$ relate different gauging procedures $\algA^+$ and $\algA^-$ connecting $\Dchiral{p_1/N_1} \otimes ... \otimes \Dchiral{p_k/N_k}$ to $\Dchiral{p_1/N_1 + ... + p_k/N_k}$:
 \bea \label{eq: 2morphismspic}
\begin{tz}
 \draw[slice] (0,0) -- (6,0) -- (6,4) -- (0,4) -- cycle;
 \draw[wire] (3,0) node[below] {$\text{Mod}_{\algA^+}$} to (3,4) node[above] {$\text{Mod}_{\algA^-}$}; 
 \node at (3,2) {$\bullet$};
 \node[right] at (3,2) {$\mmor_{\algA^+ , \, \algA^- }$};
 \node[fill=white] at (0.5,1) {\small$ \lls \Dchiral{p_1/N_1} \otimes ... \otimes \Dchiral{p_k/N_k} \rrs$};
 \node[fill=white] at (5,1) {\small$ \lls \Dchiral{p_1/N_1 + ... + p_k/N_k} \rrs$};
\end{tz}
 \eea
 We explicitly describe the two-morphisms as Dirichlet boundary conditions for the associator $\AF_{\Dchiral{p_1/N_1}, \, \Dchiral{p_2/N_2}, \, \Dchiral{p_3/N_3}}$ one-morphism  by folding the diagram \eqref{eq: 2morphismspic}. 
The coupling to the bulk one-form symmetry can be made explicit as follows: the 2d TFT for the associator can be decomposed as a direct sum over idempotents $\pi_i$, with $i$ and index labelling simple local operators in the corresponding (multi)fusion category satisfying (see \cite{Moore:2006dw,Huang:2021zvu}):
\be
\pi_i \otimes \pi_j = \delta_{ij} \, \pi_i \, .
\ee
A 2d TFT $\boldsymbol{T}$ then splits into invertible theories $\boldsymbol{T}_{\pi_i}$:
\be
\boldsymbol{T} = \bigoplus_i \boldsymbol{T}_{\pi_i} \, ,
\ee
with lines operators acting as domain walls between these theories. Idempotents $\pi_i$ also describe topological boundary conditions for $\boldsymbol{T}$, obtained by blowing up a $\pi_i$ insertion into an empty circle.
Because of the coupling to the bulk one-form symmetry, the insertion of the idempotent also covers the 2d-worldvolume of the defect with a bulk one-form symmetry defect $\Uonef_{q_i/N}$, $N=\text{lcm}(N_1, N_2, N_3)$ and $q_i$ is the one-form symmetry charge of $\pi_i$ seen as a boundary condition $B_i$ for the associator TFT (a line defect).\footnote{\ An example might be helpful. In 2d $\bZ_N$ gauge theory the idempotents are related to the basis of vertex operators $V_k = e^{\frac{2 \pi i k}{N} \phi}$ by Fourier transform $\pi_j = \frac{1}{N} \sum_{k=0}^{N-1} V_k \, e^{\frac{2 \pi i k}{N}} $.
The insertion of $\pi_k$ in the theory is a delta function setting $\phi= j$. The one-form symmetry generator $\Uonef_{1/N}$ couples to the theory via $\left(\Uonef_{1/N} \right)^\phi$. Thus inserting the idempotent leaves behind an operator $\Uonef_{j/N}$.
} 
\bea
\begin{tz}
\filldraw[color=white!85!blue, opacity=0.5] (0,0.25) to (2,0.25) to (3,2.25) to (1,2.25) -- cycle; 
\draw[slice] (0,0) to (2,0) node[below] {$\AF$} to (3,2) to (1,2) -- cycle;    
\node[dot,label={[label distance=0.2]-90:$\pi_i$}] at (1.5,1) {};
\node at (4,1 ) {\Huge$ \overset{\text{\small Open up}}{\leadsto}$};
\node at (0.5,2.25) {$\Uonef_{\frac{q_i}{N}}$};
\begin{scope}[shift={(5,0)}]
\filldraw[color=white!85!blue, opacity=0.5] (0,0.25) to (2,0.25) to (3,2.25) to (1,2.25) -- cycle; 
\draw[slice] (0,0) to (2,0) node[below] {$\AF$} to (3,2) to (1,2) -- cycle;        
\filldraw[wire, fill=white, rotate around={-45:(1.5,1)}] (1.5,1) node {$B_i$} ellipse (0.4 and 0.5);
\node at (0.5,2.25) {$\Uonef_{\frac{q_i}{N}}$};
\end{scope}
\end{tz}
\eea
The expansion then reads:
\be
\AF[\algA^+ \, \algA^-]_{\Dchiral{p_1/N_1}, \, \Dchiral{p_2/N_2}, \, \Dchiral{p_3/N_3}} = \bigoplus_{q = 0}^{N-1} \left( \, \bigoplus_{i: q_i=q} \boldsymbol{T}_{\pi_i}  \right) \otimes \Uonef_{q/N} \, .
\ee
The information contained in a single $\boldsymbol{T}_{\pi_i}$ theory is a relative Euler counterterm, which is given in terms of the quantum dimension of the corresponding boundary condition\cite{Huang:2021zvu}:
\be
\langle \boldsymbol{T}_{\pi_i} \rangle_{\Sigma_g} = \dim(B_i)^{2-2g} \, .
\ee
When the coefficients $\cZ_q = \biggl< \left( \, \bigoplus_{i: q_i=q} \boldsymbol{T}_{\pi_i}  \right) \biggr>$ are all equal $\cZ_q = \cZ_{q'}$ the TFT is Witt equivalent to a condensation defect $\cond^{(1)}_N$. 

We then conclude by describing how the associator can be physically detected by dynamical monopole lines, turning it into a physical object.

\subsection{Non-invertible Chiral symmetry in four dimensions}\label{sec:review}

Consider a $D=3+1$ dimensional theory that has a conserved 1-form symmetry $U(1)^{(1)}$ with a conserved 2-form current $d \ast j^{(2)} = 0$ as well as  1-form $j^{(1)}_{\chi}$ that satisfies a modified conservation equation of the form
\be\label{eq:chirala}
d \ast j_{\chi}^{(1)} = \ast j^{(2)} \wedge \ast j^{(2)}\,.
\ee
It was recently demonstrated \cite{Choi:2022jqy,Cordova:2022ieu} that such $D=3+1$ dimensional theory has a non-invertible symmetry parametrized by $p/N \in \mathbb Q / \mathbb Z$ that can be exploited to define selection rules for the quantum numbers associated to the charge $Q_\chi(\Sigma_3) = \int_{\Sigma_3} \ast j_{\chi}^{(1)}$. Because of \eqref{eq:chirala} in order to obtain a conserved symmetry defect for the charge $Q_\chi$ we need to stack to the standard symmetry operator
\be
{\Uchiral}_\alpha = \text{exp}\left(2 \pi i \, \alpha \, Q_\chi \right) \, , \ \ \ \alpha =  \frac{p}{N} \, , \ \ \ \gcd(p,N)=1 \, ,
\ee
a copy of the 3d minimal abelian TFT $\mathcal A^{N,p}$ \cite{Hsin:2018vcg} coupled to $ \ast j^{(2)}$. As a spin theory $\cA^{N, p}$ has a spectrum of invertible line operators with a unique generator $L$ of spin:
\be
\theta_L = \exp\left(\frac{2 \pi i p}{2 N} \right)\,,
\ee
and $L^N = 1$.\footnote{\ This is true as a spin TFT, in general $L^N$ can be a transparent fermion since its spin is $\theta_{L^N} = (-)^{N p}$} We give more details about this family of theories in Appendix \ref{app: minimal}.
This procedure gives a family of topological defects with charges $p/N \in \mathbb Q/\mathbb Z$. The resulting defect is
\be
\Dchiral{p/N} \equiv \Uchiral_{p/N} \otimes \mathcal A^{N,p}\left[\frac{1}{N} \ast j^{(2)}\right],
\ee
where the minimal TFT couples with the bulk by regarding $\frac{1}{N}*j^{(2)}$ as its symmetry background for the one-form symmetry generated by $L$.
Many interesting examples either realise such a setup \cite{Choi:2022jqy,Cordova:2022ieu} or a more limited $\bZ_N$ subgroup of it\cite{Kaidi:2021xfk}.
Following our discussion in Section \ref{sec:witteq}, the stacking of the minimal TFT is in fact a gauge choice of a representative for the equivalence class $\lls \Dchiral{p/N} \rrs$. Any other TFT with the same one-form symmetry anomaly is also a valid choice. The minimal TFT, as the name suggests, gives a representative with the minimal possible total quantum dimension \cite{Hsin:2018vcg}.  

\paragraph{A first sketch of the chiral symmetry category.} In the case of chiral symmetry category the objects are the operators $\Dchiral{p/N}$ which we have discussed in Section \ref{sec:review} together with the condensate defects associated to the $U(1)^{(1)}$ form symmetry (which correspond to the trivial equivalence class). 

We can think of the magnetic $U(1)$ one-form symmetry as one-endomorphisms of the identity defect:
\be
U(1)^{(1)} \subset \text{End}_{(1)}(\unit) \, .
\ee
Following our previous discussion we must also include condensation defects for $U(1)^{(1)}$. To complete the chiral symmetry category we will only need condensations defects for the discrete $\bZ_K \subset U(1)$ subgroups. Such defects are supported on either codimension one or two surfaces and we denote them by $\cond_{K, \, \alpha}^{(0)} \, , \  \cond_K^{(1)} $ accordingly. The label
$\alpha$ is a choice of discrete torsion, classified by \cite{Cordova:2017vab,Choi:2022jqy}:
\be
\Hom\left( \text{Tors}(\Omega_3^{Spin}(B\bZ_K)),U(1) \right) = \begin{cases}
    \bZ_K \, , \ \ \ &K= 1 \mod 2\, , \\
    \bZ_{2K} \times \bZ_2 \, , \ \ \ &K = 0 \mod 4 \\
    \bZ_{4K} \, , \ \ \ & K = 2 \mod 4 \, .
\end{cases}
\ee
while no such choice is present in codimension two. They can be explicitly realized as sums over one-form symmetry defects:
\begin{align}
&\cond_{K, \, \alpha}^{(0)}(\Sigma) = \sum_{a \, \in \, H^1(\Sigma, \, \bZ_K)} \, \exp\left( \frac{2 \pi i}{K} \int a^* \alpha + \frac{2 \pi i}{K} \int a \cup [\star j^{(2)}] \right) \, , \\
&\cond_{K}^{(1)}(\Sigma) = \sum_{\phi \, \in \, H^0(\Sigma, \, \bZ_K)} \exp\left( \frac{2 \pi i}{K} \int \phi \cup [\star j^{(2)}] \right) \, .
\end{align}
We now move on to examine the categorical structure underlying these objects and its interplay with the tensor product operation.

\subsection{Morphisms of the chiral symmetry category} \label{ss: morphisms}
We now give a thorough description of junctions involving at most one chiral symmetry defect. As noted in \cite{Choi:2022jqy,Cordova:2022ieu} these include junctions between external one-form symmetry surfaces and the chiral defect $\Dchiral{p/N}$. We also uncover a slightly richer structure induced by interfaces between the chiral defect and condensation defects, which give rise to 1-endomorphisms (codimension one interfaces on $\Dchiral{p/N}$) for the chiral symmetry.

\bigskip

Let us start with the simpler structures.  The 1-form symmetry defects $\Uonef \in U(1)^{(1)}$ fuse according to the group implying that
\be
\text{Hom}_{(2)}(\Uonef_\alpha \otimes \Uonef_\beta, \, \Uonef_{\gamma}) = \boldsymbol{\delta}_{\alpha + \beta}^{\gamma} \, L_{\alpha, \beta}^\gamma \, , \ \ \ L_{\alpha, \, \beta}^\gamma \simeq \text{Vec} \, .
\ee
The operators $\Uonef_\alpha$ can also end on a chiral symmetry defect $\Dchiral{p/N}$, provided they are in the $\bZ_N$ subgroup $\alpha = \frac{ k p}{N}$ to ensure their coupling to the minimal theory $\cA^{N, p}$.
This junction defines a two-morphism $L$ (by folding the chiral symmetry defect), which we identify with the generating line of the minimal TFT:
\be
\Hom_{(2)}\left(\Uonef_{\frac{k p}{N}}, \unit^{(1)}_{\Dchiral{p/N}}\right) = L^k \, .
\ee
As the lines $L$ live in a three-dimensional ambient defect $\Dchiral{p/N}$ they form a braided category, which is described by the minimal TFT $\cA^{N, \, p}$. This means that we have a three-morphism $b: L \otimes L' \to L' \otimes L$ implementing the braiding. Its gauge-invariant information is encoded in the braiding matrix
\be
\boldsymbol{B}_{L^k, \, L^{k'}} = \exp\left( \frac{2 \pi i p}{N} k \, k'  \right) \, .
\ee
We now move onto the study of the chiral defects $\Dchiral{p/N}$ themselves. We first notice that objects $\Dchiral{p/N}$ in $\scatchi{(0)}$ are graded in terms of the invertible operators $\Uchiral$. This strongly constraints the structure of the morphisms between these objects. In particular,
\be \label{eq: zeromorph}
\text{Hom}\left(\Dchiral{p/N}, \Dchiral{p'/N'}\right) = \begin{cases} \text{End}(\cA^{N,p}) &\text{if } p/N = p'/N' \text{ mod } 1\\ \emptyset &\text{otherwise} 
\end{cases}
\ee
immediately follows from this.\footnote{ \
To be precise, the line $L^k$ is on the boundary of $\Uonef_{\frac{kp}{N}}$, immersed onto the worldvolume of $\Dchiral{\frac{p}{N}}$, and thus there are objects in $\End(\Dchiral{p/N})$ (which should be a fusion 2-category) corresponding to the immersed one-form symmetry generators. Such a fusion 2-category can be constructed by first regarding the modular tensor category of $\mathcal{A}^{N,p}$ as a $\mathbb{Z}_N$-crossed braided fusion category, where the part with grading $k\in\mathbb{Z}_N$ contains the only object $L^k$, and then applying the construction in \cite[Example 2.1.23]{douglas2018fusion} (See also \cite{Cui:2016bmd}). This category has objects corresponding to elements of $\mathbb{Z}_N$, which we identify with $\Uonef_{\frac{kp}{N}}$. Although this fusion 2-category is equivalent to the trivial 2-category as fusion 2-category, we expect that the equivalence is not consistent with the ambient fusion 3-category structure. Instead, we treat $\End(\Dchiral{p/N})$ as $\End(\mathcal{A}^{N,p})$, which is justified in the main text.
}
Above we have denoted by $\text{End}(\cA^{N,\,p})$ the set of topological interfaces for the minimal theories. These operators implement a generalized symmetry of the minimal theory $\cA^{N,p}$, which are in correspondence with one-gaugings \cite{Roumpedakis:2022aik} of subgroups $\salgA$ in $\cA^{N,p}$. As reviewed in Section \ref{sec:category} this is a true fact for any 3d MTC. Endomorphisms are classified by symmetric Frobenius algebras \cite{Fuchs:2002cm}. This was first understood in \cite{Fuchs:2002cm} and given a nice physical interpretation in \cite{Kapustin:2010hk,Kapustin:2010if}.
We hence will use $\cV_{\salgA}$ to denote the symmetry defect realized by gauging $\salgA$ onto a codimension-one surface in the $\cA^{N,p}$ theory.
A genuine interface must also preserve the $\bZ_N$ one-form symmetry charge on the two sides, as the charge mismatch must be compensated by a bulk one-form symmetry defect. Since the minimal TFT $\cA^{N, \, p}$ has exactly one line for each value of the charge $q \in \bZ_N$, the only one-form symmetry-preserving interface is the identity.
The space \eqref{eq: zeromorph} should instead be thought as arising from a junction between the chiral defect $\Dchiral{p/N}$ and a condensation defect $\cC_K^{(0)}$, for some $K$ dividing $N$. This defines a \emph{twisted} interface, generated by condensing an algebra $\salgA$ which is charged under the one-form symmetry.
Alternatively, since all bulk condensates are endable, we can compose $\cC_K^{(0)}$ with the morphism $\boldsymbol{\tau} : \cond_K^{(0)} \to \unit$ and deform this onto the minimal theory, giving rise to a genuine interface (which is explicitly coupled to the bulk one-form symmetry).
\begin{figure}[t]
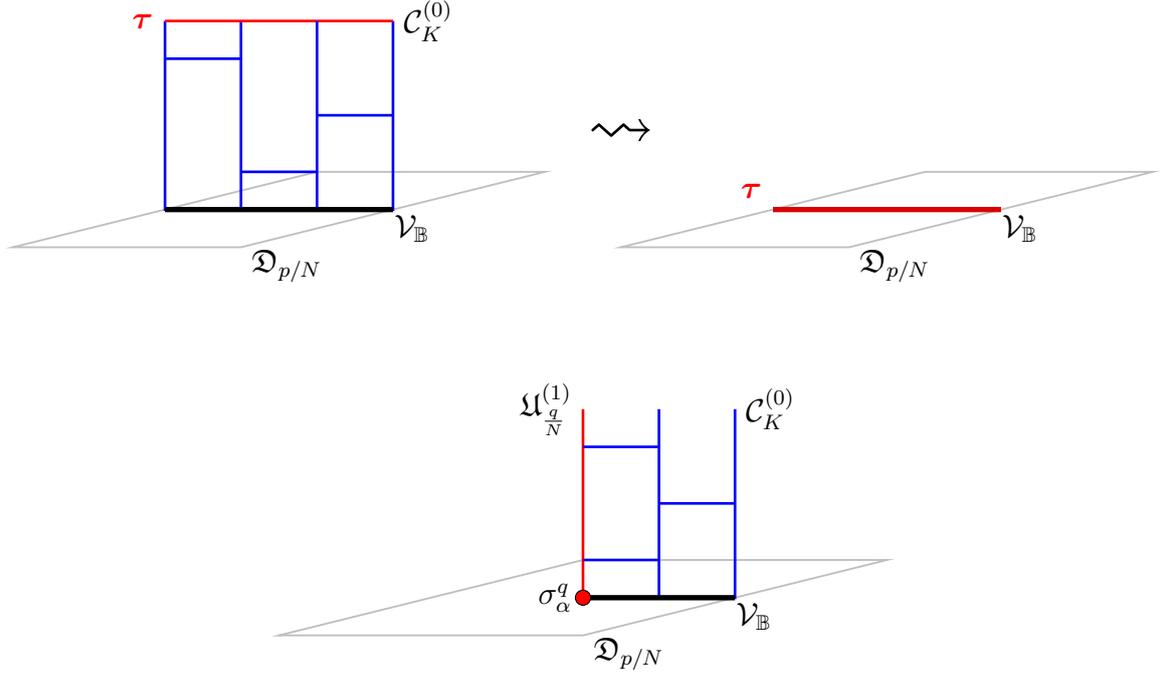

\label{fig:twistedsec}
\begin{center}
\begin{tz}
\draw[slice] (0,0) -- (3,0) -- (7,1) -- (4,1) -- cycle;
\draw[color=blue, line width= 1] (2,3) -- (2,0.5) -- (5,0.5) -- (5,3); \draw[color=blue, line width= 1] (3,3) -- (3,0.5); 
\draw[color=red, line width=1] (2,3) node[left] {$\boldsymbol{\tau}$} -- (5,3);
\draw[color=blue, line width= 1] (4,3) -- (4,0.5); \draw[color=blue, line width= 1] (2,2.5) -- (3,2.5); \draw[color=blue, line width= 1] (3,1) -- (4,1); \draw[color=blue, line width= 1] (4,1.75) -- (5,1.75);  
\draw[color=black, line width= 2] (2,0.5) -- (5,0.5);
\node[right] at (3,-0.25) {$\Dchiral{p/N}$};
\node at (5.25,0.25) {$\cV_\salgA$}; 
\node[right] at (5,3) {$\cC_K^{(0)}$};
\node at (8,1.5) {\Huge$\leadsto$};
\begin{scope}[shift={(8,0)}]
    \draw[slice] (0,0) -- (3,0) -- (7,1) -- (4,1) -- cycle;
\draw[color=black!15!red, line width= 2] (2,0.5) -- (5,0.5);
\node[right] at (3,-0.25) {$\Dchiral{p/N}$};
\node at (5.25,0.25) {$\cV_\salgA$}; 
\node[left, color=red] at (2,0.75) {$\boldsymbol{\tau}$};
\end{scope}
\end{tz} \\[2.5em]
\begin{tz}
\draw[slice] (-1,0) -- (3,0) -- (7,1) -- (3,1) -- cycle;
\draw[color=blue, line width= 1]  (3,0.5) -- (5,0.5) -- (5,3); \draw[color=red, line width= 1] (3,3) -- (3,0.5); 
\draw[color=blue, line width= 1] (4,3) -- (4,0.5); \draw[color=blue, line width= 1] (3,1) -- (4,1); \draw[color=blue, line width= 1] (3,2.5) -- (4,2.5); \draw[color=blue, line width= 1] (4,1.75) -- (5,1.75);  
\draw[color=black, line width= 2] (3,0.5) -- (5,0.5);
\draw[fill=red] (3,0.5) circle (0.1);
\node[right] at (3,-0.25) {$\Dchiral{p/N}$};
\node at (5.25,0.25) {$\cV_\salgA$}; 
\node[right] at (5,3) {$\cC_K^{(0)}$};
\node[left] at (3,3) {$\Uonef_{\frac{q}{N}}$};
\node[left] at (3,0.5) {$\sigma_\alpha^q$};
\end{tz}
\end{center}
\caption{Above: condensation defect terminating on a twisted topological interface for $\Dchiral{p/N}$ and the deformation into a local twisted interface using the twist map. Below, a twist defect $\sigma_\alpha^q$ obtained by terminating the same configuration inside $\Dchiral{p/N}$.}
\end{figure}
Charged line operators can be condensed on the boundary of $\cC_K^{(0)}$, allowing for a nontrivial symmetry action. This is consistent since the one-form symmetry charge is only conserved $\mod N/K$ in going through the defect, due to the presence of the condensate.

\bigskip

From these defects we can construct a large class of two-morphisms $\sigma_\alpha^q(\salgA, \, \salgA')$ which are interfaces between $\cV_\salgA$ and $\cV_{\salgA'}$. By folding this interface around $\sigma_\alpha^q$ we can reduce their study to that of twist defects $\sigma_\alpha^q$ for the symmetry interfaces $\cV_\salgA$:

The simplest example for both symmetry defects and twist operators is the charge conjugation symmetry in $\cA^{2n, p}$. It can be associated \cite{Fuchs:2002cm,Roumpedakis:2022aik} to the condensation of the even subgroup:
\be
\salgA = \bigoplus_{r \, \in \, \bZ_n} L^{2r} \, 
\ee
on a codimension one surface.
As the lines forming $\salgA$ all have even charges the defect $C$ must sit at the end of the bulk $\bZ_n$ condensate $\cond_{n}^{(0)}$. The twisted sector is composed by two objects $\sigma_\pm$ \cite{Barkeshli:2014cna} of dimension $\sqrt{n}$. In the absence of symmetry fractionalization the only consistent fusion rules for these lines:
\bea
&\sigma_+ \otimes \sigma_+ = \sigma_- \otimes \sigma_- = \bigoplus_{r \, \in \, \bZ_n} L^{2r} \, , \\
&\sigma_+ \otimes \sigma_- = \bigoplus_{r \, \in \, \bZ_n} L^{2r +1} \, .
\eea
Thus one twisted sector (say $\sigma_+$) has charge $0 \mod 2$, while the other has charge $1 \mod 2$. The assignment of twisted sector $\sigma_\alpha^q$ should be thought of as a further categorical datum defining the symmetry. It would be interesting to understand which bulk observables are sensible to such a choice.

\subsection{One-morphisms for chiral symmetry}\label{ss: junctions}
Given two objects $\Dchiral{p/N}$ and $\Dchiral{p'/N'}$ the corresponding fusion channels are constrained by the grading inherited from the $\Uchiral$ operators. In other words, the equivalence classes $\lls\Dchiral{p/N}\rrs$ are fused according to the group $\mathbb{Q}/\mathbb{Z}$. 
To realize this nice composition law in terms of the defects themselves, instead of the equivalence classes, we construct the following 1-morphism
\be \label{eq: 1morphisms}
\framebox{ $\phantom{\Big|} \begin{aligned}   \Dchiral{p/N} \otimes \Dchiral{p'/N'} \xrightarrow{\quad\mor_{L'}\quad} &\cA^{M/M', \, (p^{-1} K_1 + p'^{-1} K_2)/M'} \; \; \Dchiral{{(p K_2 + p' K_1)/M' \over K/M'}}  \\ \equiv \, &\cC_{\frac{p}{N} \, \frac{p'}{N'}}^{\frac{p}{N} + \frac{p'}{N'}}  \; \; \Dchiral{p/N + p'/N'}
 \end{aligned} \phantom{\Big|} $ } 
\ee
where $K=\text{lcm}(N, N')$, $M=\gcd(N,N')$, $N= K_1 M$, $N'=K_2 M$ and $M' = \gcd(p K_2 + p' K_1, L)$. The morphism $\mor_{M'}$ is a gauging interface for a $\bZ_{M'}$ one-form symmetry of the $\Dchiral{p/N} \otimes \Dchiral{p'/N'}$ theory. 
Note that the morphism $\mor_{M'}$ is not an isomorphism between the objects, but induces the equality on the equivalence classes. As anticipated the equivalence classes
\be
\lls \Dchiral{p/N} \rrs \otimes \lls \Dchiral{p'/N'} \rrs = \lls \Dchiral{p/N + p'/N'} \rrs
\ee
fuse in an invertible way according to the group $\bQ/\bZ$.

\medskip

As an illustrative example let us consider the fusion $\Dchiral{1/4} \otimes \Dchiral{1/4}$. We denote by $L$ and $L'$ the generators of the two $\cA^{4,1}$ factors. Both lines can arise at the end of bulk $\Uonef_{1/4}$ one-form symmetry surfaces. A naive application of the $\bQ/\bZ$ composition would imply that after the fusion we have a defect $\Dchiral{2/4}$ supporting an $\cA^{4,2}$ theory. Such theory is however ill-defined, as the square of its generator is an undetectable (transparent) line.
This can be traced back, in the starting tensor product, to the presence of a bosonic line which is furthermore neutral under the bulk one-form symmetry, namely $L^2 \otimes (L')^2$.
This $\bZ_2$ symmetry can thus be gauged. The group of lines after the gauging --- modulo fusion with $\unit + L^2 \otimes (L')^2$ --- is generated by $\cL = L \otimes L'$ and $\cK = L \otimes (L')^3$. The two are mutually local (they braid trivially between each other) and have spin and one-form symmetry charge:
\bea
&\theta_\cL = i \, ,  \ \ \ &&\theta_{\cK} = -i \, , \\
&q(\cL) = 2 \, , \ \ \ &&q(\cK) = 0 \, .
\eea
Thus they form a $\cA^{2,1}\left[\frac{1}{2} \star j^{(2)}\right] \otimes \cA^{2,-1}$ theory, which only couples to the $\bZ_2$ subgroup of the original $\bZ_4$. We have thus derived, taking into account the grading by $\Uchiral$:
\be
\Dchiral{1/4} \otimes \Dchiral{1/4} \xrightarrow{\quad \mor_2 \quad} \ \cA^{2,-1} \ \Dchiral{1/2} \, .
\ee
The decoupled TFT $\cA^{2,-1} \simeq \cA^{2,1}$ should be interpreted as a TFT-valued fusion coefficient. Notice that the morphism implements the equivalence relation of Section \ref{sec:witteq}:
\be
\lls \Dchiral{1/4} \rrs \otimes \lls \Dchiral{1/4} \rrs = \lls \Dchiral{1/2} \rrs \, .
\ee
The fate of the original coupling to the $\bZ_4$ subgroup is clarified by this description. Before $\mor_2$ is performed we can freely end a $\bZ_4$ one-form symmetry surface onto either defect, generating a line (say) $L$. This line is not neutral under the symmetry generated by $L^2 \otimes (L')^2$ and thus remains stuck on the gauging interface $\mor_2$ together with the one-form symmetry generator.\footnote{\ In a more formal language from $L$ we can induce a module with respect to the algebra $\algA = \unit \oplus (L^2 \otimes (L')^2)$, but not a local one.} From the point of view of the gauged theory the line $L$ is fractionally charged, and thus it is an excitation confined onto the gauging wall.
\bea
\begin{tz}
\draw[slice] (0.25,-1) to[out=up, in = \dr] (0,0); \draw[slice] (-0.25,-1) to[out=up, in = \dl] (0,0); 
\node[below] at (0,-1) {$\Dchiral{1/4} \otimes \Dchiral{1/4}$};
\draw[slice] (0,0) to (0,1) node[above] {$\Dchiral{1/2}$};
\node[dot, label={[label distance=0.2]45:$\mor_2$} ] at (0,0) {};
\draw[slice, color=blue] (-1.5,-0.75) node[left] {$\Uonef_{\frac{1}{4}}$} to (-0.25,-0.75) node[dot, label={[label distance=0.2]135:$L$}] {};
\node at (1.75,0) {$\simeq$};
\begin{scope}[shift={(5,0)}]
\draw[slice] (0.25,-1) to[out=up, in = \dr] (0,0); \draw[slice] (-0.25,-1) to[out=up, in = \dl] (0,0); 
\node[below] at (0,-1) {$\Dchiral{1/4} \otimes \Dchiral{1/4}$};
\draw[slice] (0,0) to (0,1) node[above] {$\Dchiral{1/2}$};
\node[dot, label={[label distance=0.2]45:$\mor_2$} ] at (0,0) {};
\draw[slice, color=blue] (-1.5,0) node[left] {$\Uonef_{\frac{1}{4}}$} to (0,0) node[dot, label={[label distance=0.2]135:$L$}] {};    
\end{scope}
\end{tz}
\, .
\eea
\bigskip
On the other hand the surface $\Uonef_{\frac{1}{2}}$ can end onto the line $L \otimes L'$, which can be passed through the gauging interface to become the generator $\cL$ of $\Dchiral{1/2}$:
\bea
\begin{tz}
\draw[slice] (0.25,-1) to[out=up, in = \dr] (0,0); \draw[slice] (-0.25,-1) to[out=up, in = \dl] (0,0); 
\node[below] at (0,-1) {$\Dchiral{1/4} \otimes \Dchiral{1/4}$};
\draw[slice] (0,0) to (0,1) node[above] {$\Dchiral{1/2}$};
\node[dot, label={[label distance=0.2]45:$\mor_2$} ] at (0,0) {};
\draw[slice, color=blue] (-1.5,-0.75) node[left] {$\Uonef_{\frac{1}{2}}$} to (-0.25,-0.75) node[dot, label={[label distance=0.2]135:$L$}] {} to (0.25,-0.75) node[dot, label={[label distance=0.2]45:$L'$}] {};
\node at (1.75,0) {$\simeq$};
\begin{scope}[shift={(5,0)}]
\draw[slice] (0.25,-1) to[out=up, in = \dr] (0,0); \draw[slice] (-0.25,-1) to[out=up, in = \dl] (0,0); 
\node[below] at (0,-1) {$\Dchiral{1/4} \otimes \Dchiral{1/4}$};
\draw[slice] (0,0) to (0,1) node[above] {$\Dchiral{1/2}$};
\node[dot, label={[label distance=0.2]45:$\mor_2$} ] at (0,0) {};
\draw[slice, color=blue] (-1.5,0.5) node[left] {$\Uonef_{\frac{1}{2}}$} to (0,0.5) node[dot, label={[label distance=0.2]135:$\cL$}] {};    
\end{scope}
\end{tz}
    \, .
\eea

Let us now pass onto the general derivation. We start from the case $N=N'$
\be
\Dchiral{p/N} \otimes \Dchiral{q/N} = {\Uchiral}_{(p + q)/N} \otimes \left( \cA^{N, \, p} \otimes \cA^{N, \,q} \right)\left[\frac{1}{N} \star j^{(2)}\right] \, ,
\ee
to lighten the notation we will define $b = \star j^{(2)} \in H^2(X,\bZ_N)$. Let us denote by $L_{(p)}$ and $L_{(q)}$ the generators of the two minimal TFTs. The lines generated by \footnote{\ The inverses are multiplicative inverses taken in $\bZ_N^\times$, which exist since both $p$ and $q$ are coprime with $N$.}
\be
\cK = \left(L_{(p)}\right)^{p^{-1}} \otimes \left(L_{(q)}\right)^{-q^{-1}} \, ,
\ee
are neutral under the diagonal $\bZ_N$ one-form symmetry and form a decoupled TFT in the fusion product.\footnote{\ The conventional way to assign one-form symmetry charge to the lines of $\cA^{N, \, p}$ is to give $L_{(p)}$ charge $p$. We will follow this convention throughout our manipulations. Different conventions are related by a relabeling of the bulk field $b$.} Such decoupled theory is self-consistent if and only if $\gcd(p+q,N)=1$. This follows from the spins
\be
\theta\left[\cK\right] = \exp\left(\frac{ \pi i p^{-1} q^{-1} }{N} (p + q) \right)
\ee 
If $\gcd(p+q,N)=M \neq 1$ the lines $\cK^{N/M}$ are completely transparent objects and thus the TFT is inconsistent \cite{Hsin:2018vcg}.
However, when $\gcd(p+q, N)=M$, lines $\cK^{N/M}$ form an anomaly-free decoupled $\bZ_M$ subgroup. Following the same logic as in our previous example we can gauge this symmetry to define the interface $\mor_M$. As this group is neutral under the bulk one-form symmetry the interface is truly local (it needs no attachment of condensation defects from the bulk). A quick computation shows that\footnote{\
Lines neutral under the $\bZ_M$ are of the form $\left(L_{(p)} \right)^r \otimes \left(L_{(q)}\right)^s$ with
\be
(r , \, s) = \begin{cases}
    (\ell, \, \ell) \mod N/M \\
    (p^{-1} \ell', \, -q^{-1} \ell') \mod N/M
\end{cases}
\ee
The two sets are mutually local and only the first is charged under the diagonal $\bZ_N$ symmetry.
}
\be
\frac{\left( \cA^{N, p} \otimes \cA^{N,q} \right) [b]}{\bZ_M}  = \cA^{N/M, \, (p^{-1} + q^{-1})/M} \otimes \cA^{N/M , \, (p + q)/M}[M b]
\ee
Where we emphasize the coupling to the $\bZ_{N/M}$ subgroup of the bulk one-form symmetry, as all lines in the second factor have charges multiple of $M$. 
Reinstating the grading by the chiral symmetry
\be
\Dchiral{p/N} \otimes \Dchiral{q/N} \xrightarrow{\quad\mor_M\quad}  \cA^{N/M, \, (p^{-1} + q^{-1})/M} \; \;  \Dchiral{\frac{(p+q)/M}{N/M}} \, , \ \ \ M = \gcd(p+q,N) \, .
\ee
And again the decoupled TFT $\cA^{N/M, \, (p^{-1} + q^{-1})/M}$ is interpreted as a ``TFT-valued" fusion coefficient. The case $p=-q$ deserves special attention. This fusion rule is usually written as
\be
\Dchiral{p/N} \otimes \overline{\Dchiral{p/N}} = (\bZ_N)_{Np}\left[\frac{1}{N} \star j^{(2)}\right] = \cond_N^{(0)} \left[\frac{1}{N} \star j^{(2)}\right] \, .
\ee
Using the twist morphism $\boldsymbol{\tau} : \cond_N^{(0)} \to \unit$ to terminate the condensate this is correctly reproduced by our computation.

\bigskip

The general case can be treated similarly. The fusion defect is
 \be
\Dchiral{p/N} \otimes \Dchiral{p'/N'} = {\Uchiral}_{(p N'+ p' N)/N N'} \otimes \left( \cA^{N, \, p} \otimes \cA^{N', \, p'} \right) \left[\frac{1}{K} \star j^{(2)}\right] \, , \ \ \ K = \text{lcm}(N,N') \, .
\ee
Let $M=\gcd(N,N')$ and $N=M K_1$, $M= M K_2$, $K= M K_1 K_2$. Uncharged lines now form a $\bZ_M$ subgroup, and are generated by:
\be
\cK = \left(L_{(p)}\right)^{K_1 p^{-1}} \otimes \left(L_{(p')}\right)^{-K_2 p'^{-1}}\, , \ \ \ \ \cK^M = 1\, ,
\ee
where the inverses are taken in $\bZ_M^\times$. They define a non-degenerate theory only if:\footnote{\ This follows from
\be
\theta\left[ K\right] = \exp\left(\pi i (pp')^{-1} \frac{p K_2 + p' K_1}{M} \right) \, .
\ee
}
\be
M' \equiv \gcd(p K_2 + p' K_1, M) = 1 \, .
\ee
When $M'\neq 1$ defining a consistent morphism requires us to gauge the anomaly-free decoupled $\bZ_{M'}$ symmetry. After gauging
\be
\frac{\left( \cA^{N, \, p} \otimes \cA^{N', \,p'} \right)[b]}{\bZ_{M'}} = \cA^{M/M', \, (p^{-1} K_1 + p'^{-1} K_2)/M'} \otimes \; \cA^{K/M', \, (p K_2 + p' K_1)/M'}\left[M' b \right]
\ee
From which equation \eqref{eq: 1morphisms} follows.

\paragraph{Twisted interfaces} As we have seen in Section \ref{ss: morphisms}, a $\Dchiral{p/N}$ defect often admits nontrivial 1-endomorphisms which are topological interfaces of the minimal $\cA^{N,p}$ TFT. We also concluded that, in the chiral symmetry category, these must be accompanied by a nontrivial bulk one-form symmetry dressing.

\bigskip

It is natural to ask how this construction extends to the fusion product
\be
\Dchiral{p/N} \otimes \Dchiral{p'/N'} \, .
\ee
Suppose that the two theories are such that $M' = \gcd(p K_2 + p' K_1, M)>1$, so that gauging a  neutral $\bZ_{M'}$ subgroup $\algA$ describes the one-morphism. This also implies that $\text{End}(\cA^{N,p} \otimes \cA^{N', p'})$ is nontrivial as $N, \, N'$ cannot be prime numbers.
An interface $\cV_\salgA \; \in \; \text{End}( \cA^{N, \, p} \otimes \cA^{N', \, p'})$ can be used to induce a new algebra $\cV_\salgA \left( \algA \right)$ from the neutral one $\algA$. 
This is implemented graphically by surrounding $\algA$ by a $\cV_\salgA$ surface:
\bea
\begin{tikzpicture}[scale=0.55]
\draw[dashed, color=black!30!green] (0,0) arc (180:0: 1 and 0.5); \draw[color=black!30!green] (2,3) arc (0:180: 1 and 0.5);
\draw[color=blue, line width= 1.5] (1,-1) -- (1,4);
\node[above] at (1,4) {$\algA$};
\filldraw[color=blue, fill=white!90!blue, opacity=0.85] (0,0) arc (-180:0: 1 and 0.5) -- (2,3) arc (0:-180: 1 and 0.5) -- cycle;
\node at (2.5, -0.25) {$\cV_\salgA$};
\node at (3,1.5) {$=$} ;
\draw[color=blue, line width= 1.5] (5,-1) -- (5,4);
\node[above] at (5,4) {$\cV_\salgA \left( \algA \right)$};
\end{tikzpicture}
\eea
We can define a new morphism:
\be
\overline{\cV_\salgA}\;  \circ \; \mor_{\algA} \; \in \; \Hom\left( \Dchiral{p/N} \otimes \Dchiral{p'/N'} , \, \Dchiral{(p/N + p'/N')}\right)
\ee
which amounts to gauging $\cV_\salgA\left( \algA \right)$ instead. This interface is charged under a $\bZ_M$ subgroup of the bulk one-form symmetry and thus comes attached to a $\bZ_M$ condensation defect.

\medskip

As an example consider $\Dchiral{1/8} \otimes \Dchiral{3/8}$. From our discussion this can be mapped to $\Dchiral{1/2}$ by gauging a $\bZ_4$ subgroup generated by the line $(L_1)^2 \otimes (L_2)^{2}$ which has integer spin and trivial charge under the diagonal $\bZ_8$ subgroup of the bulk one-form symmetry.
The $\cA^{1/8} \otimes \cA^{3/8}$ theory has an invertible $\bZ_2^C \times \bZ_2^C$ zero-form symmetry with both factors corresponding to charge conjugation. The generator of a single $\bZ_2^C$ is given by a condensation of the algebra $\salgA = \unit \oplus L^2 \oplus L^4 \oplus L^6$ in either of the two theories and must be supplemented by a bulk $\bZ_4$ condensation defect $\cond_4^{(0)}$.\footnote{\ There are also two non-invertible defects obtained by gauging $\salgA = \unit \oplus L^4$ which acts as a projector onto the $\cA^{2,1}$ sub-algebra generated by $\salgA$ and $L^2 \oplus L^6$. It will be discussed more thoroughly in Section \ref{sec: associatorchiral}.} 

\medskip

While $C \otimes C$ leaves the algebra $\algA$ invariant (and preserves the diagonal one-form symmetry charge), $\cV = \unit \otimes C$ does not. $\cV\left( \algA\right)$ corresponds to $(L_1)^2 \otimes (L_2)^{-2}$, which has integer spin but charge $4 \mod 8$ under the bulk $\bZ_8$.  
After gauging $\algA$ the remaining lines are the $\algA$ orbits of $L_1 \otimes L_2$ and $L_1 \otimes (L_2)^{-3}$, which generate the topological lines of the chiral defect $\Dchiral{1/2}$ and of the decoupled $\cA^{2, \,1}$ TFT respectively. Acting with $\cV_{\unit \otimes \salgA} = \unit \otimes C$ shows that these come as the images of $L_1 \otimes (L_2)^{-1}$ and $L_1 \otimes (L_2)^3$ which have the same spins and charge $\mp 2 \mod 8$. This assignment is still consistent since the charge conjugation interface can screen one-form symmetry charge mod 2. The morphism $\Tilde{\mor}_4 = \cV_{\unit \, \otimes \, \salgA} \circ \mor_4$ is still a map:
\be
\Dchiral{1/8} \otimes \Dchiral{3/8} \xrightarrow{\quad \Tilde{\mor}_4 \quad} \Dchiral{1/2} \, ,
\ee
which however only preserves one-form symmetry charge modulo 2.

\subsection{The associator for the chiral symmetry category}\label{sec: associatorchiral}
We now set out to explicitly describe the associator one-morphisms $\AF_{\ob_1, \ob_2, \ob_3}$ in the chiral symmetry category. 
We will do so by viewing them as interface TFTs:
\be
\AF_{\ob_1, \ob_2, \ob_3} : ((\ob_1 \otimes \ob_2) \otimes \ob_3) \to (\ob_1 \otimes (\ob_2 \otimes \ob_3)) \, .
\ee
and use the information encoded in the morphisms $\mor_K$ to describe them explicitly.
From our previous discussion the associator $\AF_{\Dchiral{p_1/N_1} , \, \Dchiral{p_2/N_2} , \, \Dchiral{p_3/N_3}}$ takes the diagrammatic form
\be \label{eq: associator}
\begin{gathered}
\begin{tz}[td,scale=2.5]
\begin{scope}[xyplane=0, on layer=superfront]
\draw[slice] (0,0) to [out=up, in=\dl] (1,2) to (1,4);
\draw[slice] (1,0) to [out=up, in=\dl] (1.5,1) to [out=up, in=\dr] (1,2);
\draw[slice] (2,0) to [out=up, in=\dr] (1.5,1);
\draw[slice] (0,0) to [out=down, in=\ul] (0.5,-1) to [out=down, in=\ul] (1,-2) to (1,-4);
\draw[slice] (1,0) to [out=down, in=\ur] (0.5,-1);
\draw[slice] (2,0) to [out=down, in=\ur] (1,-2);
\end{scope}
\begin{scope}[xzplane=2]
\draw[wire,short] (1,0) node[dot] {} to (1,\h) node[dot]{};
\end{scope}
\begin{scope}[xzplane=1]
\draw[wire,short] (1.5,0) node[dot] {} to (1.5,\h) node[dot]{};
\end{scope}
\begin{scope}[xyplane=0.3, on layer=superfront]
\node[right] at (1,4.5) {$\Dchiral{p/N}$};
\node[right] at (1.5,2.3) {$\Dchiral{p_{23}/N_{23}}$}  ;
\node[right] at (-1,0.75)  {$\Dchiral{p_1/N_1}$};
\node[right] at (1.9,0.75) {$\Dchiral{p_3/N_3}$};
\node[right] at (-0.3,-1.5) {$\Dchiral{p_{12}/N_{12}}$};
\node[right] at (1,-3.75) {$\Dchiral{p/N}$};
\node[right] at (0.25,0.4) {$\Dchiral{p_2/N_2}$};
\node at (2*\h,-0.25) {$\overline{\mor}_{K_{(12)3}}$};
\node at (2*\h,1.5*\h) {$\mor_{K_{1(23)}}$};
\node at (2*\h + 0.5,\h) {$\mor_{K_{2 3}}$};
\node at (2*\h - 0.5 ,0.5*\h -0.6) {$\overline{\mor}_{K_{1 2}}$};
\end{scope}
\begin{scope}[xyplane=\h, on layer=superfront]
\draw[slice] (0,0) to [out=up, in=\dl] (1,2) to (1,4);
\draw[slice] (1,0) to [out=up, in=\dl] (1.5,1) to [out=up, in=\dr] (1,2);
\draw[slice] (2,0) to [out=up, in=\dr] (1.5,1);
\draw[slice] (0,0) to [out=down, in=\ul] (0.5,-1) to [out=down, in=\ul] (1,-2) to (1,-4);
\draw[slice] (1,0) to [out=down, in=\ur] (0.5,-1);
\draw[slice] (2,0) to [out=down, in=\ur] (1,-2);
\end{scope}
\begin{scope}[xzplane=-1]
\draw[wire,short] (0.5,0) node[dot]{} to (0.5,\h) node[dot]{};
\end{scope}
\begin{scope}[xzplane=-2]
\draw[wire,short] (1,0) node[dot]{} to (1,\h) node[dot]{};
\end{scope}
\end{tz}
\end{gathered}
\ee
The two sides of the ``bubble" can be understood as defining two different ways of gauging a symmetry inside $\Dchiral{p_1/N} \otimes \Dchiral{p_2/N} \otimes \Dchiral{p_3/N}$ to reach $\Dchiral{p/N}$ (up to possibly different choices for decoupled TFTs). The gauging procedure is performed sequentially by:
\be
\mor_{K_{(12)3}} \circ \mor_{K_{12}} \ \ \ \text{or} \ \ \  \mor_{K_{1(23)}} \circ \mor_{K_{23}} \, .
\ee
These gauging operations can be described by giving two commutative symmetric Frobenius algebras $\algA^+ =    \algA_{12} \triangleright \algA_{(12)3} $ and $\algA^- =  \algA_{23} \triangleright \algA_{1(23)}$. We give a quick review of the relevant definitions below.
Shrinking the bubble gives rise to a codimension-one interface on $\Dchiral{p/N}$, which we dub the associator 1-morphism:
\bea
\begin{tz}[scale=0.5]
\draw[slice] (0,-0.5) to [out=up, in=\dl] (1,1);
\draw[slice] (2,-0.5) to [out=up, in=\dr] (1,1);
\draw[slice] (1,1) to [in=\dl, out=up] (2,2.5);
\draw[slice] (4,-0.5) to [out=up, in=\dr] (2,2.5) to (2,3.5);
\draw[slice] (0,-0.5) to [in=\ul, out=down] (2,-3.5) to (2,-4.5);
\draw[slice] (2,-0.5) to [out=down, in = \ul] (3,-2);
\draw[slice] (4,-0.5) to [out=down, in = \ur] (3,-2);
\draw[slice] (3,-2) to [out=down, in = \ur] (2,-3.5);
\node[dot, label={[label distance=-0.1cm]0:$\algA_{12}$}] at (1,1) {}; \node[dot, label={[label distance=-0.2cm]45:$\algA_{(12)3}$}] at (2,2.5) {}; \node[dot,label={[label distance=-0.1cm]180:$\algA_{23}$}] at (3,-2) {}; \node[dot, label={[label distance=-0.2cm]-135:$\algA_{1(23)}$}] at (2,-3.5) {};
\node[left] at (2,-0.5) { $\frac{p_2}{N_2}$};
\node[left] at (0,-0.5) { $\frac{p_1}{N_1}$};
\node[right] at (4,-0.5) { $\frac{p_3}{N_3}$};
\node[left] at (1.5,2.25) { $\frac{p_{12} }{N_{12} }$};
\node[right] at (2.5,-3.25) { $\frac{p_{23} }{N_{23} }$};
\node[above] at (2,3.5) { $\frac{p}{N}$};
\node[below] at (2,-4.5) { $\frac{p}{N}$};
\node at (6,-0.5) {$=$};
\draw[slice] (8,-4.5) to (8,3.5);
\node[above] at (8,3.5) { $\frac{p}{N}$};
\node[below] at (8,-4.5) { $\frac{p}{N}$};
\draw[fill=blue] (8,-0.5) circle (0.1); \node[right] at (8.5,-0.5) {$\ds \AF \begin{bmatrix} \algA_{12} & \algA_{(12)3} \\ \algA_{23} & \algA_{1(23)} \end{bmatrix}_{p_1/N_1, \, p_2/N_2, \, p_3/N_3} $};
\end{tz}
\eea
Alternatively, it can be represented  as a slab $\Sigma_2 \times I$ with topological boundary conditions on the two ends implementing the condensation of $\algA^+$ and $\algA^-$ respectively.
\bea \label{eq: slab}
\begin{tz}
\draw[slice] (0,2) to (1,2) to (1,0) to (0,0);
\filldraw[slice, color=white!95!blue] (1,2) to (4,2) to (4,0) to (1,0) --cycle;
\draw[slice] (5,2) to (4,2) to (4,0) to (5,0);
\draw[slice, color=blue, line width=1] (1,2) to (1,0) node[below] {${}_{\algA^-}\text{Mod}$};
\draw[slice, color=blue, line width=1] (4,2) to (4,0) node[below] {$\text{Mod}_{\algA^+}$};
\node at (0,1) {$\frac{p}{N}$};
\node at (5,1) {$\frac{p}{N}$};
\node at (2.5,1) {$\frac{p_1}{N_1} \otimes \frac{p_2}{N_2} \otimes \frac{p_3}{N_3}$};
\node at (6.5,1) {\Huge$ \overset{\text{\small Shrink}}{\leadsto}$};
\begin{scope}[shift={(8.5,0)}]
\draw[slice] (0,2) to (2,2); \draw[slice] (0,0) to (2,0);
\draw[slice, color=blue, line width=1] (1,0) node[below] {$\AF_{\Dchiral{p_1/N_1} , \Dchiral{p_2/N_2} , \Dchiral{p_3/N_3}}$} to (1,2);
\node at (0,1) {$\frac{p}{N}$};
\node at (2,1) {$\frac{p}{N}$};
\end{scope}
\end{tz}
\eea
The module categories $\text{Mod}_{\algA^+}$ and ${}_{\algA^-}\text{Mod}$ of right-$\algA^+$ (and left $\algA^-$)-modules describe the gauging interfaces for $\algA^+$ and $\algA^-$ (see \eg \cite{Fuchs:2002cm} for an extensive review).
Operators confined to the interface form an explicitly computable category. Notice that we do not consider deconfined operators, which can be lifted off the interface and into the bulk.\footnote{\ More specifically, for a single interface a line operator $L$ is deconfined if it is in the image of the inclusion map $\iota: \cC^{\text{loc}}_\algA \xrightarrow{\quad} \text{Mod}_\algA$. With $\cC^{\text{loc}}_\algA$ being the category of local $\algA$ modules describing the theory after condensation of $\algA$.} The 2d TFT also contains various local vertex operators. Some of them only exist as interfaces between lines on the two sides. We also do not consider them in our description.
We introduce the following notation:
\bea\label{2dTFTdata}
&\fM &&= \biggl\{\text{Lines living on the F-symbol TFT} \biggr\}\,,\\
&\fM_{c} &&= \biggl\{ \text{Lines confined on the F-symbol TFT} \biggr\}\,, \\
&\mathfrak{V} &&= \biggl\{\text{Local vertex operators on the F-symbol TFT} \biggr\} \, .
\eea

Confined line operators $M \in \fM_c$ couple in general to a larger bulk one-form symmetry $\bZ_{\text{lcm}(N_1, N_2, N_3)}$ than $\bZ_N$ and form a condensation defect living on the associator morphism (interface). In Section~\ref{sec:monopoleline}, we show that this can be detected by passing a dynamical monopole line $T$ though the associator interface, thus making it a physical observable.

As we have seen in Section \ref{ss: morphisms} the category of endomorphisms (topological interfaces) of the chiral defect $\Dchiral{p/N}$ is described by the symmetry surface defects $\cV_{\salgA}$ for the corresponding minimal TFT. Thus the associator 1-morphism must act as a particular endomorphism $\cV_{\salgA}$, which we will explain in more detail below.

\paragraph{Structure of the associator}  Before delving into explicit examples, let us comment the general structure which may appear in the non-invertible chiral symmetry example. A given associator $\AF$ implements a topological interface between $ \cA_+ \otimes \Dchiral{p/N} $ and $\cA_- \otimes \Dchiral{p/N}$, where $\cA_+$ and $\cA_-$ are decoupled TFT depending on the choice of the gauged algebras $\algA^+$ and $\algA^-$. It can be shown \cite{Kong:2013aya} that the two sides in this case are always Witt equivalent, which in particular implies that the associator induces a well defined endomorphism for the class $\lls \Dchiral{p/N} \rrs$. This is labelled by a zero-form symmetry generator $\cV_{\salgA}$ for the minimal representative $\cA^{N, \, p}$.
If $\AF$ is an \textit{untwisted} interface --- i.e. it does not live at the end of a bulk condensate --- then it must act trivially on charged lines.

\medskip

On the other hand, if we allow for \textit{twisted} interfaces (i.e. attached to bulk condensates), then the F-symbol TFT can acts as a nontrivial endomorphism $\cV_{\salgA} \; \in \; \text{End}\left( \cA^{N, \, p} \right) $ on the charged lines.
In practice this happens whenever any of the algebras $\algA_{12}, \, \algA_{23}, \, \algA_{1(23)}, \, \algA_{(12)3}$ to be charged under the bulk one-form symmetry. Notice that $\cV_{\salgA}$ could be either invertible or non-invertible. 
We can equivalently regard these interfaces as arising from the insertion of a zero-form symmetry defect in the middle of the slab. 
To show this consider folding the configuration in \eqref{eq: slab} in the middle:
\bea \label{eq: Witteq}
\begin{tz}
\filldraw[slice, color=white!95!blue] (1,2) to (3,2) to (3,0) to (1,0) --cycle; 
\node at (2,1) { $ \small \cC \boxtimes \overline{\cC}$};
\draw[slice] (5,2) to (3,2) to (3,0) to (5,0);
\node at (4.5,1) { $\small \cC/\algA^+ \boxtimes \overline{\cC/\algA^-}$};
\draw[slice, color=blue, line width=1] (3,2) to (3,0) node[below] {$\text{Mod}_{\algA^+ \boxtimes \overline{\algA^-}}$};
\draw[slice, color=blue, line width=1] (1,2) to (1,0) node[below] {$\text{Dir}$};
\node at (8,1) {\Huge$ \overset{\text{\small Shrink}}{\leadsto}$};
\begin{scope}[shift={(9.5,0)}]
 \draw[slice] (3,2) to (1,2) to (1,0) to (3,0); 
\draw[slice, color=blue, line width=1] (1,2) to (1,0) node[below] {$\AF$}; 
\node at (2.5,1) {$ \small \cC/\algA^+ \boxtimes \overline{\cC/\algA^-}$};
\end{scope}
\end{tz}
\eea
Here $\cC$ is the shorthand for the category of lines describing $\Dchiral{p_1/N_1} \otimes \Dchiral{p_2/N_2} \otimes \Dchiral{p_3/N_3} $ and we have chosen the canonical Dirichlet boundary condition corresponding to the condensation of the diagonal algebra in $\cC \boxtimes \overline{\cC}$. Different choices of gapped boundary conditions on the left lead to different associators $\AF$ after shrinking the slab. Such choices are indeed known to be in one-to-one correspondence with zero-form symmetry surface operators $\cV_{\salgA}$ in $\cC$ \cite{Kapustin:2010if}.

\bigskip

Let us describe the generic structure of the interface for uncharged algebras $\algA^+$ and $\algA^-$. Two interface lines $M, \, M' \in \fM$ in \eqref{2dTFTdata} should be identified under left fusion with $\algA^-$ and right fusion with $\algA^+$.\footnote{\ This is just the statement that it must be simultaneously a left $\algA^-$ module and a right $\algA^+$ module. See \cite{Fuchs:2002cm} for a review of the category of modules over an algebra.}
Such invariant lines can be labelled by elements of the double quotient:
\be
M \in \algA^- \backslash \cC / {\algA^+} \, ,
\ee
and compose in a group-like manner. Let us introduce the short-hand $\algA^{+-} = \algA^+ \cap \, \algA^-$.
In terms of the simple objects $L \in \cC$, the physically distinct interface lines $M$ in the F-symbol TFT can be constructed as orbits:
\be\label{defML}
M(L) = \frac{1}{|\algA^{+ -}|} \algA^- \otimes L \otimes \algA^+ \, .
\ee
A line $M$ generally carries charge under a $\bZ_{N_{123}}$ bulk one-form symmetry, with $N_{123}= \text{lcm}(N_1, N_2, N_3)$. Of this only a $\bZ_N$ subgroup couples consistently with the theory after the gauging. Lines charged under the quotient:
\be
\bZ_K = \bZ_{N_{123}}/\bZ_N \, , \
\ee
must thus remain confined onto the interface.
Likewise we can consider local vertex operators $V \in \mathfrak{V}$ in the F-symbol TFT. These can be described by lines $H \in \cC$ stretching between the two sides of the slab:
\bea \label{eq: vertexop}
\begin{tz}
\draw[slice] (0,2) to (1,2) to (1,0) to (0,0);
\filldraw[slice, color=white!95!blue] (1,2) to (4,2) to (4,0) to (1,0) --cycle;
\draw[slice] (5,2) to (4,2) to (4,0) to (5,0);
\draw[slice, color=blue, line width=1] (1,2) to (1,0) node[below] {${}_{\algA^-}\text{Mod}$};
\draw[slice, color=blue, line width=1] (4,2) to (4,0) node[below] {$\text{Mod}_{\algA^+}$};
\node at (0,1) {$\cC/\algA^-$};
\node at (5,1) {$\cC/\algA^+$};
\draw[wire] (1,1) node[dot] {} to (4,1) node[dot] {};
\node[below] at (2.5,1) {$H$};
\node at (6.5,1) {\Huge$ \overset{\text{\small Shrink}}{\leadsto}$};
\begin{scope}[shift={(9,0)}]
\draw[slice] (-1,2) to (3,2); \draw[slice] (-1,0) to (3,0);
\draw[fill=blue] (1,1) circle (0.05) node[right] {$V(H)$}  ;
\draw[slice, color=blue, line width=1] (1,0) node[below] {$\AF_{\Dchiral{p_1/N_1} , \Dchiral{p_2/N_2} , \Dchiral{p_3/N_3}}$} to (1,2);
\node at (-1,1) {$\cC/\algA^-$};
\node at (3,1) {$\cC/\algA^+$};
\end{scope}
\end{tz}
\eea
The left and right junctions describe maps between $H$ and the identity lines $\unit_{\algA^+}$ and $\unit_{\algA^-}$ of the condensed theory on the two sides. They can be thought of as inclusions:
\be
\iota^+ : H \xrightarrow{\quad} \algA^+ \, , \ \ \ \iota^-: H \xrightarrow{\quad} \algA^- \, ,
\ee
implying that $H$ must be an element of the intersection:
\be
H \in \algA^{+-} \, .
\ee
The vertex operator $V(H)$ is acted upon in a well defined way by the lines $M(L)$ \eqref{defML}:
\bea
\begin{tz}
\draw[fill=blue] (0,0) circle (0.05) node[below] {$V(H)$};
\draw[wire] (1,-1) node[below] {$M(L)$} to (1,1) ;
\node[right] at (1.5,0) {$= \ \ \boldsymbol{B}_{H,L} $};
\begin{scope}[shift={(4,0)}]
\draw[fill=blue] (1,0) circle (0.05) node[below] {$V(H)$};
\draw[wire] (0,-1) node[below] {$M(L)$} to (0,1) ;    
\end{scope}
\end{tz}
\eea
The relation follows from describing the following topological manipulation in the 3d slab:
\bea
\begin{tz}
\draw[slice] (0,2) to (1,2) to (1,0) to (0,0);
\filldraw[slice, color=white!95!blue] (1,2) to (4,2) to (4,0) to (1,0) --cycle;
\draw[slice] (5,2) to (4,2) to (4,0) to (5,0);
\draw[slice, color=blue, line width=1] (1,2) to (1,0) node[below] {${}_{\algA^-}\text{Mod}$};
\draw[slice, color=blue, line width=1] (4,2) to (4,0) node[below] {$\text{Mod}_{\algA^+}$};
\draw[braid wire] (3,0) to (3,2) node[above] {$L$};
\draw[braid wire] (1,1) node[dot] {} to (4,1) node[dot] {};
\node[below] at (2.5,1) {$H$};
\node at (6.5,1) {\Huge$ \overset{\text{\small Slide over}}{\leadsto}$};
\begin{scope}[shift={(9,0)}]
\node at (-0.5,1) {$\boldsymbol{B}_{H  \, L}$};
\draw[slice] (0,2) to (1,2) to (1,0) to (0,0);
\filldraw[slice, color=white!95!blue] (1,2) to (4,2) to (4,0) to (1,0) --cycle;
\draw[slice] (5,2) to (4,2) to (4,0) to (5,0);
\draw[slice, color=blue, line width=1] (1,2) to (1,0) node[below] {${}_{\algA^-}\text{Mod}$};
\draw[slice, color=blue, line width=1] (4,2) to (4,0) node[below] {$\text{Mod}_{\algA^+}$};
\draw[braid wire] (1,1) node[dot] {} to (4,1) node[dot] {};
\draw[braid wire] (1.5,0) to (1.5,2) node[above] {$L$};
\node[below] at (2.5,1) {$H$};
\end{scope}    
\end{tz}
\eea
The braiding is well defined on classes $M(L)$ due to the fact that $H$ has trivial braiding with all lines in both $\algA^+$ and $\algA^-$ (as both algebras are commutative).
This immediately leads to:\footnote{\ A deconfined line $M(L)$ lies in the image of the projector $\pi_{\algA^+}$ mapping the elements of the double quotient into the category of local $\algA^+$ modules. More explicitly, $M(L)$ contains a closed $\algA^+$ orbit which braids trivially with $\algA^+$ (the same is valid for $\algA^-$).
} 
\be
\boldsymbol{B}_{H,L} = 1 \, , \ \ \ \forall \, H \, \in \algA^{+-} \ \text{if} \ M(L) \ \text{is not confined} \, .
\ee
The converse is also true: the number $\frac{|\cC|}{|\algA^+||\algA^-|}$ of $M(L)$ lines transparent to all vertex operators is also the number of interface lines which are deconfined.\footnote{\ The counting is as follows: the number of neutral lines under $\algA^{+-}$ is $\frac{|\cC|}{|\algA^{+-}|}$, taking equivalence with respect to the double quotient divides this by a factor $\frac{|\algA^+||\algA^-|}{|\algA^{+-}|}$. To count deconfined lines we must find how many lines are neutral w.r.t. either $\algA^+$ or $\algA^-$ and then take the quotient. The former number is $\frac{|\cC|}{|\algA^+|} \frac{|\cC|}{|\algA^-|}/\frac{|\cC| |\algA^{+-}|}{|\algA^+| |\algA^-|} = \frac{|\cC|}{|\algA^{+-}|}$, while taking quotients divides this by a factor $\frac{|\algA^+||\algA^-|}{|\algA^{+-}|}$ giving the desired result.
}
Since $|\fM| = \frac{|\cC| |\algA^{+-}|}{|\algA^+| |\algA^-|}$ we have that the equivalence,
\be
\fM_c \simeq \mathfrak{V} \, ,
\ee
as sets. We can then restrict the pairing ${\bm B}_{H,L}$ to a nondegenerate pairing $\tilde{\boldsymbol{B}}_{V,M}$ on $\mathfrak{V} \times \fM_c$. Such a pairing can be used to describe idempotents:
\be
\pi_M = \frac{1}{|\fM_c|} \sum_{V \, \in \, \mathfrak{V} } \, \check{\boldsymbol{B}}_{V, M} \, V \, , \ \ \ \  \pi_M \otimes \pi_{M'} = \delta_{M \, M'} \, \pi_M \, ,
\ee
which further satisfy:
\bea
\begin{tz}
\draw[fill=blue] (0,0) circle (0.05) node[below] {$\pi_M$};
\draw[wire] (1,-1) node[below] {$M'$} to (1,1) ;
\node[right] at (1.5,0) {$= $};
\begin{scope}[shift={(2.75,0)}]
\draw[fill=blue] (1,0) circle (0.05) node[below] {$\pi_{M + M'}$};
\draw[wire] (0,-1) node[below] {$M'$} to (0,1) ;    
\end{scope}
\end{tz}
\eea
Thus the confined lines $M'$ are domain walls between 2d vacua $\pi_M$ and $\pi_{M + M'}$. Clearly all boundary conditions have the same quantum dimension here and the bulk coupling for the associator only depends on the degeneracy of charged boundary conditions. Due to the quotient by decoupled lines (which must lie at the end of operators $\Uonef_{\frac{r K}{N_{123}}}$) the F-symbol does not couple locally to the full $\bZ_{N_{123}}$ but only to its $\bZ_K$ quotient.\footnote{\ In the following we assume that $\algA^+$ and $\algA^-$ are both neutral.} As insertions of $\pi_\unit$ amount to insertions of trivial $\bZ_K$ one-form symmetry surfaces, we can assign to $\pi_M$ the same charge as $M$ modulo $K$. We can thus expand:
\be
\AF[\algA^+ \ \algA^-] = \cV_\salgA \otimes \bigoplus_{q=0}^{K-1} d_q \ \Uonef_{\frac{q}{N_{123}}} \, ,
\ee
where $d_q$ is the number of charge $q \mod K$ confined lines $M \in \fM_c$ and
we implicitly mod out by $\Uonef_{K/N_{123}} \simeq \unit$.
As charged lines are equally distributed in the minimal TFTs we conclude that $d_q = d_0$ and the F-symbol is always a $\bZ_K$ condensate (up to an overall decoupled TFT).
Notice that, even if the intersection of the gauged algebras $\algA^{+-}$ is trivial, the F-symbol can still carry information corresponding to a non-trivial endomorphism of $\Dchiral{p/N}$.

\bigskip

It is natural to ask how the bulk theory detects lines confined to the associator interface $\AF$. The answer is that we can use non-topological monopole lines $T^N$ as probes. We will show at the end of this Section that the associator TFT acts on them as a projector, thus making it into a physical observable.

\paragraph{Junctions}
After having described the category of confined lines living on the F-symbol $\AF$ we would like to further characterize the interface TFT through its action on operators. The action of $\AF_{p_1/N_1, \, p_2/N_2 , \, p_3/N_3}$ defines a space of topological junctions between lines $L$ and $L'$ on the two sides
\bea
\begin{tikzpicture}
    \node[right] at (5,0.75) {$ \ds \mor_{L, L'} \; \in \; \Hom_{\AF_{p_1/N_1, \, p_2/N_2 , \, p_3/N_3}} \left( L, \, L' \right) \, .$};
          \draw[line width=0.75] (1,0.75) -- (2.5,1.75);
         
        \filldraw[fill=white!95!blue, opacity=0.5] (0,0) -- (2,-0.5) -- (2,1.5) -- (0,2) -- cycle;
       
        \draw[line width=0.75] (-0.5,-0.25) -- (1,0.75);
        \draw[fill=blue] (1,0.75) circle (0.05);
       \node at (1.45,0.3) {\small $\mor_{L, L'}$};
       \node[below] at (-0.5,-0.25) {$L$};
       \node[above] at (2.5,1.75) {$L'$};
       \node at (2.25,-0.75) {$\AF$};
    \end{tikzpicture}
\eea
The structure of the allowed junctions $\mor_{L, L'}$ is quite constrained. Since $L$ and $L'$ can both be brought onto the interface the tensor product
\be
L \otimes M \, , \ \ \ M \; \in \; \fM \, ,
\ee
induces a new (generally reducible) interface line. At the level of the quotient this induces a tensor product operation $L \otimes_{\algA^+} M$.
The same is valid for the map $M \otimes L'$, which induces a line $M \otimes_{\algA^-} L'$. We can decompose this into irreducible components
\be
L \otimes_{\algA^+} M = \bigoplus_{M'} \; n_L(M, M') \; M' \, , \ \ \ n_L(M, M') = \dim \left( \Hom_{\AF_{p_1/N_1, \, p_2/N_2 , \, p_3/N_3}} \left( L \otimes M, M'\right) \right) \, ,
\ee
such decomposition correspond to the existence of junctions
\bea
 \begin{tikzpicture}
        \filldraw[fill=white!95!blue, opacity=0.5] (0,0) -- (2,-0.5) -- (2,1.5) -- (0,2) -- cycle;
        \draw[line width=0.75] (-0.5,-0.5) -- (1,0.5);
        \draw[color=blue, line width=1] (1,0.5) -- (1,-0.25);
           \draw[color=blue, line width=1] (1,0.5) -- (1,1.75);
           \draw[fill=blue] (1,0.5) circle (0.05);
       \node[right] at (1,0.75) {$\mor_L(M,M')$};
       \node[below] at (-0.5,-0.5) {$L$};
       \node[below] at (1,-0.25) {$M$};
       \node[above] at (1,1.75) {$M'$};
              \node at (2.25,-0.75) {$\AF$};
       \node[right] at (5,0.75) {$\ds \mor_L(M, M') \; \in \; \Hom_{\AF_{p_1/N_1, \, p_2/N_2 , \, p_3/N_3}} \left( L \otimes M, M'\right) \, .$};
    \end{tikzpicture}
\eea
Which leads to:
\bea
\begin{tikzpicture}
\begin{scope}[shift={(-5,0)}]
   \draw[line width=0.75] (1,0.75) -- (2.5,1.75);
        \filldraw[fill=white!95!blue, opacity=0.5] (0,0) -- (2,-0.5) -- (2,1.5) -- (0,2) -- cycle;
        \draw[line width=0.75] (-0.5,-0.25) -- (1,0.75);
        \draw[fill=blue] (1,0.75) circle (0.05);
       \node at (1.5,0.25) {$\mor_{L, L'}$};
       \node[below] at (-0.5,-0.25) {$L$};
       \node[above] at (2.5,1.75) {$L'$};
       \node at (2.25,-0.75) {$\AF$};  
       \node at (3.5,0.75) {$=$};
\end{scope}
     \draw[line width=0.75] (1,1) -- (2.5,2);
        \filldraw[fill=white!95!blue, opacity=0.5] (0,0) -- (2,-0.5) -- (2,1.5) -- (0,2) -- cycle;
        \draw[line width=0.75] (-0.5,-0.5) -- (1,0.5);
        \draw[color=blue, line width=1,dashed] (1,0.5) -- (1,-0.25);
           \draw[color=blue, line width=1] (1,0.5) -- (1,1);
           \draw[fill=blue] (1,0.5) circle (0.05);
         \draw[fill=blue] (1,1) circle (0.05);
       \draw[color=blue, line width=1,dashed] (1,1) -- (1,1.75);
       \node[right] at (1,0.25) {$\mor_L(\unit,M)$};
      \node[left] at (1,1.25) {$\overline{\mor}_{L'}(M,\unit)$};
      \node[left] at (1,0.75) {$M$};
       \node[below] at (-0.5,-0.5) {$L$};
       \node[below] at (1,-0.25) {$\unit$};
       \node[above] at (1,1.75) {$\unit$};
              \node at (2.25,-0.75) {$\AF$};
      \node[above] at (2.5,2) {$L'$};
  \node[right] at (4,0.75) {$\ds \mor_{L, L'} = \mor_L(\unit, M) \circ \overline{\mor}_{L'} (M, \unit)$}; 
\end{tikzpicture}
\eea
We suppress a possible sum over the intermediate $M$, which will be absent in the cases we consider. A topological interface between MTCs must implement an automorphism of the modular data. This implies that the spin of a line is conserved before and after the action of $\AF_{p_1/N_1, \, p_2/N_2 , \, p_3/N_3}$:\footnote{When working with spin TFTs the action preserves $\theta$ modulo a sign.}
\be
\theta\left[ \AF_{p_1/N_1, \, p_2/N_2 , \, p_3/N_3}\left( L \right)  \right] \equiv \theta[L'] = \theta\left[ L \right]
\ee
This condition strongly constrains the possible allowed junctions. We now give some concrete examples for the computation of F-symbols. 
\paragraph{Example a)} 
\be
\unit \xrightarrow{\quad \overline{\mor}_N \quad} \Dchiral{p/N} \otimes \Dchiral{-p/N} \xrightarrow{\quad \mor_N \quad} \unit
\ee
Describing a ``bubble'' of $\Dchiral{p/N}$.
Clearly $\algA^+ = \algA^- = \algA $ and $\algA$ is spanned by diagonal lines $(L, \overline{L})$. The double quotient reduces to a single one, whose representatives can be taken to be:
\be
M(L \otimes \unit) \, . 
\ee
On the other hand we have vertex operators
\be
V(L\otimes\overline{L}) \, ,
\ee
described by stretching an $\algA$ line through the slab. Using $V(L^{p^{-1}} \otimes \overline{L}^{p^{-1}}) \equiv V$ and $M(L \otimes \unit) \equiv M$ as generators it is clear that they braid canonically and that:
\be
\pi_k = \frac{1}{N} \sum_{\ell=0}^{N-1} e^{\frac{2\pi i \ell}{N}} \, V^\ell \, ,
\ee
give the idempotent basis. The F-symbol becomes a 2-d condensate:
\be
\AF_{p/N, \, -p/N} = \bigoplus_{k = 0}^{N-1} \Uonef_{k/N} = \cond^{(1)}_N \, .
\ee

\paragraph{Example b)}
\be\cA^{2, \, 1} \; \unit \xrightarrow{\quad \overline{\mor}_{4} \quad } \Dchiral{1/4} \otimes \Dchiral{1/4} \otimes \Dchiral{1/2} \xrightarrow{\quad \mor_{2} \; \circ \; \mor_2 \quad} \cA^{2, \, 1} \; \unit \, . \ee
Which defines an F-symbol\footnote{\ We will denote the algebras $\algA_i$ by their generators for simplicity.}
\be
\AF\begin{bmatrix}
 (L_1)^2 \otimes (L_2)^2 & L_1 \otimes L_2 \otimes L_3 \\
 \unit & L_1 \otimes L_2 \otimes L_3
\end{bmatrix}_{1/4, \, 1/4, \, 1/2} 
\ee
This describes the $\bZ_4$ gauging either as a sequential gauging of $\bZ_2 \triangleright \bZ_2$ or as a one-step gauging of $\bZ_4$. We still have $\algA^+ = \algA^-=\algA$. The double quotient has eight elements generated by:
\bea
&M \equiv M(L_1 \otimes \unit \otimes \unit) \, , \ \ \ &&M^4 = 1 \, , \\
&\Tilde{M} \equiv M(L_1^2 \otimes \unit \otimes L_3) \, , \ \ \ &&\Tilde{M}^2 = 1 \, .
\eea
The line $\Tilde{M}$ has vanishing one-form symmetry charge modulo 4 and spin $1/4$ modulo $1/2$. Indeed it can be lifted in the bulk to describe the generator of the $\cA^{2,1}$ theory. Vertex operators are generated by:
\be
V \equiv V(L_1 \otimes L_2 \otimes L_3) \, , \ \ \ V^4 = 1 \, ,
\ee
are neutral to $\Tilde{M}$ and have canonical pairing with $M$. Again we conclude that the F-symbol is a condensate:
\be
\AF\begin{bmatrix}
 (L_1)^2 \otimes (L_2)^2 & L_1 \otimes L_2 \otimes L_3 \\
 \unit & L_1 \otimes L_2 \otimes L_3
\end{bmatrix}_{1/4, \, 1/4, \, 1/2} = \bigoplus_{k=0}^3 \Uonef_{k/4} = \cond^{(1)}_4 \, .
\ee

\paragraph{Example c)} 
\be\Dchiral{1/4} \xrightarrow{\quad \overline{\mor}_4 \quad} \Dchiral{1/4} \otimes \Dchiral{1/4} \otimes \Dchiral{3/4}  \xrightarrow{\quad \mor_2 \quad} \left( \cA^{2, \,1} \right)^2 \; \Dchiral{1/4} \ee
In this case the choice of algebra is not unique due to the charge conjugation symmetry of $\Dchiral{1/4}$.
We start with the neutral interface:
\be
\AF \begin{bmatrix}
(L_1)^2 \otimes (L_2)^2  & \unit \\
L_2 \otimes L_3  & \unit 
\end{bmatrix}_{1/4, \, 1/4, \, 3/4}
\ee
$\algA^+ = (L_1)^2 \otimes (L_2)^2 $ and $\algA^- = L_2 \otimes L_3$ now have an empty intersection. 
Lines after gauging $\algA^+$ are described by:
\footnote{\ Notice that we always think in terms of spin theories, in particular $\cA^{2, \,1} = \cA^{2, \, -1}$. This is taken into account also in the list below.}
\bea
\cM &= M_{\algA^+} \left( L_1 \otimes L_2 \otimes (L_3)^2 \right) \, , \ \ \ \cM^2 &&=1 , \ \ &&&\cA^{2,1} \, , \\
\cM' &= M_{\algA^+} \left( L_1 \otimes (L_2)^3 \otimes \unit \right) \, , \ \ \ (\cM')^2&&=1, \ \ \ &&&\cA^{2,1} \, , \\
\cL &= M_{\algA^+} \left( L_1 \otimes L_2 \otimes L_3 \right) \, , \ \ \ \ \ \ \ \cL^4&&=1 , \ \ &&&\Dchiral{1/4} \, ,
\eea
where $M_{\algA}(L)= L \otimes \algA$. While in the case of $\algA^-$
\be
\cL'= M_{\algA^-} \left( L_1 \otimes \unit \otimes \unit \right) \, , \ \ \ (\cL')^4=1 , \ \ \ \Dchiral{1/4} \, .
\ee
On the other hand the double quotient is generated by:
\bea
&M \equiv M(L_1 \otimes \unit \otimes \unit) \, , \ \ \ &&M^4 = 1 \, , \\
& \Tilde{M}\equiv M ( L_1 \otimes \unit \otimes L_3 ) \, , \ \ \ &&\Tilde{M}^2=1 \, .
\eea
Since
\be
\cL \; \otimes_{\algA^+} \; M(\unit) = M(\unit) \; \otimes_{\algA^-} \; \cL' = M
\ee
the interface acts as the identity on the $\Dchiral{1/4}$ lines. Notice that there are no vertex operators and all lines can be lifted into the bulk. The interface is thus trivial:\footnote{\ Notice that the line $\Tilde{M}$ serves as a boundary condition for the (fermionic) DW theory $(A^{2,1})^2$ and can be lifted in the bulk.} 
\be
\AF \begin{bmatrix}
(L_1)^2 \otimes (L_2)^2  & \unit \\
L_2 \otimes L_3  & \unit 
\end{bmatrix}_{1/4, \, 1/4, \, 3/4} = \unit \, .
\ee

\bigskip

Let us now consider twisting the interface by the zero-form symmetry $\unit \otimes \unit \otimes C$. We can make it act on $\algA^-$, giving a new algebra generated by $L_2 \otimes (L_3)^{-1}$ which has charge $2 \mod 4$. We have the interface
\be
\AF \begin{bmatrix}
(L_1)^2 \otimes (L_2)^2  & \unit \\
L_2 \otimes (L_3)^{-1}  & \unit 
\end{bmatrix}_{1/4, \, 1/4, \, 3/4}
\ee
The structure is the same as before, but using the new algebra we now find
\be
\cL \otimes_{\algA^+} \; M(\unit) = M(\unit) \otimes_{\algA^-} (\cL')^3 \, ,
\ee
signalling that the F-symbol
\be
\AF \begin{bmatrix}
(L_1)^2 \otimes (L_2)^2  & \unit \\
L_2 \otimes (L_3)^{-1}  & \unit 
\end{bmatrix}_{1/4, \, 1/4, \, 3/4} = C \, ,
\ee
implements charge conjugation.

\paragraph{Example d): non invertible interfaces}
This construction can be used to generate F-symbols which involve non invertible interfaces. The first example is the fusion $\Dchiral{1/8} \otimes \Dchiral{3/8} \otimes \Dchiral{5/8}$. All of the defects have a non invertible interface $P$ which acts as a projector on the lines $\unit \oplus L^4$ and $L^2 \oplus L^6$.\footnote{\ In CFT language this corresponds to the fact that $U(1)_8$ has a chiral subalgebra $U(1)_2$} $P$ can be described as an higher gauging:
\be
P = \cV_{ \unit \oplus L^4  } \, 
\ee
where the condensed algebra has lines of charge $0, 4 \mod 8$ and must be viewed as the boundary of a bulk $\cond^{(0)}_{2}$. In this case the F-symbol involving $P$ can be found e.g. by choosing
\be
\algA^+ = \bigoplus_{r = 0}^4 L_1^{2r} \otimes L_2^{2 r}  \, , \ \ \ \algA^- = \left(\bigoplus_{s =0}^4 L_2^{2 s} \otimes L_3^{2 s} \right) \otimes \left(\unit \oplus L_3^4 \right) \, .
\ee
Indeed $\algA^-$ is the image under $\unit \otimes \unit \otimes P$ of the neutral algebra ${\algA'}^- = \bigoplus_{r=0}^8 L_2^r \otimes L_3^{- 3 r}$. Notice that there too the algebras have trivial intersection.
A similar computation as in the previous example gives:
\be
\AF[\algA^+ \ \algA^-]_{1/8, \, 3/8, \, 5/8} = P \, .
\ee

\subsection{Two-morphisms from the associator}
Having described the associator TFT there is a rather simple way to extract the two-morphisms $\mmor_{\algA^+ , \, \algA^-}$ from it. Consider starting from \eqref{eq: 2morphismspic} and ``bending" the picture as follows:
\bea
\begin{tz}
 \draw[slice] (0,0) -- (6,0) -- (6,4) -- (0,4) -- cycle;
\draw[wire] (2,0) to[out=up, in=left] (3,2) to[out=right, in=up] (4,0);
 \node at (3,2) {$\bullet$};
 \node[above] at (3,2) {$\mmor_{\algA^+ , \, \algA^- }$};
 \node at (1,3) {\small$ \lls \Dchiral{p/N} \rrs$};   
 \node at (3,-0.75) {\small $\lls \Dchiral{p_1/N_1} \otimes \Dchiral{p_2/N_2} \otimes \Dchiral{p_3/N_3} \rrs$};
 \node[rotate=-90] at (3,0) {\Large $\leadsto$};
\node[left] at (2,1) {$\text{Mod}_{\algA^+}$};
\node[right] at (4,1) {$\text{Mod}_{\algA^-}$};
\node at (6.5,2) {$=$};
\begin{scope}[shift={(7,0)}]
\draw[slice] (0,0) -- (6,0) -- (6,4) -- (0,4) -- cycle;
\draw[wire, color=blue] (3,0) node[below] {$\AF_{p_1/N_1, \, p_2/N_2, \, p_3/N_3}$} to (3,2);    
\node at (3,2) {$\bullet$};
\node[above] at (3,2) {$\mmor_{\algA^+ , \, \algA^- }$};
\end{scope}
\end{tz}
\eea
Thus we can extract the two-morphism $\eta_{\algA^+ , \, \algA^-}$ from the corresponding associator by considering gapped (topological) boundary conditions for the associator TFT. More precisely, given our choice of basis for the associators $\AF\begin{bmatrix} \algA_{12} & \algA_{(12)3} \\ \algA_{23} & \algA_{1(23)} \end{bmatrix}_{p_1/N_1, \, p_2/N_2, \, p_3/N_3}$ we can can define the two-morphisms $\mmor\begin{bmatrix} \algA_{12} & \algA_{(12)3} \\ \algA_{23} & \algA_{1(23)} \end{bmatrix}$ as Dirichlet boundary conditions for $\AF$. These are isomorphic to the set of idempotents $\pi_M$  we have described above.

\subsection{The pentagonator}
\begin{figure}[!htb]
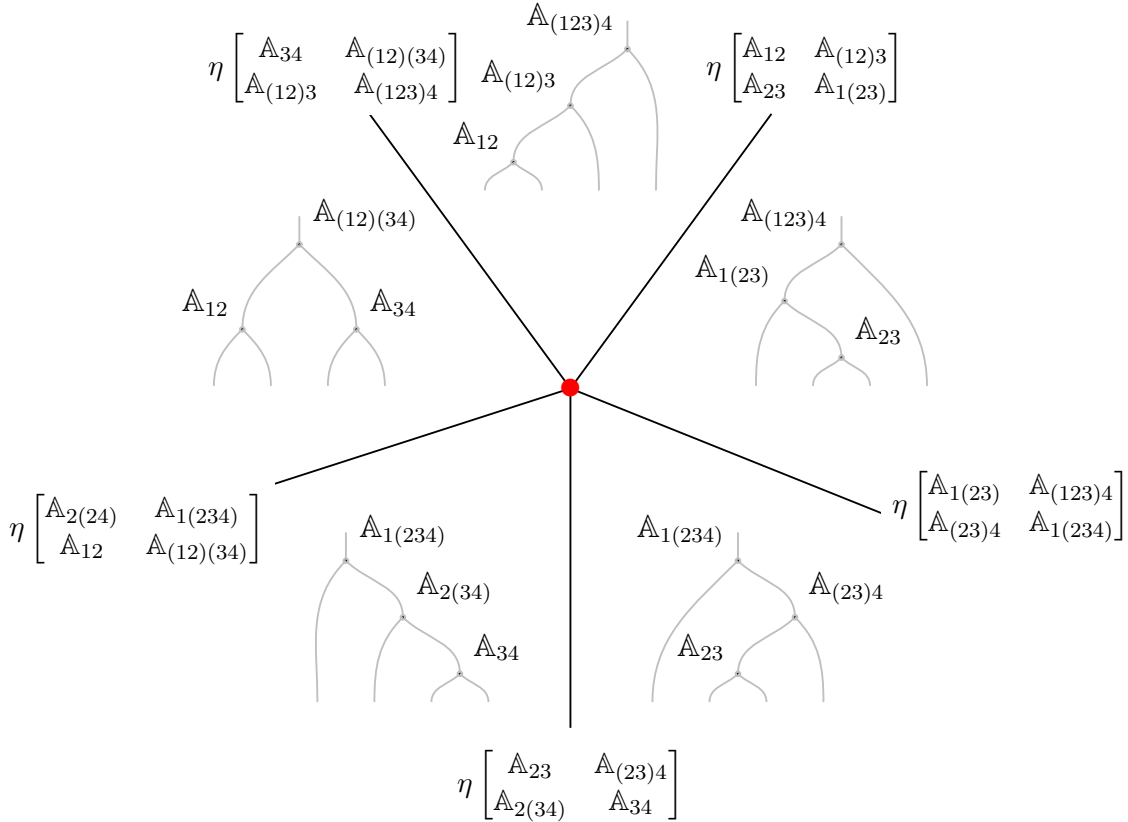

\label{fig:pentagonator}
\begin{tz}[scale=0.75]
\begin{scope}[shift={(-1.5,-1.5)}]
\begin{scope}[shift={(90:5)}]
\draw[slice] (0,0) to[out=up, in=\dl] (0.5,0.5) node[dot, label={[label distance=0.2]135:$\algA_{12}$}] {} to[out=up, in=\dl] (1.5,1.5) node[dot, label={[label distance=0.2]135:$\algA_{(12)3}$}] {} to[out=up, in=\dl]  (2.5,2.5) node[dot, label={[label distance=0.2]135:$\algA_{(123)4}$}] {} to (2.5,3);    
\draw[slice] (1,0) to[out=up, in=\dr] (0.5,0.5);
\draw[slice] (2,0) to[out=up, in=\dr] (1.5,1.5);
\draw[slice] (3,0) to[out=up, in=\dr] (2.5,2.5);
\end{scope}
\begin{scope}[shift={(18:5)}]
\draw[slice] (1,0) to[out=up, in=\dl] (1.5,0.5) node[dot, label={[label distance=0.2]45:$\algA_{23}$}] {} to[out=up, in=\dr] (0.5,1.5) node[dot, label={[label distance=0.2]135:$\algA_{1(23)}$}] {} to[out=up, in=\dl]  (1.5,2.5) node[dot, label={[label distance=0.2]135:$\algA_{(123)4}$}] {} to (1.5,3);    
\draw[slice] (0,0) to[out=up, in=\dl] (0.5,1.5);
\draw[slice] (2,0) to[out=up, in=\dr] (1.5,0.5);
\draw[slice] (3,0) to[out=up, in=\dr] (1.5,2.5);    
\end{scope}
\begin{scope}[shift={(-54:5)}]
\draw[slice] (1,0) to[out=up, in=\dl] (1.5,0.5) node[dot, label={[label distance=0.2]135:$\algA_{23}$}] {} to[out=up, in=\dl] (2.5,1.5) node[dot, label={[label distance=0.2]45:$\algA_{(23)4}$}] {} to[out=up, in=\dr]  (1.5,2.5) node[dot, label={[label distance=0.2]135:$\algA_{1(234)}$}] {} to[out=up, in=down] (1.5,3);    
\draw[slice] (0,0) to[out=up, in=\dl] (1.5,2.5);
\draw[slice] (2,0) to[out=up, in=\dr] (1.5,0.5);
\draw[slice] (3,0) to[out=up, in=\dr] (2.5,1.5);    
\end{scope}
\begin{scope}[shift={(-126:5)}]
\draw[slice] (2,0) to[out=up, in=\dl] (2.5,0.5) node[dot, label={[label distance=0.2]45:$\algA_{34}$}] {} to[out=up, in=\dr] (1.5,1.5) node[dot, label={[label distance=0.2]45:$\algA_{2(34)}$}] {} to[out=up, in=\dr]  (0.5,2.5) node[dot, label={[label distance=0.2]45:$\algA_{1(234)}$}] {} to[out=up, in=down] (0.5,3);    
\draw[slice] (0,0) to[out=up, in=\dl] (0.5,2.5);
\draw[slice] (1,0) to[out=up, in=\dl] (1.5,1.5);
\draw[slice] (3,0) to[out=up, in=\dr] (2.5,0.5); 
\end{scope}
\begin{scope}[shift={(162:5)}]
\draw[slice] (0,0) to[out=up, in=\dl] (0.5,1) node[dot, label={[label distance=0.2]135:$\algA_{12}$}] {} to[out=up, in=\dl] (1.5,2.5) node[dot, label={[label distance=0.2]45:$\algA_{(12)(34)}$}] {} to (1.5,3);
\draw[slice] (2,0) to[out=up, in=\dl] (2.5,1) node[dot, label={[label distance=0.2]45:$\algA_{34}$}] {} to[out=up, in=\dr] (1.5,2.5);
\draw[slice] (3,0) to[out=up, in=\dr] (2.5,1);        
\draw[slice] (1,0) to[out=up, in=\dr] (0.5,1);
\end{scope}
\end{scope}
\draw[wire] (0,0) to (54:6); \node at (54:7) {\small$\eta\begin{bmatrix}
    \algA_{12} & \algA_{(12)3} \\ 
    \algA_{23} & \algA_{1(23)}
\end{bmatrix}$};
\draw[wire] (0,0) to (-22:6);
\node[fill=white] at (-15:8) {\small$\eta\begin{bmatrix}
    \algA_{1(23)} & \algA_{(123)4} \\ 
    \algA_{(23)4} & \algA_{1(234)}
\end{bmatrix}$};
\draw[wire] (0,0) to (-90:6);
\node[fill=white] at (-90:7) {\small$\eta\begin{bmatrix}
    \algA_{23} & \algA_{(23)4} \\ 
    \algA_{2(34)} & \algA_{34}
\end{bmatrix}$};
\draw[wire] (0,0) to (126:6);
\node[fill=white] at (126:7) {\small$\eta\begin{bmatrix}
    \algA_{34} & \algA_{(12)(34)} \\ 
    \algA_{(12)3} & \algA_{(123)4}
\end{bmatrix}$};
\draw[wire] (0,0) to (198:6);
\node[fill=white] at (198:8) {\small$\eta\begin{bmatrix}
    \algA_{2(24)} & \algA_{1(234)} \\ 
    \algA_{12} & \algA_{(12)(34)}
\end{bmatrix}$};
\node[red] at (0,0) {\Large$\bullet$};
\end{tz}
\caption{Schematic representation of the pentagonator for the chiral symmetry category. The black lines represent codimension-two interfaces between the different configurations. These can be deformed into each other in the third direction giving rise to a five-valent junction supporting the pentagonator 3-morphism $\mmmor$.}
\end{figure}

The two-morphisms $\mmor\begin{bmatrix} \algA_{12} & \algA_{(12)3} \\ \algA_{23} & \algA_{1(23)} \end{bmatrix}$ are not all independent. Given a four-fold fusion $\Dchiral{p_1/N_1} \otimes \Dchiral{p_2/N_2} \otimes \Dchiral{p_3/N_3} \otimes \Dchiral{p_4/N_4}$ we can use the two morphisms to relate different sequential gauging procedures as displayed in Figure \ref{fig:pentagonator}.

\subsection{Detecting the associator}
\label{sec:monopoleline}
A natural question concerns our ability to detect the associator $\AF_{\ob_1 , \, \ob_2 , \, \ob_3}$. To do so we might try to understand its action on the dynamical degrees of freedom of the theory. These are either local operators $\cO$, which carry chiral charge, or monopole lines $T$ which are charged under the magnetic $U(1)^{(1)}$.
According to the analysis in \cite{Cordova:2022ieu, Choi:2022jqy} a monopole line $T$ passing through the chiral symmetry defect $\Dchiral{p/N}$ is mapped into a non-genuine line operator:
\be
T W^{p/N} \, ,
\ee
where $W$ is the fundamental Wilson line of the gauge theory. This operator is attached to a surface $\Uonef_{\frac{p}{N}}$ which connects $T W^{p/N}$ to the generator $L$ of $\Dchiral{p/N}$.
To analyze the interplay between the F-symbol and the monopole lines $T$ we pass the topological surface defining the F-symbol through $T$. 
\begin{figure}[t]
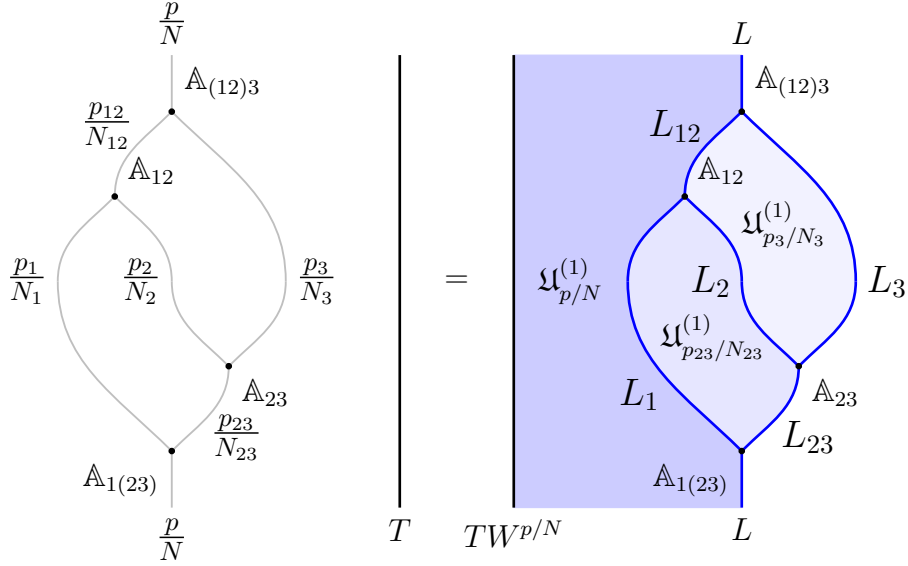

\label{eq: generalpassing}
\begin{center}
\begin{tz}[scale=0.75]
\draw[slice] (0,-0.5) to [out=up, in=\dl] (1,1);
\draw[slice] (2,-0.5) to [out=up, in=\dr] (1,1);
\draw[slice] (1,1) to [in=\dl, out=up] (2,2.5);
\draw[slice] (4,-0.5) to [out=up, in=\dr] (2,2.5) to (2,3.5);
\draw[slice] (0,-0.5) to [in=\ul, out=down] (2,-3.5) to (2,-4.5);
\draw[slice] (2,-0.5) to [out=down, in = \ul] (3,-2);
\draw[slice] (4,-0.5) to [out=down, in = \ur] (3,-2);
\draw[slice] (3,-2) to [out=down, in = \ur] (2,-3.5);
\node[dot, label={[label distance=0.2]45:$\algA_{12}$}] at (1,1) {}; \node[dot, label={[label distance=0.1]45:$\algA_{(12)3}$}] at (2,2.5) {}; \node[dot, label={[label distance=0.2]-45:$\algA_{23}$}] at (3,-2) {}; \node[dot, label={[label distance=0.1]-135:$\algA_{1(23)}$}] at (2,-3.5) {};
\node[left] at (2,-0.5) {\large $\frac{p_2}{N_2}$};
\node[left] at (0,-0.5) {\large $\frac{p_1}{N_1}$};
\node[right] at (4,-0.5) {\large $\frac{p_3}{N_3}$};
\node[left] at (1.5,2.25) {\large $\frac{p_{12} }{N_{12} }$};
\node[right] at (2.5,-3.25) {\large $\frac{p_{23} }{N_{23} }$};
\node[above] at (2,3.5) {\large $\frac{p}{N}$};
\node[below] at (2,-4.5) {\large $\frac{p}{N}$};
\draw[color=black, line width=1]  (6,-4.5) node[below] {$T$} to (6,3.5);
\node at (7,-0.5) {$=$};
\begin{scope}[shift={(10,0)}]
\filldraw[color=white!80!blue] (2,-4.5) -- (-2,-4.5) -- (-2,3.5) -- (2,3.5) -- cycle; 
\filldraw[color=white!90!blue] (0,-0.5) to [out=up, in=\dl] (1,1) to [out=\dr, in=up] (2,-0.5) to [out=down, in = \ul]  (3,-2) to [out=down, in = \ur] (2,-3.5) to [in=down, out=\ul] (0,-0.5);
\filldraw[color=white!95!blue] (1,1) to [in=\dl, out=up] (2,2.5)  to [in=up, out=\dr] (4,-0.5) to [out=down, in = \ur] (3,-2) to [in=down, out = \ul] (2,-0.5) to [out=up, in=\dr] (1,1);
\draw[slice, color=blue, line width=1] (0,-0.5) to [out=up, in=\dl] (1,1);
\draw[slice ,color=blue, line width=1] (2,-0.5) to [out=up, in=\dr] (1,1) ;
\draw[slice , color=blue, line width=1] (1,1) to [in=\dl, out=up] (2,2.5);
\draw[slice, color=blue, line width=1] (4,-0.5) to [out=up, in=\dr] (2,2.5) to (2,3.5);
\draw[slice, color=blue, line width=1] (0,-0.5) to [in=\ul, out=down] (2,-3.5) to (2,-4.5);
\draw[slice, color=blue, line width=1] (2,-0.5) to [out=down, in = \ul] (3,-2);
\draw[slice, color=blue, line width=1] (4,-0.5) to [out=down, in = \ur] (3,-2);
\draw[slice, color=blue, line width=1] (3,-2) to [out=down, in = \ur] (2,-3.5);
\node[dot, label={[label distance=0.1]45:$\algA_{12}$}] at (1,1) {}; \node[dot, label={[label distance=0.2]45:$\algA_{(12)3}$}] at (2,2.5) {}; \node[dot, label={[label distance=0.2]-45:$\algA_{23}$}] at (3,-2) {}; \node[dot, label={[label distance=0.1]-135:$\algA_{1(23)}$}] at (2,-3.5) {};
\draw[color=black, line width=1]  (-2,-4.5) node[below] {$T W^{p/N}$} to (-2,3.5);
\node at (-1,-0.5) {$\Uonef_{p/N}$};
\node at (1.5,-1.5) {$\Uonef_{ p_{23}/N_{23}}$};
\node at (2.75,0.5) {$\Uonef_{ p_3/N_3}$};
\node[left] at (2,-0.5) {\large $L_2$};
\node[left] at (0.75,-2.5) {\large $L_1$};
\node[right] at (4,-0.5) {\large $L_3$};
\node[left] at (1.5,2.25) {\large $L_{12}$};
\node[right] at (2.5,-3.25) {\large $L_{23}$};
\node[below] at (2,-4.5) {$L$};
\node[above] at (2,3.5) {$L$};
\end{scope}
\end{tz}
\end{center}
\caption{A monopole line "passing through" the F-symbol. Here we draw the relevant 2d slice of the 4d process. Passing the monopole line through turns on nontrivial line operators in the F-symbol bubble.}
\end{figure}
Each time the monopole line $T$ passes through a chiral symmetry surface\footnote{\ Since the line $T$ is not topological it would be more precise to say that the chiral symmetry surface passes through $T$.} it leaves behind the generating line of the corresponding minimal TFT and is attached to a one-form symmetry defect ending on a fractional Wilson line. Above is a slice of the final configuration, with the one-form symmetry surfaces localized on it. Shrinking the second figure gives rise to a correlator on the 2d interface $\AF_{\ob_1, \ob_2, \ob_3}\begin{bmatrix} \algA_{12} & \algA_{(12)3} \\ \algA_{23} & \algA_{1(23)} \end{bmatrix} $. Such correlator can detect the passing of the monopole line and makes the associator physical.
Notice that we could have also chosen to slide the monopole line through in the orthogonal direction. This would lead to the correlation function of a line operator in the associator TFT. Such configuration is also interesting on its own but we do not study it here.

\bigskip

\begin{figure}
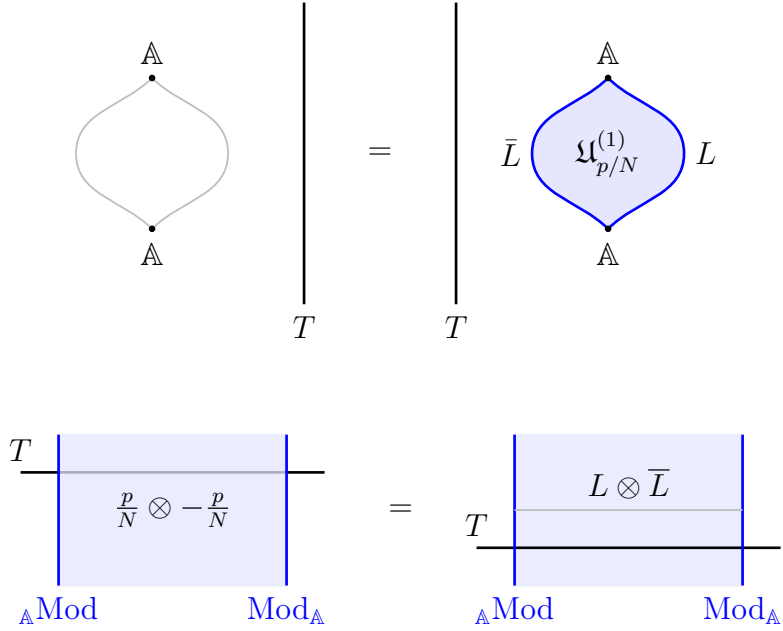

    \centering
\begin{tz}
\draw[slice] (0,0) to [in=\dl, out=up] (1,1) to [in=up, out=\dr] (2,0);
\draw[slice] (0,0) to [in=\ul, out=down] (1,-1) to [in=down, out=\ur] (2,0);
\draw[color=black, line width=1] (3,-2) node[below] {$T$}-- (3,2);
\node[dot, label={[label distance=0.2]90:$\algA$}] at (1,1) {};
\node[dot, label={[label distance=0.2]-90:$\algA$}] at (1,-1) {};
\node at (4,0) {$=$};
\begin{scope}[shift={(6,0)}]
\filldraw[color=blue, fill=white!90!blue,  line width =1]   (0,0) to [in=\dl, out=up] (1,1) to [in=up, out=\dr] (2,0) to [in=\ur, out=down] (1,-1) to [in=down, out=\ul] (0,0);  
\node[dot, label={[label distance=0.2]90:$\algA$}] at (1,1) {};
\node[dot, label={[label distance=0.2]-90:$\algA$}] at (1,-1) {};
\node[left] at (0,0) {$\bar{L}$}; 
\node[right] at (2,0) {$L$};
\node at (1,0) {$\Uonef_{p/N}$};
\draw[color=black, line width=1] (-1,-2) node[below] {$T$}-- (-1,2);
\end{scope}
\end{tz} \\[2.5em]
\begin{tz}
\draw[color=black, line width=1] (0.5,1.5) node[above] {$T$} -- (4.5,1.5);
    \filldraw[slice, color=white!90!blue, opacity=0.75] (1,2) to (4,2) to (4,0) to (1,0) --cycle;
\draw[slice, color=blue, line width=1] (1,2) to (1,0) node[below] {${}_{\algA}\text{Mod}$};
\draw[slice, color=blue, line width=1] (4,2) to (4,0) node[below] {$\text{Mod}_{\algA}$};
\node at (2.5,1) {$\frac{p}{N} \otimes -\frac{p}{N} $};
\node at (5.5,1) {$=$};
\begin{scope}[shift={(6,0)}]
    \filldraw[slice, color=white!90!blue, opacity=0.75] (1,2) to (4,2) to (4,0) to (1,0) --cycle;
    \draw[color=black, line width=1] (0.5,0.5) node[above] {$T$} -- (4.5,0.5);
\draw[slice, color=blue, line width=1] (1,2) to (1,0) node[below] {${}_{\algA}\text{Mod}$};
\draw[slice, color=blue, line width=1] (4,2) to (4,0) node[below] {$\text{Mod}_{\algA}$};
\draw[slice] (1,1) to (2.5,1) node[above] {$L \otimes \overline{L}$} to  (4,1);
\end{scope}
\end{tz}
    \caption{Passing a monopole line through the $\Dchiral{p/N} \otimes \overline{\Dchiral{p/N}}$ interface. On the left is a vertical slice of the process, while on the right it is represented in the slab picture.}
    \label{fig:bubble1}
\end{figure}

We consider configurations for which the monopole line cannot be detected by the initial and final configurations. This assures that only degrees of freedom confined on the interface can be activated.
As a warm-up set $N_1 = N = N_2$, $N_3=0$ and $p_1= p = - p_2$, leading to Figure \ref{fig:bubble1}.
The bubble on the right-hand side computes a one-point function in the associator TFT on $\Sigma_2 \times S^1$. From the lower part of Figure \ref{fig:bubble1} we see that this is precisely the un-normalized\footnote{\ The normalized one-point function is obtained by dividing this by the associator partition function $\langle \AF[\algA \ \algA]_{p/N, \, -p/N} \rangle_{\Sigma_2}$.} one-point function of the vertex operator $V(L \otimes \overline{L})$, which vanishes identically.
Alternatively consider the explicit expression for the associator:
\be
\AF[\algA \ \algA]_{p/N, \, -p/N}= \cC_N^{(1)} = \bigoplus_{k=0}^{N-1} \Uonef_{k/N}(\Sigma_2) \, ,
\ee
Passing the monopole line through produces nontrivial phases as $T$ links with $\Uonef_{r/N}$:
\be
\AF[\algA \ \algA]_{p/N, \, -p/N} \otimes T = T \otimes  \sum_{k=0}^{N-1} e^{\frac{2 \pi i k}{N}} \, \Uonef_{k/N}(\Sigma_2) \, .
\ee 
Shrinking $\Uonef_{k/N}(\Sigma_2)$ gives a sum:
\be
\sum_{k=0}^{N-1} e^{\frac{2 \pi i k}{N}} = 0 = \sum_{k=0}^{N-1} e^{\frac{2 \pi i k}{N}} \langle \pi_k \rangle_{\Sigma_2} =  \langle V \rangle_{\Sigma_2} \, .
\ee
Where the r.h.s. is the vev of the vertex operator on $\Sigma_2$ expanded in an idempotent basis.

\bigskip

We can also consider a generic F-symbol, but we instead pass through a 't Hooft line $T^N$, which undergoes a standard (non-fractional) Witten effect passing through $\Dchiral{p/N}$:
\bea
\begin{tz}[scale=0.75]
\draw[slice] (0,-0.5) to [out=up, in=\dl] (1,1);
\draw[slice] (2,-0.5) to [out=up, in=\dr] (1,1);
\draw[slice] (1,1) to [in=\dl, out=up] (2,2.5);
\draw[slice] (4,-0.5) to [out=up, in=\dr] (2,2.5) to (2,3.5);
\draw[slice] (0,-0.5) to [in=\ul, out=down] (2,-3.5) to (2,-4.5);
\draw[slice] (2,-0.5) to [out=down, in = \ul] (3,-2);
\draw[slice] (4,-0.5) to [out=down, in = \ur] (3,-2);
\draw[slice] (3,-2) to [out=down, in = \ur] (2,-3.5);
\node[dot, label={[label distance=0.2]45:$\algA_{12}$}] at (1,1) {}; \node[dot, label={[label distance=0.1]45:$\algA_{(12)3}$}] at (2,2.5) {}; \node[dot, label={[label distance=0.2]-45:$\algA_{23}$}] at (3,-2) {}; \node[dot, label={[label distance=0.1]-135:$\algA_{1(23)}$}] at (2,-3.5) {};
\node[left] at (2,-0.5) {\large $\frac{p_2}{N_2}$};
\node[left] at (0,-0.5) {\large $\frac{p_1}{N_1}$};
\node[right] at (4,-0.5) {\large $\frac{p_3}{N_3}$};
\node[left] at (1.5,2.25) {\large $\frac{p_{12} }{N_{12} }$};
\node[right] at (2.5,-3.25) {\large $\frac{p_{23} }{N_{23} }$};
\node[above] at (2,3.5) {\large $\frac{p}{N}$};
\node[below] at (2,-4.5) {\large $\frac{p}{N}$};
\draw[color=black, line width=1]  (6,-4.5) node[below] {$T^N$} to (6,3.5);
\node at (7,-0.5) {$=$};
\begin{scope}[shift={(10,0)}]
\filldraw[color=white!90!blue] (0,-0.5) to [out=up, in=\dl] (1,1) to [out=\dr, in=up] (2,-0.5) to [out=down, in = \ul]  (3,-2) to [out=down, in = \ur] (2,-3.5) to [in=down, out=\ul] (0,-0.5);
\filldraw[color=white!95!blue] (1,1) to [in=\dl, out=up] (2,2.5)  to [in=up, out=\dr] (4,-0.5) to [out=down, in = \ur] (3,-2) to [in=down, out = \ul] (2,-0.5) to [out=up, in=\dr] (1,1);
\draw[slice, color=blue, line width=1] (0,-0.5) to [out=up, in=\dl] (1,1);
\draw[slice ,color=blue, line width=1] (2,-0.5) to [out=up, in=\dr] (1,1) ;
\draw[slice , color=blue, line width=1] (1,1) to [in=\dl, out=up] (2,2.5);
\draw[slice, color=blue, line width=1] (4,-0.5) to [out=up, in=\dr] (2,2.5); \draw[slice] (2,2.5) to (2,3.5);
\draw[slice, color=blue, line width=1] (0,-0.5) to [in=\ul, out=down] (2,-3.5); \draw[slice] (2,-3.5) to (2,-4.5);
\draw[slice, color=blue, line width=1] (2,-0.5) to [out=down, in = \ul] (3,-2);
\draw[slice, color=blue, line width=1] (4,-0.5) to [out=down, in = \ur] (3,-2);
\draw[slice, color=blue, line width=1] (3,-2) to [out=down, in = \ur] (2,-3.5);
\node[dot] at (1,1) {}; \node[dot] at (2,2.5) {}; \node[dot] at (3,-2) {}; \node[dot] at (2,-3.5) {};
\draw[color=black, line width=1]  (-2,-4.5) node[below] {$T^N W^{p}$} to (-2,3.5);

\node[rotate=-45] at (1.25,-1.5) {\small $\Uonef_{p_{23} N/N_{23}}$};
\node[rotate=-45] at (3,0.5) {\small$\Uonef_{p_3 N/N_3}$};
\node[left] at (2,-0.25) {\large $L_2^N$};
\node[left] at (0.75,-2.5) {\large $L_1^N$};
\node[right] at (4,-0.5) {\large $L_3^N$};
\node[left] at (1.5,2.25) {\large $L_{12}^N$};
\node[right] at (2.5,-3.25) {\large $L_{23}^N$};
\end{scope}
\end{tz}
\eea
If $N = \text{lcm}(N_1, N_2, N_3) \equiv N_{123}$ then all the lines $L_i^N$ are trivial and the F-symbol is blind to the monopole line. If instead $N< N_{123}$ (some of) the lines $L_i^N$ are nontrivial and the configuration once again describes the one-point function of a vertex operator $V(L_1^N \otimes L_2^N \otimes L_3^N) \equiv V_N$.
According to our construction we expand:
\be
\AF[\algA^+ \ \algA^-]_{p_1/N_1, \, p_2/N_2, \, p_3/N_3} = d_0 \, \bigoplus_{q=0}^{K-1} \Uonef_{\frac{q}{N_{123}}} \, ,
\ee
and we can compute:
\be
\AF[\algA^+ \ \algA^-]_{p_1/N_1, \, p_2/N_2, \, p_3/N_3} \otimes T^N = T^N \otimes d_0 \, \bigoplus_{q=0}^{K-1} e^{\frac{2 \pi i q}{K}} \, \Uonef_{\frac{q}{N_{123}}} \, .
\ee
Again shrinking the $\Uonef$ surface leads to a factor:
\be
d_0 \sum_{q=0}^{K-1} e^{\frac{2 \pi i q}{K}} = 0 = \sum_{M \in \fM_c} \tilde{\boldsymbol{B}}_{V_N \, M} \, \langle \pi_M \rangle_{\Sigma_2} = \langle V_N \rangle_{\Sigma_2} \, .
\ee

\section{Ward identities from higher structure and surgery} \label{sec: Ward}

In this section we explore consequences of the higher structure from generalized symmetry in the form of Ward identities satisfied by the partition functions of the QFT with insertions of (extended non-topological) operators.
 
As observed in \cite{Harlow:2018tng,Choi:2022jqy,Cordova:2022ieu}, the non-invertibility of the chiral symmetry does not play a role in a correlator of local operators on a flat space.
This is because the non-invertibility comes only as a condensation of lower dimensional operators, which acts trivially on local operators. 
In addition, the trivial spacetime topology allows the defect to pass through the whole spacetime without causing self-intersection.
In this section we consider the Ward identity stemmed from the non-invertible chiral symmetry in a correlator on a non-trivial spacetime topology. This can be thought of as a global version of the discussion in Section~\ref{sec:monopoleline} which concerns local moves of the topological network in the presence of a charged non-topological defect. 

Before the explicit calculation, let us try to guess the Ward identity.
In massless QED, the consequence of the ABJ-anomaly is that a chiral rotation is absorbed by a shift of the theta angle. Explicitly, this boils down to the following relation between correlation functions, 
\begin{equation}
    \left\langle\prod_i \mathcal{O}_i(x_i) \right\rangle_{X,\theta}^{\text{QED}} =
    e^{\mathrm{i}\alpha \sum_i q(\mathcal{O}_i)}\left\langle\prod_i \mathcal{O}_i(x_i) \right\rangle_{X,\theta-\alpha}^{\text{QED}},
\end{equation}
where $X$ is the spacetime manifold, $\theta$ is the theta angle, $\alpha$ is the chiral rotation parameter, and $q(\mathcal{O}_i)$ is the chiral charge of the local operator $\mathcal{O}_i$.

What is a natural generalization of this equation for a general (possibly non-Lagrangian) QFT with non-invertible chiral (possibly finite) symmetry?
To guess it, we recall that the theta angle shift in an abelian gauge theory can be achieved by gauging its magnetic one-form symmetry with a Dijkgraaf-Witten (DW)-twist (i.e. a discrete theta angle). This motivate us to write the following equation, independent of a Lagrangian description\cite{Apte:2022xtu}:
\begin{equation}
    \left\langle\prod_i \mathcal{O}_i(x_i) \right\rangle_{X} =
    \mathcal{N} e^{2\pi\mathrm{i}\frac{p}{N}\sum_i q(\mathcal{O}_i)}\sum_{b\in H^2(X,\mathbb{Z}_N)}e^{2\pi\mathrm{i}\frac{(p)^{-1}_{\gamma(N)N}}{2N}\int_X \mathfrak{P}(b)} \left\langle\prod_i \mathcal{O}_i(x_i) \right\rangle_{X,b},
\end{equation}
where $b$ is the magnetic one-form symmetry background, $\mathfrak{P}$ is the Pontryagin square (we follow the conventions of \cite{Apte:2022xtu}), $\mathcal{N}$ is an normalization factor,$\gamma(N)=\gcd(N,2)$, and $(p^{-1})_n$ the multiplicative inverse of $p$ in $\mathbb{Z}_n^{\times}$. We will henceforth drop the subscript $n$ from the multiplicative inverse.
Without insertions, this is nothing but the definition of the KW-duality under gauging the $\mathbb{Z}_N^{(1)}$ symmetry with a twist.
In the rest of this section we derive the equation using the definition of chiral symmetry by topological operators for a large class of 4-manifolds $X$.

We begin by discussing explicitly the example of the spacetime manifold $X=S^2 \times S^2$. We proceed by schematically review the handle-body decompositions for general 4-manifolds, which allows a convenient presentation of them in terms of successive surgeries of various kinds of handles over a 4-ball. From this perspective a compact 4-manifold can always be viewed as a null bordism either of $S^3$ or of a connected sum $\#^k S^1 \times S^2$. We conclude exploiting this description of 4-manifolds to derive a family of general Ward identities for the non-invertible chiral symmetries for the case of 4-manifolds that are constructed out of 2-handles only.\footnote{\ A basic example of such 4-manifold is given by $S^2 \times S^2$ which is the first non-trivial example to which our results apply.} 

\medskip

We stress that in the discussion below we never make use of a Lagrangian formulation for the theory, therefore the derivation extends to all theories with a non-invertible chiral higher symmetry. Moreover, the strategy to derive Ward identities we outline in this section extends to more general non-invertible symmetries straightforwardly. Topology can be exploited to generate non-trivial self-intersections of the defect worldvolumes which, when the symmetry is non-invertible, produce non-trivial higher associator relations and thus identities between correlators.
\subsection{Ward identities for $S^2 \times S^2$}\label{sec:cosmology}
\paragraph{Partition Function on $S^2 \times S^2$} As a first example, let us consider putting the system on $S^2 \times S^2$. There are various ways of visualising $S^2 \times S^2$. One option is to view one of the two $S^2$'s as an $S^1$ fibration over an interval where the $S^1$ is shrinking on the two ends. This gives a fibration of $S^2 \times S^2$ over an interval with fiber $S^2 \times S^1$. We can now proceed to visualise the fiber as $S^2$ times an interval with the two endpoints identified:
\bea\label{eq:s2s1}
\begin{tikzpicture}[scale=0.75]
\shade[ball color = gray!40, opacity = 0.4] (0,0) circle (0.7cm);
  \draw (0,0) circle (0.7cm);
  \shade[ball color = gray!40, opacity = 0.4] (0,0) circle (2cm);
  \draw (0,0) circle (2cm);
  \draw (-2,0) arc (180:360:2 and 0.6);
  \draw[dashed] (2,0) arc (0:180:2 and 0.6);
  \fill[fill=black] (0,0) circle (1pt);
\end{tikzpicture}
\eea
In the picture above the inner sphere and the outer sphere are identified. For our analysis in this section it is sufficient to consider the projection to a transverse plane that intersects the inner and the outer $S^2$'s in \eqref{eq:s2s1} along their equators:
\bea\begin{gathered}
\begin{tikzpicture}[scale=0.75]
  \draw (1.5,1.5) circle (0.7);
  \draw (1.5,1.5) circle (2);
\end{tikzpicture}
\end{gathered}
\eea
First let us consider the effect of nucleating a bubble of the $\Dchiral{p/N}$ defect from the vacuum on a contractible $S^3$. In the $S^2 \times S^1$ fiber this would look like an $S^2$, which along the base of the fibration shrinks at the two sides and spans an $S^3$. In our projection, the $S^2$ is represented by a blue circle. Inserting the defect  gives an overall inverse factor of $\Dchiral{p/N}(S^3) = 1/\sqrt{N}$\bea\begin{gathered}
\begin{tikzpicture}[scale=0.75]
  \draw (1.5,1.5) circle (0.7);
  \draw (1.5,1.5) circle (2);
\end{tikzpicture}
\end{gathered} \quad =\quad  \sqrt{N} \, \begin{gathered}
\begin{tikzpicture}[scale=0.75]
  \draw (1.5,1.5) circle (0.7);
  \node at (-0.3,1.5) {{\tiny\textcolor{blue}{{$\vee$}}}};
  \draw[thick,color=blue] (0.2,1.5) circle (0.5);
  \node at (0.6,2.4) {$\Dchiral{p/N}$};
  \draw (1.5,1.5) circle (2);
\end{tikzpicture}\end{gathered} \quad = \quad  \sqrt{N} \,  \begin{gathered}
\begin{tikzpicture}[scale=0.75]
  \draw (1.5,1.5) circle (0.7);
  \draw[thick,color=blue] (-0.45,2) arc (60:-60:1.0 and 0.6);
    \node at (0.05,1.5) {{\tiny\textcolor{blue}{{$\wedge$}}}};
  \node at (0.45,1.5) {{\tiny\textcolor{blue}{{$\vee$}}}};
  \draw[thick,color=blue] (1.02,2) arc (115:253:1.0 and 0.6);
  \node at (0.6,2.4) {$\Dchiral{p/N}$};
  \draw (1.5,1.5) circle (2);
\end{tikzpicture}
\end{gathered}
\label{eq:s2s2w1}
\eea
where in the r.h.s. we have slided the defect to the left and exploited the non-trivial topology of the configuration to bring it back to touching itself. At this point we can perform a surgery when the two sides of the defect $S^3$ meet. The surgery operation is obtained by carving out a solid torus $D^2 \times S^1$ from the $S^3$ and then gluing it back by filling in the other circle, namely $S^1 \times D^2$ where the $T^2$ is fixed (like an $S$ move on the fillings). This leads to a topology for our defect which is $S^2 \times S^1$. While performing surgery we use the morphism $\mor_N: \Dchiral{p/N}  \otimes \overline{\Dchiral{p/N}}  \to \unit$ in the solid torus. This leaves behind a duality defect $\Dchiral{p/N}$ on $S^2 \times S^1$, with the $S^1$ contractible in the bulk and a condensation defect $\cC_N$ on the solid torus:
\be
\cC_N =  \sum_{b \in H^1(D^2 \times S^1, \, \bZ_N)} \Uonef(\text{PD}(b)) \, ,
\ee
where we use $\text{PD}$ to denote the Poincaré-Lefschetz dual. The only non-contractible cycle is a two-cycle spanning the disk $D^2$ and ending on the boundary $S^1$ (this is a cycle in relative homology $H_2(D^2 \times S^1, T^2, \, \bZ_N)$, which is the correct dual description on the gauge fields living in $H^1(D^2 \times S^1, \, \bZ_N)$). The latter projects to the red line in \eqref{eq:s2s2w2} below, while the disk extends inside the region between the inner and the outer $S^2$'s drawn in equation \eqref{eq:s2s1}. On the boundary of this cycle we have a line operator $L^{p^{-1} b}$ of the minimal TFT corresponding to the element $b \in H^2(D^2 \times S^1, \, \bZ_N)$. The inverse power $p^{-1}$ appears because the generator $L$ is taken to have charge $p$ under the one-form symmetry.
 In the transverse plane we are drawing this line gets projected to a pair of black dots, one directed into the page and one outgoing:
\bea\label{eq:s2s2w2}
\eqref{eq:s2s2w1} = \sqrt{N} \, \begin{gathered}
\begin{tikzpicture}[scale=0.95]
  \draw (1.5,1.5) circle (0.7);
  \draw[thick,color=blue] (-0.48,2) -- (1.05,2);
  \draw[thick,color=blue] (-0.48,1) -- (1.05,1);
  \draw[thick, color=red] (0.7,2) --  (0.7,1);
  \draw[fill=black] (0.7,2) circle (0.05);
  \draw[fill=black] (0.7,1) circle (0.05);
  \node at (0.7,0.6) {$L$};
  \node at (0.4,1.5) {\textcolor{red}{$\cC_N$}};
  \node at (0.6,2.4) {\textcolor{blue}{$\Dchiral{p/N}$}};
  \draw (1.5,1.5) circle (2);
\end{tikzpicture}
\end{gathered}=\sqrt{N}\, \begin{gathered}
\begin{tikzpicture}[scale=0.95]
  \draw (1.5,1.5) circle (0.7);
  \draw[thick,color=blue] (-0.48,2) -- (1.05,2);
  \draw[thick,color=blue] (-0.48,1) -- (1.05,1);
  \draw[thick, color=red] (0.7,2) --  (0.7,1);
  \draw[fill=black] (0.7,2) circle (0.05);
  \draw[fill=black] (0.7,1) circle (0.05);
  \node at (0.7,0.7) {$L$};
  \node at (0.4,1.5) {\textcolor{red}{$\cC_N$}};
  \draw (1.5,1.5) circle (2);
  \draw[fill=white, color=white] (0.2,2) circle (0.2);
  \draw[fill=white, color=white] (0.2,1) circle (0.2);
  \draw [blue,thick,domain=-161:161] plot ({1.5+1.6*cos(\x)}, {1.5+1.6*sin(\x)});
  \draw [blue,thick,domain=-158:158] plot ({1.5+1.25*cos(\x)}, {1.5+1.25*sin(\x)});
   \draw[thick, color=violet] (2.75,1.5) --  (3.1,1.5);
  \draw[fill=black] (2.75,1.5) circle (0.05);
  \draw[fill=black] (3.1,1.5) circle (0.05);
    \node[right] at (3.6,1.5) {\textcolor{violet}{$\cC_N$}};
     \node at (2.4,1.5) {$L'$};
 \end{tikzpicture}
 \end{gathered}
\eea
Next on the RHS of equation \eqref{eq:s2s2w2} we have repeated the same move by bringing the defect around the anulus in the transverse plane and performing another time the same surgery operation. Notice however that this time while the projection of the disk is still a line in the plane, the disks extends in along the fibration interval. To distinguish this cycle from the other we have drawn it in violet in the Figure. Then exploiting the topology of the configuration and the identification of the inner with the outer $S^2$'s in equation \eqref{eq:s2s1}, we see that by this second surgery operation our defect worldvolume is back to the topology of an $S^3$ which we can shrink back to zero size: 
\bea
\eqref{eq:s2s2w2} = \sqrt{N} \, \begin{gathered}
\begin{tikzpicture}[scale=0.95]
  \draw (1.5,1.5) circle (0.7);
  \draw[thick,color=blue] (-0.2,1.9) -- (0.55,1.9);
  \draw[thick,color=blue] (-0.2,1.1) -- (0.55,1.1);
  \draw[thick, color=red] (0.3,1.9) --  (0.3,1.1);
  \draw[fill=black] (0.3,1.9) circle (0.05);
  \draw[fill=black] (0.3,1.1) circle (0.05);
  \node at (0.3,0.5) {$L$};
  \node at (0.0,1.5) {\textcolor{red}{$\cC_N$}};
  \draw (1.5,1.5) circle (2);
  \draw [blue,thick,domain=-165:165] plot ({1.5+1.8*cos(\x)}, {1.5+1.8*sin(\x)});
  \draw [blue,thick,domain=-158:158] plot ({1.5+1.1*cos(\x)}, {1.5+1.1*sin(\x)});
   \draw[thick, color=violet] (3.3,1.5) --  (3.5,1.5);
      \draw[thick, color=violet] (2.6,1.5) --  (2.2,1.5);
  \draw[fill=black] (2.6,1.5) circle (0.05);
  \draw[fill=black] (3.3,1.5) circle (0.05);
    \node at (4,1.5) {\textcolor{violet}{$\cC_N$}};
     \node at (3,1.5) {$L'$};
 \end{tikzpicture}
 \end{gathered}= \sqrt{N}  \begin{gathered}
\begin{tikzpicture}[scale=0.95]
  \draw (1.5,1.5) circle (0.7);
  \draw (1.5,1.5) circle (2);
  \draw [red,thick,domain=16:345] plot ({1.5+1.5*cos(\x)}, {1.5+1.5*sin(\x)});
  \draw[thick,color=blue] (2.95,1.5) circle (0.37);
   \draw[thick, color=violet] (3.3,1.5) --  (3.5,1.5);
      \draw[thick, color=violet] (2.6,1.5) --  (2.2,1.5);
  \draw[fill=black] (2.6,1.5) circle (0.05);
  \draw[fill=black] (3.3,1.5) circle (0.05);
     \node at (1.9,1.5) {$L'$};
     \draw[fill=black] (2.95,1.85) circle (0.05);
  \draw[fill=black] (2.95,1.15) circle (0.05);
  \node at (2.95,0.5) {$L$};
      \node[right] at (4,1.5) {\textcolor{violet}{$\cC_N$}};
  \node at (1.5,0.5) {\textcolor{red}{$\cC_N$}};
 \end{tikzpicture}
 \end{gathered}\label{eq:s2s2w3}
\eea 
However notice that this comes at a price: the lines $L^{p^{-1} b}$ and $(L')^{p^{-1} c}$ form an Hopf link on the $S^3$, so un-linking them gives rise to the phase $\exp\left(\frac{2\pi i p^{-1}}{N} \, b c\right)$:
 \bea
 \eqref{eq:s2s2w3} = 
 \frac{1}{N} \sum_{b, c \, \in \, \bZ_N} \ \ e^{\frac{2 \pi i p^{-1}}{N} b \, c} \, \ \ \begin{gathered}
\begin{tikzpicture}[scale=0.95]
  \draw (1.5,1.5) circle (0.7);
  \node at (1.5,-0.9) {\textcolor{red}{$\Uonef_{2\pi b/N}$}};
  \draw (1.5,1.5) circle (2);
    \draw [red,thick] (1.5,1.5) circle (1.3);
      \draw[thick, color=violet] (2.2,1.5) --  (3.5,1.5);
    \node[right] at (4,1.5) {\textcolor{violet}{$\Uonef_{2\pi c/N}$}};
 \end{tikzpicture}
 \end{gathered}
\eea
Above, on the r.h.s. we have recollected all factors of $\sqrt{N}$ and exploited the definition of the 1-form symmetry condensates. We then notice that the overall phase is just a rewriting of the Pontryagin square $p^{-1} \, \fP$ on $S^2 \times S^2$. We have proved that the $S^2 \times S^2$ partition function is thus equivalent to its gauging with discrete torsion $p^{-1} \, \fP(B)$, this is our desired Ward identity.

\medskip

\paragraph{Correlators} One can consider inserting local operators $\mathcal O_i(x_i)$ at points $x_i$ along the $S^2 \times S^2$ background. If these operators have a nontrivial chiral symmetry charge $q(\mathcal O_i)$ -- such as chiral fermions would with respect to the usual axial $U(1)$ -- then when we perform the topological manipulation leading to the Ward identity we derived above, this would lead to additional phases for the corresponding $S^2 \times S^2$ correlators. More precisely, an insertion of the bubble of $\Dchiral{p/N}$ produces an overall factor
\be
\prod_{i} \exp\left(\frac{2 \pi i p}{N} q(\mathcal O_i) \right) \,.
\ee
Hence these operators behave exactly like configurations with chiral fermion insertions: this phase is exactly the one that would have been obtained by a $\frac{2 \pi p}{N}$ shift of the theory's theta angle in the Lagrangian example. Notice however that our derivation above does not depend on any Lagrangian description of the QFT, and so it should be seen as an alternative derivation of the familiar axial selection rule on chiral fermions valid also for systems that do not have an ordinary Lagrangian formulation. In particular, it applies for all the class S theories with non-invertible duality defects recently discussed in \cite{Bashmakov:2022jtl,Bashmakov:2022uek,Antinucci:2022cdi}.

\paragraph{1-form symmetry backgrounds} It is straightforward to extend the above analysis in the presence of 1-form symmetry background fields. Let us denote the corresponding generators as  $B$ and $C$ in $H^2(S^2\times S^2,\mathbb Z_N)$, that correspond to the insertion of $\Uonef_{2 \pi B/N}$ on one of the $S^2$'s and $\Uonef_{2 \pi C/N}$ on the other. With the same notations and conventions above, we consider
\bea\label{eq:s2s2back1}
\begin{gathered}
\begin{tikzpicture}[scale=0.95]
  \draw (1.5,1.5) circle (0.7);
  \node at (1.5,-0.9) {\textcolor{red}{$\Uonef_{2\pi B/N}$}};
  \draw (1.5,1.5) circle (2);
    \draw [red,thick] (1.5,1.5) circle (1.3);
      \draw[thick, color=violet] (2.2,1.5) --  (3.5,1.5);
    \node[right] at (4,1.5) {\textcolor{violet}{$\Uonef_{2\pi C/N}$}};
 \end{tikzpicture}
 \end{gathered} =  \sqrt{N} \, \begin{gathered}
\begin{tikzpicture}[scale=1.15]
  \draw (1.5,1.5) circle (0.7);
  \draw (1.5,1.5) circle (2);
  \draw [red,thick,domain=16:345] plot ({1.5+1.5*cos(\x)}, {1.5+1.5*sin(\x)});
  \draw[thick,color=blue] (2.95,1.5) circle (0.37);
   \draw[thick, color=violet] (3.3,1.5) --  (3.5,1.5);
      \draw[thick, color=violet] (2.6,1.5) --  (2.2,1.5);
  \draw[fill=black] (2.6,1.5) circle (0.05);
  \draw[fill=black] (3.3,1.5) circle (0.05);
     \node at (1.7,1.5) {$L^{p^{-1} C}$};
     \draw[fill=black] (2.95,1.85) circle (0.05);
  \draw[fill=black] (2.95,1.15) circle (0.05);
  \node at (2.35,0.75) {$L^{p^{-1} B}$};
      \node[right] at (4,1.5) {\textcolor{violet}{$\Uonef_{2\pi C/N}$}};
  \node at (1.5,-0.9) {\textcolor{red}{$\Uonef_{2\pi B/N}$}};
 \end{tikzpicture}
 \end{gathered}
\eea
where on the r.h.s. we have bubbled up an insertion of the defect $\Dchiral{p/N}$ along an $S^3$ (in blue in the equation). We can now perform the same manipulations as in the previous paragraphs but carefully sliding the lines $L^{p^{-1}B}$ and $L^{p^{-1} C}$ away from the locations of the surgeries produces an overall phase $e^{\frac{2 \pi i p^{-1}}{N} \, B \, C}$ from un-linking them.  This result in a final bubble configuration that looks as follows:
\bea
\sqrt{N} e^{\frac{2 \pi i p^{-1}}{N} \, B \, C} \times
\begin{gathered}
\begin{tikzpicture}[scale=1.25]
 \draw  (3,0) -- (3,3);
 \draw (0,0) -- (0,3);
  \draw[thick, color=blue] (1.5,1.5) circle (1);
  \draw[thick, color=violet] (0,1.5) -- (0.5,1.5);   
  \draw[thick, color=violet] (3,1.5) -- (2.5,1.5);   
  \draw[thick, color=red] (1.5,0) -- (1.5,0.5);    
  \draw[thick,color=red] (1.5,3) -- (1.5,2.5); 
   \draw[fill=black] (0.5,1.5) circle (0.05);  
   \draw[fill=black] (2.5,1.5) circle (0.05);  
   \draw[fill=black] (1.5,0.5) circle (0.05); 
   \draw[fill=black] (1.5,2.5) circle (0.05);
   \node at (0.25,1.75) {$L'$};    \node at (1.15,2.75) {$L$};
    \node[above] at (1.5,3) {$\cond_N$}; \node[right] at (3,1.5) {$\cond_N$};
\draw[thick, color=red] (1.5,0.5) ++ (15:0.5) arc (15:165:0.5 and 0.5); \draw[fill=black] (1.5,0.5) ++ (15:0.5) circle (0.05) node[below] {$L^{p^{-1} B}$}; 
\draw[fill=black] (1.5,0.5) ++ (165:0.5) circle (0.05); 
\draw[thick, color=violet] (2.5,1.5) ++ (105:0.5) arc (105:255:0.5 and 0.5); \draw[fill=black] (2.5,1.5) ++ (105:0.5) circle (0.05); \draw[fill=black] (2.5,1.5) ++ (255:0.5) circle (0.05);  \node[fill=white] at (3,2.25) {$L^{p^{-1} C}$};
    \node[above] at (1.5,0.85) {\tiny $\Uonef_{2 \pi B/N}$}; \node[left] at (2.25,2.05) {\tiny$\Uonef_{2 \pi C/N}$};
    \end{tikzpicture}
    \end{gathered}
\eea
where we have included the phase from the linking of the background lines and the normalization factor but we have omitted drawing the ambient space -- the lines $L$ and $L'$ are junctions with condensates extending in the annulus as in the r.h.s. of \eqref{eq:s2s2w3}. As a final result, shrinking the $S^3$ to a point and picking a phase from each Hopf link we obtain
\bea
\eqref{eq:s2s2back1} =  \frac{1}{N} e^{\frac{2 \pi i p^{-1}}{N} \, B \, C}\, \sum_{b, \, c \, \in \, \bZ_N} \ e^{\frac{2 \pi i p^{-1}}{N} (B c + b C + b c) } \ \ 
 \begin{gathered}
\begin{tikzpicture}[scale=0.95]
  \draw (1.5,1.5) circle (0.7);
  \node at (1.5,-0.9) {\textcolor{red}{$\Uonef_{2\pi b/N}$}};
  \draw (1.5,1.5) circle (2);
    \draw [red,thick] (1.5,1.5) circle (1.3);
      \draw[thick, color=violet] (2.2,1.5) --  (3.5,1.5);
    \node[right] at (4,1.5) {\textcolor{violet}{$\Uonef_{2\pi c/N}$}};
 \end{tikzpicture}
 \end{gathered}
\eea
Introducing a generic background field $\textsf{B} \in H^2(S^2 \times S^2, \, \bZ_N)$ for the one-form symmetry and the partition function $\cZ[\textsf{B}]$ in the one-form symmetry background the above identity reads:
\be
\cZ[\textsf{B}] = \frac{\exp\left( \frac{2 \pi i p^{-1}}{2 N} \int \fP(\textsf{B}) \right)}{\sqrt{|H^2(S^2 \times S^2, \, \bZ_N)|}} \, \sum_{b \, \in \, H^2(S^2 \times S^2, \, \bZ_N)} \exp\left( \frac{2 \pi i p^{-1}}{2 N} \int \fP(b)  + \frac{2 \pi i p^{-1} }{ N} \int \textsf{B} \cup b   \right) \, \cZ[b] \, .
\ee
And expressed the self-duality of the partition function $\cZ$ under gauging a $\bZ_N$ subgroup of the one-form symmetry with discrete torsion $p^{-1} \, \fP(b)$, dual coupling $p^{-1} \textsf{B} \cup b$ and a background counterterm $p^{-1} \fP(\textsf{B})$. 

\medskip

It is simple to see that the only $\bZ_N$ TFT which is invariant under this operation is the trivially gapped theory $\cZ[\textsf{B}]=1$.

\subsection{4-manifolds, surgery, and more general Ward identities}

Our readers familiar with surgery theory and the handlebody decomposition of 4-manifolds might have recognized in the discussion above that the Hopf link formed by the lines $L$ and $L'$ is precisely the surgery diagram for the handlebody decomposition of $S^2 \times S^2$. This is no coincidence: one can indeed view the compact 4-manifold $S^2 \times S^2$ as a null bordism
\be
\boldsymbol{\Omega}_{S^2\times S^2}:\emptyset \to S^3 \xrightarrow{H_2^{1}} S^1 \times S^2 \xrightarrow{H_2^{2}} S^3 \to \emptyset
\ee
where the maps $H_2^i$ for $i=1,2$ are 4-dimensional 2-handle-gluing operations. We can associate a Ward identity to $\boldsymbol{\Omega}_{S^2\times S^2}$ by nucleating the symmetry defect $\Dchiral{p/N}$ on the left side of the bordism and then shrink it onto the other side after passing through the surgery maps. In what follows we explain this terminology and, in light of this remark, we revisit the above derivation and formulate a generalization of our Ward identity to more general 4-manifolds $X^4$ viewing them as bordisms $\boldsymbol{\Omega}_{X^4}$ thanks to their handlebody decompositions.

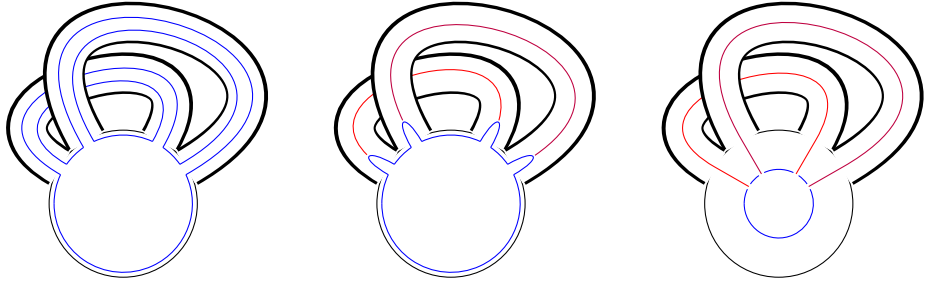
\begin{figure}[t]
    \centering
\begin{tikzpicture}[scale=.65]
    \pgfmathsetmacro{\R}{1.5}
    \pgfmathsetmacro{\r}{1.4}
    \pgfmathsetmacro{\Rm}{.04}
    \pgfmathsetmacro{\RR}{3}
    \pgfmathsetmacro{\RRR}{5}
    \pgfmathsetmacro{\rr}{\RR+\R-\r}
    \pgfmathsetmacro{\rrr}{\RRR+\R-\r}
    \pgfmathsetmacro{\rrm}{.4}
    \pgfmathsetmacro{\widone}{.25cm}
    \coordinate (O) at (0,0);
    \coordinate (0) at ($(O)+(180:\R)$);
    \coordinate (0b) at ($(O)+(180:\r)$);
    \coordinate (1a) at ($(O)+(155:\r)$);
    \coordinate (1b) at ($(O)+(145:\r)$);
    \coordinate (1c) at ($(O)+(150:\R)$);
    \coordinate (1d) at ($(O)+(150:\R-\Rm)$);
    \coordinate (2a) at ($(O)+(125:\r)$);
    \coordinate (2b) at ($(O)+(115:\r)$);
    \coordinate (2c) at ($(O)+(120:\R)$);
    \coordinate (2d) at ($(O)+(120:\R-\Rm)$);
    \coordinate (3a) at ($(O)+(65:\r)$);
    \coordinate (3b) at ($(O)+(55:\r)$);
    \coordinate (3c) at ($(O)+(60:\R)$);
    \coordinate (3d) at ($(O)+(60:\R-\Rm)$);
    \coordinate (4a) at ($(O)+(35:\r)$);
    \coordinate (4b) at ($(O)+(25:\r)$);
    \coordinate (4c) at ($(O)+(30:\R)$);
    \coordinate (4d) at ($(O)+(30:\R-\Rm)$);
    \draw (O) circle(\R);
    \draw[line width=.55cm,overlay] (1c) ..controls ++(150:\RR) and ++(60:\RR) .. (3c);
    \draw[line width=.47cm,draw=white,overlay] (1d) ..controls ++(150:\RR+\Rm) and ++(60:\RR+\Rm) .. (3d);
    \draw[blue,overlay] (1a) ..controls ++(150:\rr+\rrm) and ++(60:\rr+\rrm) .. (3b);
    \draw[blue,overlay] (1b) ..controls ++(150:\rr-\rrm) and ++(60:\rr-\rrm) .. (3a);
    \draw[line width=.55cm,overlay] (2c) ..controls ++(120:\RRR) and ++(30:\RRR) .. (4c);
    \draw[line width=.47cm,draw=white,overlay] (2d) ..controls ++(120:\RRR+\Rm) and ++(30:\RRR+\Rm) .. (4d);
    \draw[blue,overlay] (2a) ..controls ++(120:\rrr+\rrm) and ++(30:\rrr+\rrm) .. (4b);
    \draw[blue,overlay] (2b) ..controls ++(120:\rrr-\rrm) and ++(30:\rrr-\rrm) .. (4a);
    \foreach \x/\a/\b in {0b/180/155,1b/145/125,2b/115/65,3b/55/35,4b/25/-180}{
    \draw[blue] (\x) arc(\a:\b:\r);
    }

    \node at (0:2*\R) {};
    \node at (180:2*\R) {};
    \node at (90:2*\R) {};
\end{tikzpicture}
\begin{tikzpicture}[scale=.65]
    \pgfmathsetmacro{\R}{1.5}
    \pgfmathsetmacro{\r}{1.4}
    \pgfmathsetmacro{\Rm}{.04}
    \pgfmathsetmacro{\RR}{3}
    \pgfmathsetmacro{\RRR}{5}
    \pgfmathsetmacro{\rr}{\RR+\R-\r}
    \pgfmathsetmacro{\rrr}{\RRR+\R-\r}
    \pgfmathsetmacro{\rrm}{.4}
    \coordinate (O) at (0,0);
    \coordinate (0) at ($(O)+(180:\R)$);
    \coordinate (0b) at ($(O)+(180:\r)$);
    \coordinate (1a) at ($(O)+(155:\r)$);
    \coordinate (1b) at ($(O)+(145:\r)$);
    \coordinate (1c) at ($(O)+(150:\R)$);
    \coordinate (1d) at ($(O)+(150:\R-\Rm)$);
    \coordinate (2a) at ($(O)+(125:\r)$);
    \coordinate (2b) at ($(O)+(115:\r)$);
    \coordinate (2c) at ($(O)+(120:\R)$);
    \coordinate (2d) at ($(O)+(120:\R-\Rm)$);
    \coordinate (3a) at ($(O)+(65:\r)$);
    \coordinate (3b) at ($(O)+(55:\r)$);
    \coordinate (3c) at ($(O)+(60:\R)$);
    \coordinate (3d) at ($(O)+(60:\R-\Rm)$);
    \coordinate (4a) at ($(O)+(35:\r)$);
    \coordinate (4b) at ($(O)+(25:\r)$);
    \coordinate (4c) at ($(O)+(30:\R)$);
    \coordinate (4d) at ($(O)+(30:\R-\Rm)$);
    \draw (O) circle(\R);
    \draw[line width=.55cm,overlay] (1c) ..controls ++(150:\RR) and ++(60:\RR) .. (3c);
    \draw[line width=.47cm,draw=white,overlay] (1d) ..controls ++(150:\RR+\Rm) and ++(60:\RR+\Rm) .. (3d);
    
    \draw[blue,overlay] (1a) ..controls ++(150:.7) and ++(150:.7) .. node[midway,inner sep=0] (1ab) {} (1b);
    \draw[blue,overlay] (3a) ..controls ++(60:.7) and ++(60:.7) .. node[midway,inner sep=0] (3ab) {} (3b);
    
    \draw[red,overlay] (1ab) ..controls ++(135:\RR-.9) and ++(75:\RR-.9) ..  (3ab);
    
    \draw[line width=.55cm,overlay] (2c) ..controls ++(120:\RRR) and ++(30:\RRR) .. (4c);
    \draw[line width=.47cm,draw=white,overlay] (2d) ..controls ++(120:\RRR+\Rm) and ++(30:\RRR+\Rm) .. (4d);
    
    \draw[blue,overlay] (2a) ..controls ++(120:.7) and ++(120:.7) .. node[midway,inner sep=0] (2ab) {} (2b);
    \draw[blue,overlay] (4a) ..controls ++(30:.7) and ++(30:.7) .. node[midway,inner sep=0] (4ab) {} (4b);

    \draw[purple,overlay] (2ab) ..controls ++(115:\RRR-1) and ++(35:\RRR-1) ..  (4ab);

    \foreach \x/\a/\b in {0b/180/155,1b/145/125,2b/115/65,3b/55/35,4b/25/-180}{
    \draw[blue] (\x) arc(\a:\b:\r);
    }
    
    \node at (0:2*\R) {};
    \node at (180:2*\R) {};
    \node at (90:2*\R) {};
\end{tikzpicture}
\begin{tikzpicture}[scale=.65]
    \pgfmathsetmacro{\R}{1.5}
    \pgfmathsetmacro{\r}{.7}
    \pgfmathsetmacro{\Rm}{.04}
    \pgfmathsetmacro{\RR}{3}
    \pgfmathsetmacro{\RRR}{5}
    \pgfmathsetmacro{\rr}{\RR+\R-\r}
    \pgfmathsetmacro{\rrr}{\RRR+\R-\r}
    \pgfmathsetmacro{\rrm}{.4}
    \coordinate (O) at (0,0);
    \coordinate (0) at ($(O)+(180:\R)$);
    \coordinate (0b) at ($(O)+(180:\r)$);
    \coordinate (1a) at ($(O)+(155:\r)$);
    \coordinate (1b) at ($(O)+(145:\r)$);
    \coordinate (1c) at ($(O)+(150:\R)$);
    \coordinate (1d) at ($(O)+(150:\R-\Rm)$);
    \coordinate (2a) at ($(O)+(125:\r)$);
    \coordinate (2b) at ($(O)+(115:\r)$);
    \coordinate (2c) at ($(O)+(120:\R)$);
    \coordinate (2d) at ($(O)+(120:\R-\Rm)$);
    \coordinate (3a) at ($(O)+(65:\r)$);
    \coordinate (3b) at ($(O)+(55:\r)$);
    \coordinate (3c) at ($(O)+(60:\R)$);
    \coordinate (3d) at ($(O)+(60:\R-\Rm)$);
    \coordinate (4a) at ($(O)+(35:\r)$);
    \coordinate (4b) at ($(O)+(25:\r)$);
    \coordinate (4c) at ($(O)+(30:\R)$);
    \coordinate (4d) at ($(O)+(30:\R-\Rm)$);
    \draw (O) circle(\R);
    \draw[line width=.55cm,overlay] (1c) ..controls ++(150:\RR) and ++(60:\RR) .. (3c);
    \draw[line width=.47cm,draw=white,overlay] (1d) ..controls ++(150:\RR+\Rm) and ++(60:\RR+\Rm) .. (3d);

    \draw[red,overlay] ($(O)+(150:\r)$) -- (1c) ..controls ++(150:\RR) and ++(60:\RR) ..  (3c) -- ($(O)+(60:\r)$);
    
    \draw[line width=.55cm,overlay] (2c) ..controls ++(120:\RRR) and ++(30:\RRR) .. (4c);
    \draw[line width=.47cm,draw=white,overlay] (2d) ..controls ++(120:\RRR+\Rm) and ++(30:\RRR+\Rm) .. (4d);

    \draw[purple,overlay] ($(O)+(120:\r)$) -- (2c) ..controls ++(120:\RRR) and ++(30:\RRR) ..  (4c) -- ($(O)+(30:\r)$);

    \foreach \x/\a/\b in {0b/180/155,1b/145/125,2b/115/65,3b/55/35,4b/25/-180}{
    \draw[blue] (\x) arc(\a:\b:\r);
    }
    
    \node at (0:2*\R) {};
    \node at (180:2*\R) {};
    \node at (90:2*\R) {};
\end{tikzpicture}
\caption{Passing the defect through $S^2\times S^2$ in the handlebody picture. We note that the  boundary of the handle body is diffeomorphic to $S^3$, and we cap it to create $S^2\times S^2$. We first bubble up the defect in the cap, resulting in the defect going along the boundary of the handle body (leftmost figure). Then we do the partial fusions, resulting in one-form symmetry generators though the handles (middle figure). Finally, we shrink the defect towards the center of the handlebody (rightmost figure). The one-form symmetry generators ends on the defect along the surgery diagram, which is the Hopf link. This consideration generalizes to any handlebodies with 2-handles.}
\label{fig: double handle dumbbells}
\end{figure}

\paragraph{Handlebody decompositions} A beautiful consequence of Morse theory is that it is always possible to represent $n$-dimensional orientable manifolds in terms of surgeries of $n$-dimensional handles of various types. Giving a comprehensive review of this fact is beyond the scope of this manuscript, and for this reason we will not be pedagogical here and just give a few basic definitions that will be useful for our derivation of the more general Ward identities from handlebody decompositions of 4-manifolds.  Recall that a $k$-handle in $n$-dimensions is a product of two higher dimensional discs of appropriate dimensions, $D^k \times D^{n-k}$. In the case of 4-manifolds there are 4 kinds of those -- 1-handles $D^1 \times D^3$, 2-handles $D^2 \times D^2$, 3-handles $D^3 \times D^1$ and 4-handles $D^4 \times D^0$. The latter are relevant for manifolds with boundaries -- compact 4-manifolds always admit handlebody decompositions without 4-handles. A handlebody decomposition always looks as follows
\be
D^4 = M_0 \subset M_1 \subset  M_2 \subset \cdots \subset M_{\ell-1}\subset M_\ell = X^4 
\ee
where $D^4$ is a 4-ball with boundary $\partial  D^4 = S^3$ and $M_i = M_{i-1} \cup_{H_{i}} D^{k_i} \times D^{4-k_i}$, where the above notation means that $M_i$ is obtained from $M_{i-1}$ via a surgery which is gluing in a $k_i$-handle via the embedding (attaching/surgery map) $H_i : \partial D_i^{k_i} \times D^{4-k_i} \to \partial M_{i-1}$.  One can always order the above decompositions in such a way that the smallest $i$'s correspond to the 1-handles, then followed by the 2-handles, the 3-handles, and the 4-handles. The gluings of 1-handles can be represented pictorially by just drawing on a 3-sphere the 2-spheres where these are attached by the corresponding attaching maps. For the 2-handles these can be represented by drawing the corresponding $S^1$'s together with an integer label encoding the induced framing by the twisting of the normal bundle in $S^3$, in other words for a 2-handle $H(S^1 \times \{0\})$ is a framed knot in $S^3$, which is labeled by its framing $k \in \pi_1(SO(2)) \simeq \mathbb Z$. For a four-manifold with more than one 2-handle, the corresponding $S^1$'s become various strands of a link, and one can introduce a symmetric integer valued linking pairing $A_{ij}$ where along the diagonal one enters all the framings, and the off diagonal components are linking numbers. For 4-manifolds that are built from surgeries without exploiting 1-handles, it is a theorem that the form $A_{ij}$ coincides with the intersection pairing in 2-homology $H_2(X^4) \times H_2(X^4) \to \mathbb Z$. This is a point that will be important for us later.
$3$-handles and $4$-handles of a closed $X^4$ manifold together are diffeomorphic to $\#^k (S^1 \times D^3)$ with boundary $\#^k (S^1 \times S^2)$. So these are always attached via a diffeomorphism of $\#^k (S^1 \times S^2)$.

\medskip

\paragraph{Ward identities from surgery bordism} Reacall that a bordism between two three manifolds $M_3^-$ and $M_3^+$ is a four manifold $M_4$ with $\partial M_4 = M_3^- \sqcup M_3^+$. From this perspective, closed four manifolds can be interpreted as bordisms $\emptyset \to \emptyset$ and three manifolds $M_3$ that are boundaries of four-manifolds $\partial M_4 = M_3$ are said to be null-bordant, as the latter gives a bordism $\emptyset \to M_3$. The various surgery operations we discussed above in describing the handle decomposition of a four-manifold can be interpreted as compositions of bordisms.

\medskip

\begin{figure}
\centering
\includegraphics[scale=0.5]{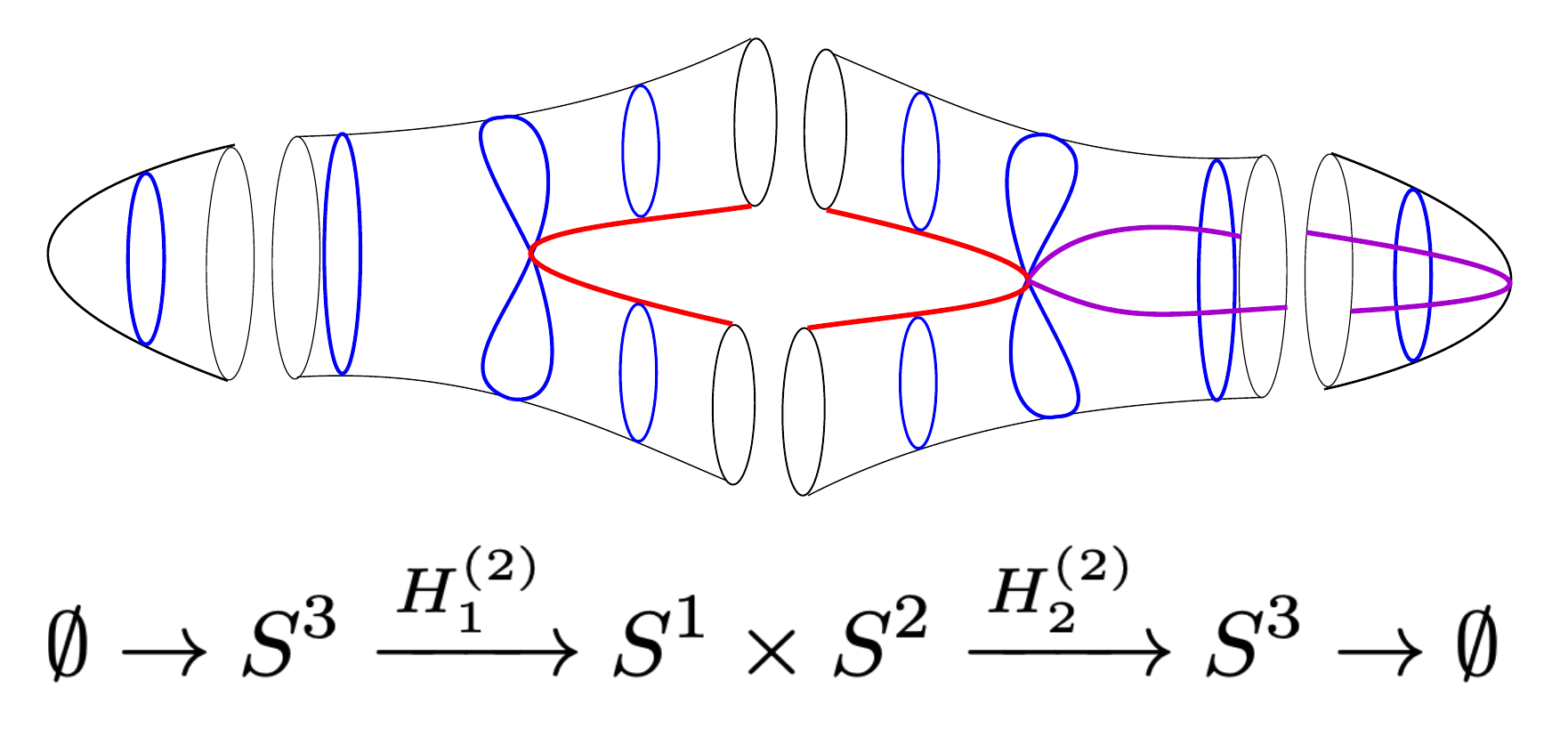}
\caption{Handle gluing and defect fusion - schematic description of the bordism description: here we are drawing just a $T^2$ section of the $S^2 \times S^2$ geometry to showcase the way in which the handle-gluing operation is causing the defect to fold on itself, leaving behind a condensate (in red and in violet). The junction of the defect and the condensate supports a nontrivial line in the 3d TFT $\mathcal A_{N,p}$, which results in the Hopf-link vev.}
\label{fig:KO1}
\end{figure}

This is a good moment to revisit the $S^2 \times S^2$ example of earlier. $S^2 \times S^2$ has a handle-body decomposition in terms of gluing of $2$-handles only. The corresponding surgery diagram is the Hopf link
\be
\begin{gathered}\begin{tikzpicture}
    \draw[braid slice] (2,0) arc (-180:0:1 and 1);
    \draw[braid slice] (3,0) arc (0:360: 1 and 1);
     \draw[braid slice] (2,0) arc (180:0:1 and 1);
     \node[below] at (2,-1) {$H_1^{(2)}$}; \node[below] at (3,-1) {$H_2^{(2)}$};
     \node[above] at (2,1) {$0$}; \node[above] at (3,1) {$0$};
\end{tikzpicture}\end{gathered}
\ee
Recall that we should think of these as the images via the two 2-handle attaching maps $H_1^{(2)}(S^1 \times \{0\})$ and $H_2^{(2)}(S^1 \times \{0\})$. The resulting intersection pairing in homology is
\be
A = \left(\begin{matrix} 0 & 1 \\ 1 & 0\end{matrix}\right)\,.
\ee
The process of nucleating the $\Dchiral{p/N}$ defect from a point can be seen as the first creation map in the bordism 
\be
\boldsymbol{\Omega}_{S^2\times S^2}:\emptyset \to S^3 \xrightarrow{\quad H_1^{(2)}} S^1 \times S^2 \xrightarrow{\quad H_2^{(2)}} S^3 \to \emptyset
\ee
then the first handle gluing (corresponding to the first surgery) forces the defect to fold against itself (see Figures \ref{fig: double handle dumbbells} and \ref{fig:KO1} as an illustration of this idea) and hence we must associate to $H_1^{(2)}$ a morphism $\mor: \Dchiral{p/N} \times \overline{\Dchiral{p/N}} \to \cond_N$. The condensation defect is  located on the attached handle. Recall that the latter must end on $\Dchiral{p/N}$ on a line which we denoted $L$ in the discussion in the previous section. This line is precisely the attaching line of the 2-handle $H_1^{(2)}(S^1 \times \{0\})$. The second handle gluing operation (corresponding to the second surgery in the discussion above) is identical similarly forces the defect to fold against itself once more, producing the second condensation supported on the second handle. Again the second condensate must end on the defect world volume along the attaching line of the second two-handle $H_2^{(2)}(S^1 \times \{0\})$ giving rise to the lines we denoted $L^\prime$ in the discussion above. This gives rise to the following elegant rewriting on the Ward identity we found in the section above:
\bea
\cZ_{S^2 \times S^2} = \frac{1}{N} \sum_{b, \, c \, \in \, \bZ_N} \frac{ \left<  \begin{tz}[scale=0.5]
 \draw[braid slice blue] (2,0) arc (-180:0:1 and 1);
    \draw[braid slice blue] (3,0) arc (0:360: 1 and 1);
     \draw[braid slice blue] (2,0) arc (180:0:1 and 1);    
      \node[left] at (1,0) {$L^{p^{-1} b}$}; \node[right] at (4,0) {$L^{p^{-1}c}$};
\end{tz} \right>_{\cA^{N, \, p}}  }{\langle \ \cdot \ \rangle_{\cA^{N, p}} } \, \cZ[b,c]_{S^2 \times S^2} \, ,
\eea
where in the last term we interpret $b, c$ as one-form symmetry backgrounds. 

\medskip

For compact 4-manifolds $X$ that are built purely by surgeries without 1-handles, the above discussion generalises straightforwardly. We have a bordism
\be
\boldsymbol{\Omega}_X \colon \emptyset \to S^3 \xrightarrow{\quad H_1^{(2)}}  S^1 \times S^2 \to \cdots \xrightarrow{\quad H_P^{(2)}} \begin{cases} S^3 \\ \#^k S^1 \times S^2\end{cases} 
 \to \emptyset
 \ee
and the 2-handle gluing operation results in a link with $P$ strands. Let us denote said link by $\mathcal L$. By the above discussion each strand of the link carries a line of the 3d TFT. Then the Ward identity reads
\be
\cZ_{X} = \frac{1}{\sqrt{H^2(X,\mathbb Z_N)}} \sum_{b \, \in \, H^2(X, \, \bZ_N)} \frac{\langle \ \mathcal L_b \, \rangle_{\cA^{N, \,p}}}{ \langle \ \cdot \ \rangle_{\cA^{N, \, p}} } \, \cZ_X(\,b\,) \, \,.
\ee
up choosing an overall Euler counterterm. Using the relation between the plumbing graph and the intersection form on $H_2$ it is simple to show that this reproduces the twist with a Pontryagin square $p^{-1} \, \fP$.

In the derivations of the above Ward identities we have used only specific F-symbols (associated to the partial fusions of the chiral symmetry defects on themselves). More in general, we conjecture that from 4-manifolds with 1-handles, more interesting topologies would arise from the associated cobordisms which we can exploit to explore Ward identities that are sensitive also to higher associators.

\section*{Acknowledgments }

We thank Mike Hopkins for illuminating discussions, Francesco Benini for collaboration in the initial stages of the project and Andrea Grigoletto for letting us know about an upcoming paper.
C.C. is supported by the ERC-COG grant NP-QFT No. 864583 ``Non-perturbative dynamics of quantum fields: from new deconfined phases of
matter to quantum black holes", as well as by the INFN ``Iniziativa Specifica ST\&FI".
The work of MDZ is supported by the European Research Council (ERC) under
the European Union’s Horizon 2020 research and innovation program (grant agreement
No. 851931). 
KO is supported by JSPS KAKENHI Grant-in-Aid No.22K13969.
MDZ and KO also acknowledges support from the Simons Foundation Grant
\#888984 (Simons Collaboration on Global Categorical Symmetries). The work of YW was
supported in part by NSF grant PHY-2210420 and by the Simons Junior Faculty Fellows program.

\appendix

\section{Minimal TFTs} \label{app: minimal}
In this appendix we recall some basic facts about minimal TFTs $\cA^{N,p}$ \cite{Hsin:2018vcg}. 
As a spin TFT $\cA^{N,p}$ is defined for all values of $p$ such that $\gcd(p,N)=1$ and $p \sim p + N$.
Its spectrum of lines is isomorphic to $\bZ_N$ and has a single generator $L$ with
\be
L^N = 1 \, , \ \ \ \theta_{L} = \exp\left(\frac{\pi i p}{N}\right) \, .
\ee
If $N p \in 2 \bZ$ then $L^N$ has integer spin and the theory is bosonic (i.e. it can be defined without reference to a spin structure). The line $L$ is taken to have charge $p$ under the one-form symmetry:
\be
q(L) = p \, .
\ee

The F-symbol $\AF_{a , \,  b , \, c}$ of $L^a$, $L^b$ and $L^c$ --- in a convenient gauge --- is
\be
\AF_{a , \,  b , \, c} = \exp\left(\frac{\pi i p}{N}(b + c - [b + c]_N) a \right) \, , \ \ \ [a]_N = a \mod N \, .
\ee
It is is trivial for even $p$, while it is valued in $\pm 1$ for odd $p$. For a bosonic theory this signals an obstruction to the 1-gauging of the one form symmetry $\bZ_N$ \cite{Roumpedakis:2022aik}.

In the same gauge the $R$ matrix implementing the braiding between lines in $\cA^{N, p}$ can be taken to be
\be
\boldsymbol{R}_{a , \, b} = \exp\left( \frac{\pi i p}{N} \, a  b \right) \, .
\ee
Which gives a braiding matrix:
\be
\boldsymbol{B}_{a b} = \exp\left( \frac{2 \pi i p}{N} \, a  b \right) \, .
\ee
The classification of unitary invertible symmetries for the minimal theories is as follows: a symmetry $\cV$ implements an automorphism of the categorical data, in particular it must act on the lines $L^k$ of $\cA^{N,p}$ as $\cV[L^k] = L^{\omega k}$, $\omega \in \bZ_N^\times$. Furthermore the spin must be preserved $\mod 1/2$. This forces:
\be
\omega^2 = 1 \mod N \, .
\ee
To be an involutive automorphism of $\bZ_N$.
Given a decomposition $N= p p'$, with $\gcd(p,p')=1$ we can find unique integers $r_0, s_0$ such that $r_0 p - s_0 p' = 1 \mod N$. Then $\omega = r_0 p + s_0 p'$ is an involution over the set of lines. If instead $p$ and $p'$ are not coprime it is possible to still construct a symmetry which will not be an automorphism (since it will have a kernel). This is represented by a non-invertible interface. See \cite{Delmastro:2019vnj} for a thorough investigation for Abelian Chern-Simons theories and \cite{Benini:2022hzx
} for an holographic application.

\section{Witt group for abelian TFTs}
Here we briefly review the results of \cite{davydov2013structure,davydov2013witt} on the Witt group for (spin) abelian TFTs. 
The Witt group $\cW$ takes the form:
\be
\cW = \left(\bigotimes_{p \ \text{prime}} \cW_p \right) \otimes \cW_{\text{spin}}
\ee
where:
\be
\cW_p = \begin{cases}
&\bZ_2 \times \bZ_2 \ \ \ \text{if} \ p=1 \mod 4 \\
&\bZ_4 \ \ \ \ \ \ \ \ \ \ \text{if} \ p=3 \mod 4 \\
& \bZ_2 \ \ \ \ \ \ \ \ \ \ \text{if} \ p=2	
\end{cases}
\ee
To this we must add a further factor $\cW_{\text{spin}} = \bZ_8$ encoding the information about invertible abelian spin TFTs.

Let us give a physical interpretation for how this comes about. First we study a theory  $\cA^{N,p}$ with $N= \prod_{n \ \text{prime}} n^{\alpha_n}$. We decompose it into minimal theories with with $\bZ_{n}$ symmetry. 
\begin{itemize}
	\item If $n$ has $\alpha_n=1$ it can be separated out, giving $\cA^{n, p_n}$, which is labelled by the Legendre symbol $l(p_n) =(p_n)^{\frac{n-1}{2}}$ (this encodes the fact that $\cA^{n, p r^2} = \cA^{n ,p}$ for $\gcd(n,r)=1$). For $n$ prime this can only take two values.
	\item If a factor appears with $\alpha_n >1$ it always has a (spin) gaugeable subgroup $\bZ_n^{\lfloor \alpha_n/2 \rfloor}$. This can be gauged in the Witt class descending to $\alpha_n = 0,1$.
\end{itemize}
A single minimal theory is then labelled by a string $\left\{ (\alpha_n, l(p_n))  \right\}$. One has to discuss fusion of different such factors. It turns out that stacking $\cA^{n, p_n}$ theories always leads to gaugeable algebras eventually. The only thing that matters is if $\cA^{n, p_n}$ with $l(p_n)=1$ is its own inverse under stacking, which means that $\cA^{n ,p_n} \times \cA^{N, p_n}$ contains a (fermionic) Lagrangian subalgebra. If this happens we have $\bZ_2 \times \bZ_2$, otherwise $\bZ_4$. $n=2$ instead only has a single Legendre symbol, so we find a $\bZ_2$.

\section{Monopole lines and associator}
In this appendix we derive the fact that monopole lines $T$ cannot be invariant under the full chiral symmetry. While for the case of the chiral symmetry the same can also be deduced as a consequence of the fusion rules for $\Dchiral{p/N}$, we believe that the derivation given below could prove to be a useful generalization for future applications. 

\bigskip

The action of generalized symmetries on various kinds of extended operators, which has been recently studied under the guise of ``higher-representations'' \cite{Bartsch:2022mpm,Bartsch:2022ytj,Bartsch:2023pzl,Bhardwaj:2023wzd} is of clear physical interest. In particular it is a valid question to ask whether a symmetry structure allows for the trivial representation on extended objects and what are the obstructions for such a possibility to be realized. We will use a well known idea by Kapustin \cite{Kapustin:2005py} which allows to map (conformal) line operators on $\bR^4$ to conformal boundary conditions on $AdS_2 \times S^2$. We will then prove that the associator constitutes an obstruction for the boundary condition to be invariant.
As a matter of fact the interplay between non-invertible symmetries and boundary conditions have been studied recently in great detail in the 2d case by \cite{Choi:2023xjw}. We refer the reader to their work for a thorough presentation.

\bigskip

Consider a configuration where a monopole line $T$ is placed in $\bR^4$. We foliate $\bR^4$ as:
\be
ds^2 = dx^2 + dr^2 + r^2 d\Omega^2 \, , 
\ee
with $d\Omega^2$ the line element on $S^2$ and $T$ stretching along $x$ at $r=0$.
Via a Weyl rescaling $ds^2 \mapsto r^{-2} ds^2$ we can map this problem to $\text{AdS}_2 \times S^2$ with $T$ on the $\text{AdS}_2$ boundary. After reducing on $S^2$ this gives a boundary condition (which we still call $T$) in 2d. The 4d one-form symmetry $U(1)^{(1)}$ descends to $U(1)^{(1)} \times U(1)^{(1)}$ in 2d. The monopole line $T^q$ becomes as a boundary condition for a universe $u_q$ satisfying:
\be \label{eq: univev}
\langle \Uonef_\alpha \rangle_{u_q} = e^{i q \alpha} \, 
\ee
\bea
\begin{tz}
    \filldraw[color=blue, line width=1, fill=white!95!blue] (0,0) circle (1); 
    \node[left] at (-1,0) {$T^q$};
    \draw[fill=black] (0,0) circle (0.1) node[below] {$\Uonef_\alpha$} ;
    \node[right] at (1.25,0) {$= \ \ \ e^{i q \alpha}$};
    \begin{scope}[shift={(4,0)}]
     \filldraw[color=blue, line width=1, fill=white!95!blue] (0,0) circle (1); 
    \node[right] at (1,0) {$T^q$};    
    \end{scope}
\end{tz}
\eea
This is just a rephrasing of the $U(1)^{(1)}$ charge of the monopole line in 4d. The chiral symmetry $\Dchiral{p/N}$ descents to a (non-invertible) zero-form symmetry. We now assume that $T^q$ is an invariant boundary condition for $\Dchiral{p/N}$. This means in particular that $T^q$ is a module for the chiral symmetry category:
\bea
\begin{tz}
\filldraw[color=white, fill=white!95!blue] (0,0) -- (0,2) -- (2,2) -- (2,0) -- cycle;  
\draw[color=blue, line width=1] (0,0) to (0,2); 
\node[left] at (0,1) {$T^q$};
\draw[slice] (0,0.5) to (2,0.5) node[right] {$\frac{p}{N}$};
\draw[slice] (0,1.5) to (2,1.5) node[right] {$\frac{q}{N}$};
\node at (3,1) {$=$};
\begin{scope}[shift={(4,0)}]
\filldraw[color=white, fill=white!95!blue] (0,0) -- (0,2) -- (2,2) -- (2,0) -- cycle;  
\draw[color=blue, line width=1] (0,0) to (0,2); 
\node[left] at (0,1) {$T^q$};
\draw[slice] (0,1) to (1,1);
\draw[slice] (1,1) node[dot] {} to[out=\ur, in = left] (2,1.5)  node[right] {$\frac{q}{N}$};
\draw[slice] (1,1) to[out=\dr, in = left] (2,0.5)  node[right] {$\frac{p}{N}$}; 
\node[below] at (0.5,1) {$\frac{p + q}{N}$};
\end{scope}
\end{tz}
\eea
We now consider a triple fusion $\frac{p_1}{N_1} \otimes \frac{p_2}{N_2} \otimes \frac{p_3}{N_3}$ on the disk:
\bea
\begin{tz}
  \filldraw[color=blue, line width=1, fill=white!95!blue] (0,0) circle (2); 
    \node[left] at (-2,0) {$T^q$};    
\draw[slice] (135:2) to[out=right, in=\ul] (157.5:1) node[dot] {};
\draw[slice] (-2,0) to[out=right, in=\dl] (157.5:1);
\draw[slice] (157.5:1)to[out=right,in=\ul] (0,0) node[dot] {};
\draw[slice] (225:2) to[out=right, in=\dl] (0,0) to (2,0);
\node[below] at (1,0) {$\frac{p}{N}$};
\node at (120:1.5) {$\frac{p_1}{N_1}$};
\node at (190:1.5) {$\frac{p_2}{N_2}$};
\node at (-110:1.5) {$\frac{p_3}{N_3}$};
\node at (100:0.5) {$\frac{p_{12}}{N_{12}}$};
\node at (3,0) {$=$};
\begin{scope}[shift={(6,0)}]
  \filldraw[color=blue, line width=1, fill=white!95!blue] (0,0) circle (2); 
\node[right] at (2,0) {$T^q$};    
\draw[slice] (2,0) to (1,0) node[dot] {} to[out=\ul, in=right] (0.5,0.5) node[dot] {};
\draw[slice]  (0.5,0.5) to[out=\ul, in =right] (0,1);
\draw[slice] (0.5,0.5) to[out=\dl, in=right] (0,0) to[out=left, in=\ur] (-0.5,-0.5);
\draw[slice] (1,0) to[out=\dl, in=right] (0,-1);
\draw[slice] (-2,0) to (-1,0) node[dot] {} to[out=\dr, in= left] (-0.5,-0.5);
\draw[slice] (-1,0) to[out=\ur, in=left] (0,1);
\draw[slice] (-0.5,-0.5) node[dot] {} to[out=\dr, in =left] (0,-1);
\node[below] at (-1.5,0) {$\frac{p}{N}$};
\node[below] at (1.5,0) {$\frac{p}{N}$};
\node[above] at (0,1) {$\frac{p_1}{N_1}$};
\node[below] at (0,-1) {$\frac{p_3}{N_3}$};
\node[below] at (0,0) {$\frac{p_2}{N_2}$};
\end{scope}
\end{tz}
\eea
On one hand, the configuration can be trivialized by using the module property on $\frac{p_1}{N_1} \otimes \frac{p_2}{N_2}$ and $\frac{p_{12}}{N_{12}} \otimes \frac{p_3}{N_3}$ instead. Shrinking the bubble gives rise to the associator $\AF$
\bea
\begin{tz}
  \filldraw[color=blue, line width=1, fill=white!95!blue] (0,0) circle (2); 
\node[left] at (-2,0) {$T^q$};    
\draw[slice] (2,0) to (1,0) node[dot] {} to[out=\ul, in=right] (0.5,0.5) node[dot] {};
\draw[slice]  (0.5,0.5) to[out=\ul, in =right] (0,1);
\draw[slice] (0.5,0.5) to[out=\dl, in=right] (0,0) to[out=left, in=\ur] (-0.5,-0.5);
\draw[slice] (1,0) to[out=\dl, in=right] (0,-1);
\draw[slice] (-2,0) to (-1,0) node[dot] {} to[out=\dr, in= left] (-0.5,-0.5);
\draw[slice] (-1,0) to[out=\ur, in=left] (0,1);
\draw[slice] (-0.5,-0.5) node[dot] {} to[out=\dr, in =left] (0,-1);  
\node[below] at (-1.5,0) {$\frac{p}{N}$};
\node[below] at (1.5,0) {$\frac{p}{N}$};
\node[above] at (0,1) {$\frac{p_1}{N_1}$};
\node[below] at (0,-1) {$\frac{p_3}{N_3}$};
\node[below] at (0,0) {$\frac{p_2}{N_2}$};
\node at (3,0) {$=$};
\begin{scope}[shift={(6,0)}]
  \filldraw[color=blue, line width=1, fill=white!95!blue] (0,0) circle (2); 
\node[right] at (2,0) {$T^q$};   
\draw[slice] (-2,0) to (2,0);
\draw[fill=black] (0,0) circle (0.1) node[below] {$\AF_{2d}$};
\node[below] at (-1.5,0) {$\frac{p}{N}$};
\node[below] at (1.5,0) {$\frac{p}{N}$};
\end{scope}
\end{tz}
\eea
Expanding the F-symbol into idempotents:
\be
\AF = \bigoplus_{q=0}^{N_{123}} \left( \bigoplus_{i: q(\pi_i)=q} \, \boldsymbol{T}_{\pi_i} \right) \otimes \Uonef_{q/N_{123}} \, ,  \ \ \ N_{123} = \text{lcm}(N_1,N_2,N_3) \, .
\ee
and reducing on $S^2$ gives a one-form symmetry defect:
\be
\AF_{2d} = \bigoplus_{q=0}^{N_{123}} \cZ_q \; \Uonef_{q/N_{123}} \, , \ \ \ \cZ_q = \sum_{i: q(\pi_i)=q} \, \cZ\left(\boldsymbol{T}_{\pi_i}\right)_{S^2} \, .
\ee
Since in our case all the $\cZ_q$ are equal pushing this to the boundary and using \eqref{eq: univev} implies that the configuration is zero unless $q = 0$.

 \bibliographystyle{ytphys}
 \bibliography{fsymbols
 }

\providecommand{\href}[2]{#2}\begingroup\raggedright\begin{thebibliography}{100}

\bibitem{Gaiotto:2014kfa}
D.~Gaiotto, A.~Kapustin, N.~Seiberg, and B.~Willett, ``{Generalized Global
  Symmetries},'' \href{http://dx.doi.org/10.1007/JHEP02(2015)172}{{\em JHEP}
  {\bfseries 02} (2015) 172}, \href{http://arxiv.org/abs/1412.5148}{{\ttfamily
  arXiv:1412.5148 [hep-th]}}.

\bibitem{Bhardwaj:2017xup}
L.~Bhardwaj and Y.~Tachikawa, ``{On finite symmetries and their gauging in two
  dimensions},'' \href{http://dx.doi.org/10.1007/JHEP03(2018)189}{{\em JHEP}
  {\bfseries 03} (2018) 189}, \href{http://arxiv.org/abs/1704.02330}{{\ttfamily
  arXiv:1704.02330 [hep-th]}}.

\bibitem{Chang:2018iay}
C.-M. Chang, Y.-H. Lin, S.-H. Shao, Y.~Wang, and X.~Yin, ``{Topological Defect
  Lines and Renormalization Group Flows in Two Dimensions},''
  \href{http://dx.doi.org/10.1007/JHEP01(2019)026}{{\em JHEP} {\bfseries 01}
  (2019) 026}, \href{http://arxiv.org/abs/1802.04445}{{\ttfamily
  arXiv:1802.04445 [hep-th]}}.

\bibitem{Komargodski:2020mxz}
Z.~Komargodski, K.~Ohmori, K.~Roumpedakis, and S.~Seifnashri, ``{Symmetries and
  strings of adjoint QCD$_{2}$},''
  \href{http://dx.doi.org/10.1007/JHEP03(2021)103}{{\em JHEP} {\bfseries 03}
  (2021) 103}, \href{http://arxiv.org/abs/2008.07567}{{\ttfamily
  arXiv:2008.07567 [hep-th]}}.

\bibitem{Lin:2019kpn}
Y.-H. Lin and S.-H. Shao, ``{Anomalies and Bounds on Charged Operators},''
  \href{http://dx.doi.org/10.1103/PhysRevD.100.025013}{{\em Phys. Rev. D}
  {\bfseries 100} no.~2, (2019) 025013},
  \href{http://arxiv.org/abs/1904.04833}{{\ttfamily arXiv:1904.04833
  [hep-th]}}.

\bibitem{Thorngren:2019iar}
R.~Thorngren and Y.~Wang, ``{Fusion Category Symmetry I: Anomaly In-Flow and
  Gapped Phases},'' \href{http://arxiv.org/abs/1912.02817}{{\ttfamily
  arXiv:1912.02817 [hep-th]}}.

\bibitem{Thorngren:2021yso}
R.~Thorngren and Y.~Wang, ``{Fusion Category Symmetry II: Categoriosities at
  $c$ = 1 and Beyond},'' \href{http://arxiv.org/abs/2106.12577}{{\ttfamily
  arXiv:2106.12577 [hep-th]}}.

\bibitem{Jacobsen:2023isq}
J.~L. Jacobsen and H.~Saleur, ``{Non-invertible symmetries and RG flows in the
  two-dimensional $O(n)$ loop model},''
  \href{http://arxiv.org/abs/2305.05746}{{\ttfamily arXiv:2305.05746
  [math-ph]}}.

\bibitem{Verlinde:1988sn}
E.~P. Verlinde, ``{Fusion Rules and Modular Transformations in 2D Conformal
  Field Theory},'' \href{http://dx.doi.org/10.1016/0550-3213(88)90603-7}{{\em
  Nucl. Phys. B} {\bfseries 300} (1988) 360--376}.

\bibitem{Petkova:2000ip}
V.~B. Petkova and J.~B. Zuber, ``{Generalized twisted partition functions},''
  \href{http://dx.doi.org/10.1016/S0370-2693(01)00276-3}{{\em Phys. Lett. B}
  {\bfseries 504} (2001) 157--164},
  \href{http://arxiv.org/abs/hep-th/0011021}{{\ttfamily arXiv:hep-th/0011021}}.

\bibitem{Fuchs:2002cm}
J.~Fuchs, I.~Runkel, and C.~Schweigert, ``{TFT construction of RCFT correlators
  1. Partition functions},''
  \href{http://dx.doi.org/10.1016/S0550-3213(02)00744-7}{{\em Nucl. Phys. B}
  {\bfseries 646} (2002) 353--497},
  \href{http://arxiv.org/abs/hep-th/0204148}{{\ttfamily arXiv:hep-th/0204148}}.

\bibitem{Aharony:2013hda}
O.~Aharony, N.~Seiberg, and Y.~Tachikawa, ``{Reading between the lines of
  four-dimensional gauge theories},''
  \href{http://dx.doi.org/10.1007/JHEP08(2013)115}{{\em JHEP} {\bfseries 08}
  (2013) 115}, \href{http://arxiv.org/abs/1305.0318}{{\ttfamily arXiv:1305.0318
  [hep-th]}}.

\bibitem{Gaiotto:2017yup}
D.~Gaiotto, A.~Kapustin, Z.~Komargodski, and N.~Seiberg, ``{Theta, Time
  Reversal, and Temperature},''
  \href{http://dx.doi.org/10.1007/JHEP05(2017)091}{{\em JHEP} {\bfseries 05}
  (2017) 091}, \href{http://arxiv.org/abs/1703.00501}{{\ttfamily
  arXiv:1703.00501 [hep-th]}}.

\bibitem{Gaiotto:2017tne}
D.~Gaiotto, Z.~Komargodski, and N.~Seiberg, ``{Time-Reversal Breaking in
  QCD$_{4}$, Walls, and Dualities in $2 + 1$ Dimensions},''
  \href{http://dx.doi.org/10.1007/JHEP01(2018)110}{{\em JHEP} {\bfseries 01}
  (2018) 110}, \href{http://arxiv.org/abs/1708.06806}{{\ttfamily
  arXiv:1708.06806 [hep-th]}}.

\bibitem{Choi:2021kmx}
Y.~Choi, C.~Cordova, P.-S. Hsin, H.~T. Lam, and S.-H. Shao, ``{Non-Invertible
  Duality Defects in 3+1 Dimensions},''
  \href{http://dx.doi.org/10.1103/PhysRevD.105.125016}{{\em Phys. Rev. D}
  {\bfseries 105} no.~12, (2022) 125016},
  \href{http://arxiv.org/abs/2111.01139}{{\ttfamily arXiv:2111.01139
  [hep-th]}}.

\bibitem{Choi:2022zal}
Y.~Choi, C.~Cordova, P.-S. Hsin, H.~T. Lam, and S.-H. Shao, ``{Non-invertible
  Condensation, Duality, and Triality Defects in 3+1 Dimensions},''
  \href{http://arxiv.org/abs/2204.09025}{{\ttfamily arXiv:2204.09025
  [hep-th]}}.

\bibitem{Kaidi:2022uux}
J.~Kaidi, G.~Zafrir, and Y.~Zheng, ``{Non-Invertible Symmetries of
  $\mathcal{N}=4$ SYM and Twisted Compactification},''
  \href{http://arxiv.org/abs/2205.01104}{{\ttfamily arXiv:2205.01104
  [hep-th]}}.

\bibitem{Gaiotto:2010be}
D.~Gaiotto, G.~W. Moore, and A.~Neitzke, ``{Framed BPS States},''
  \href{http://dx.doi.org/10.4310/ATMP.2013.v17.n2.a1}{{\em Adv. Theor. Math.
  Phys.} {\bfseries 17} no.~2, (2013) 241--397},
  \href{http://arxiv.org/abs/1006.0146}{{\ttfamily arXiv:1006.0146 [hep-th]}}.

\bibitem{Kapustin:2013qsa}
A.~Kapustin and R.~Thorngren, ``{Topological Field Theory on a Lattice,
  Discrete Theta-Angles and Confinement},''
  \href{http://dx.doi.org/10.4310/ATMP.2014.v18.n5.a4}{{\em Adv. Theor. Math.
  Phys.} {\bfseries 18} no.~5, (2014) 1233--1247},
  \href{http://arxiv.org/abs/1308.2926}{{\ttfamily arXiv:1308.2926 [hep-th]}}.

\bibitem{Kapustin:2013uxa}
A.~Kapustin and R.~Thorngren, ``{Higher symmetry and gapped phases of gauge
  theories},'' \href{http://arxiv.org/abs/1309.4721}{{\ttfamily arXiv:1309.4721
  [hep-th]}}.

\bibitem{DelZotto:2015isa}
M.~Del~Zotto, J.~J. Heckman, D.~S. Park, and T.~Rudelius, ``{On the Defect
  Group of a 6D SCFT},''
  \href{http://dx.doi.org/10.1007/s11005-016-0839-5}{{\em Lett. Math. Phys.}
  {\bfseries 106} no.~6, (2016) 765--786},
  \href{http://arxiv.org/abs/1503.04806}{{\ttfamily arXiv:1503.04806
  [hep-th]}}.

\bibitem{Sharpe:2015mja}
E.~Sharpe, ``{Notes on generalized global symmetries in QFT},''
  \href{http://dx.doi.org/10.1002/prop.201500048}{{\em Fortsch. Phys.}
  {\bfseries 63} (2015) 659--682},
  \href{http://arxiv.org/abs/1508.04770}{{\ttfamily arXiv:1508.04770
  [hep-th]}}.

\bibitem{Tachikawa:2017gyf}
Y.~Tachikawa, ``{On gauging finite subgroups},''
  \href{http://dx.doi.org/10.21468/SciPostPhys.8.1.015}{{\em SciPost Phys.}
  {\bfseries 8} no.~1, (2020) 015},
  \href{http://arxiv.org/abs/1712.09542}{{\ttfamily arXiv:1712.09542
  [hep-th]}}.

\bibitem{Cordova:2018cvg}
C.~C\'ordova, T.~T. Dumitrescu, and K.~Intriligator, ``{Exploring 2-Group
  Global Symmetries},'' \href{http://dx.doi.org/10.1007/JHEP02(2019)184}{{\em
  JHEP} {\bfseries 02} (2019) 184},
  \href{http://arxiv.org/abs/1802.04790}{{\ttfamily arXiv:1802.04790
  [hep-th]}}.

\bibitem{Hsin:2018vcg}
P.-S. Hsin, H.~T. Lam, and N.~Seiberg, ``{Comments on One-Form Global
  Symmetries and Their Gauging in 3d and 4d},''
  \href{http://dx.doi.org/10.21468/SciPostPhys.6.3.039}{{\em SciPost Phys.}
  {\bfseries 6} no.~3, (2019) 039},
  \href{http://arxiv.org/abs/1812.04716}{{\ttfamily arXiv:1812.04716
  [hep-th]}}.

\bibitem{Wan:2018bns}
Z.~Wan and J.~Wang, ``{Higher anomalies, higher symmetries, and cobordisms I:
  classification of higher-symmetry-protected topological states and their
  boundary fermionic/bosonic anomalies via a generalized cobordism theory},''
  \href{http://dx.doi.org/10.4310/AMSA.2019.v4.n2.a2}{{\em Ann. Math. Sci.
  Appl.} {\bfseries 4} no.~2, (2019) 107--311},
  \href{http://arxiv.org/abs/1812.11967}{{\ttfamily arXiv:1812.11967
  [hep-th]}}.

\bibitem{GarciaEtxebarria:2019caf}
I.~n. Garc\'\i{}a~Etxebarria, B.~Heidenreich, and D.~Regalado, ``{IIB flux
  non-commutativity and the global structure of field theories},''
  \href{http://dx.doi.org/10.1007/JHEP10(2019)169}{{\em JHEP} {\bfseries 10}
  (2019) 169}, \href{http://arxiv.org/abs/1908.08027}{{\ttfamily
  arXiv:1908.08027 [hep-th]}}.

\bibitem{Eckhard:2019jgg}
J.~Eckhard, H.~Kim, S.~Schafer-Nameki, and B.~Willett, ``{Higher-Form
  Symmetries, Bethe Vacua, and the 3d-3d Correspondence},''
  \href{http://dx.doi.org/10.1007/JHEP01(2020)101}{{\em JHEP} {\bfseries 01}
  (2020) 101}, \href{http://arxiv.org/abs/1910.14086}{{\ttfamily
  arXiv:1910.14086 [hep-th]}}.

\bibitem{Wan:2019soo}
Z.~Wan, J.~Wang, and Y.~Zheng, ``{Higher anomalies, higher symmetries, and
  cobordisms II: Lorentz symmetry extension and enriched bosonic / fermionic
  quantum gauge theory},''
  \href{http://dx.doi.org/10.4310/AMSA.2020.v5.n2.a2}{{\em Ann. Math. Sci.
  Appl.} {\bfseries 05} no.~2, (2020) 171--257},
  \href{http://arxiv.org/abs/1912.13504}{{\ttfamily arXiv:1912.13504
  [hep-th]}}.

\bibitem{Bergman:2020ifi}
O.~Bergman, Y.~Tachikawa, and G.~Zafrir, ``{Generalized symmetries and
  holography in ABJM-type theories},''
  \href{http://dx.doi.org/10.1007/JHEP07(2020)077}{{\em JHEP} {\bfseries 07}
  (2020) 077}, \href{http://arxiv.org/abs/2004.05350}{{\ttfamily
  arXiv:2004.05350 [hep-th]}}.

\bibitem{Morrison:2020ool}
D.~R. Morrison, S.~Schafer-Nameki, and B.~Willett, ``{Higher-Form Symmetries in
  5d},'' \href{http://dx.doi.org/10.1007/JHEP09(2020)024}{{\em JHEP} {\bfseries
  09} (2020) 024}, \href{http://arxiv.org/abs/2005.12296}{{\ttfamily
  arXiv:2005.12296 [hep-th]}}.

\bibitem{Albertini:2020mdx}
F.~Albertini, M.~Del~Zotto, I.~García~Etxebarria, and S.~S. Hosseini,
  ``{Higher Form Symmetries and M-theory},''
  \href{http://arxiv.org/abs/2005.12831}{{\ttfamily arXiv:2005.12831
  [hep-th]}}.

\bibitem{Hsin:2020nts}
P.-S. Hsin and H.~T. Lam, ``{Discrete theta angles, symmetries and
  anomalies},'' \href{http://dx.doi.org/10.21468/SciPostPhys.10.2.032}{{\em
  SciPost Phys.} {\bfseries 10} no.~2, (2021) 032},
  \href{http://arxiv.org/abs/2007.05915}{{\ttfamily arXiv:2007.05915
  [hep-th]}}.

\bibitem{Bah:2020uev}
I.~Bah, F.~Bonetti, and R.~Minasian, ``{Discrete and higher-form symmetries in
  SCFTs from wrapped M5-branes},''
  \href{http://dx.doi.org/10.1007/JHEP03(2021)196}{{\em JHEP} {\bfseries 03}
  (2021) 196}, \href{http://arxiv.org/abs/2007.15003}{{\ttfamily
  arXiv:2007.15003 [hep-th]}}.

\bibitem{DelZotto:2020esg}
M.~Del~Zotto, I.~Garcia~Etxebarria, and S.~S. Hosseini, ``{Higher form
  symmetries of Argyres-Douglas theories},''
  \href{http://dx.doi.org/10.1007/JHEP10(2020)056}{{\em JHEP} {\bfseries 10}
  (2020) 056}, \href{http://arxiv.org/abs/2007.15603}{{\ttfamily
  arXiv:2007.15603 [hep-th]}}.

\bibitem{Hason:2020yqf}
I.~Hason, Z.~Komargodski, and R.~Thorngren, ``{Anomaly Matching in the Symmetry
  Broken Phase: Domain Walls, CPT, and the Smith Isomorphism},''
  \href{http://dx.doi.org/10.21468/SciPostPhys.8.4.062}{{\em SciPost Phys.}
  {\bfseries 8} no.~4, (2020) 062},
  \href{http://arxiv.org/abs/1910.14039}{{\ttfamily arXiv:1910.14039
  [hep-th]}}.

\bibitem{Aasen:2020jwb}
D.~Aasen, P.~Fendley, and R.~S.~K. Mong, ``{Topological Defects on the Lattice:
  Dualities and Degeneracies},''
  \href{http://arxiv.org/abs/2008.08598}{{\ttfamily arXiv:2008.08598
  [cond-mat.stat-mech]}}.

\bibitem{Bhardwaj:2020phs}
L.~Bhardwaj and S.~Sch\"afer-Nameki, ``{Higher-form symmetries of 6d and 5d
  theories},'' \href{http://dx.doi.org/10.1007/JHEP02(2021)159}{{\em JHEP}
  {\bfseries 02} (2021) 159}, \href{http://arxiv.org/abs/2008.09600}{{\ttfamily
  arXiv:2008.09600 [hep-th]}}.

\bibitem{Apruzzi:2020zot}
F.~Apruzzi, M.~Dierigl, and L.~Lin, ``{The fate of discrete 1-form symmetries
  in 6d},'' \href{http://dx.doi.org/10.21468/SciPostPhys.12.2.047}{{\em SciPost
  Phys.} {\bfseries 12} no.~2, (2022) 047},
  \href{http://arxiv.org/abs/2008.09117}{{\ttfamily arXiv:2008.09117
  [hep-th]}}.

\bibitem{Cordova:2020tij}
C.~Cordova, T.~T. Dumitrescu, and K.~Intriligator, ``{2-Group Global Symmetries
  and Anomalies in Six-Dimensional Quantum Field Theories},''
  \href{http://dx.doi.org/10.1007/JHEP04(2021)252}{{\em JHEP} {\bfseries 04}
  (2021) 252}, \href{http://arxiv.org/abs/2009.00138}{{\ttfamily
  arXiv:2009.00138 [hep-th]}}.

\bibitem{Thorngren:2020aph}
R.~Thorngren, ``{Topological quantum field theory, symmetry breaking, and
  finite gauge theory in 3+1D},''
  \href{http://dx.doi.org/10.1103/PhysRevB.101.245160}{{\em Phys. Rev. B}
  {\bfseries 101} no.~24, (2020) 245160},
  \href{http://arxiv.org/abs/2001.11938}{{\ttfamily arXiv:2001.11938
  [cond-mat.str-el]}}.

\bibitem{DelZotto:2020sop}
M.~Del~Zotto and K.~Ohmori, ``{2-Group Symmetries of 6D Little String Theories
  and T-Duality},'' \href{http://dx.doi.org/10.1007/s00023-021-01018-3}{{\em
  Annales Henri Poincare} {\bfseries 22} no.~7, (2021) 2451--2474},
  \href{http://arxiv.org/abs/2009.03489}{{\ttfamily arXiv:2009.03489
  [hep-th]}}.

\bibitem{BenettiGenolini:2020doj}
P.~Benetti~Genolini and L.~Tizzano, ``{Instantons, symmetries and anomalies in
  five dimensions},'' \href{http://dx.doi.org/10.1007/JHEP04(2021)188}{{\em
  JHEP} {\bfseries 04} (2021) 188},
  \href{http://arxiv.org/abs/2009.07873}{{\ttfamily arXiv:2009.07873
  [hep-th]}}.

\bibitem{Yu:2020twi}
M.~Yu, ``{Symmetries and anomalies of (1+1)d theories: 2-groups and symmetry
  fractionalization},'' \href{http://dx.doi.org/10.1007/JHEP08(2021)061}{{\em
  JHEP} {\bfseries 08} (2021) 061},
  \href{http://arxiv.org/abs/2010.01136}{{\ttfamily arXiv:2010.01136
  [hep-th]}}.

\bibitem{Bhardwaj:2020ymp}
L.~Bhardwaj, Y.~Lee, and Y.~Tachikawa, ``{$SL(2,\mathbb{Z})$ action on QFTs
  with $\mathbb{Z}_2$ symmetry and the Brown-Kervaire invariants},''
  \href{http://dx.doi.org/10.1007/JHEP11(2020)141}{{\em JHEP} {\bfseries 11}
  (2020) 141}, \href{http://arxiv.org/abs/2009.10099}{{\ttfamily
  arXiv:2009.10099 [hep-th]}}.

\bibitem{DeWolfe:2020uzb}
O.~DeWolfe and K.~Higginbotham, ``{Generalized symmetries and 2-groups via
  electromagnetic duality in $AdS/CFT$},''
  \href{http://dx.doi.org/10.1103/PhysRevD.103.026011}{{\em Phys. Rev. D}
  {\bfseries 103} no.~2, (2021) 026011},
  \href{http://arxiv.org/abs/2010.06594}{{\ttfamily arXiv:2010.06594
  [hep-th]}}.

\bibitem{Gukov:2020btk}
S.~Gukov, P.-S. Hsin, and D.~Pei, ``{Generalized global symmetries of $T[M]$
  theories. Part I},'' \href{http://dx.doi.org/10.1007/JHEP04(2021)232}{{\em
  JHEP} {\bfseries 04} (2021) 232},
  \href{http://arxiv.org/abs/2010.15890}{{\ttfamily arXiv:2010.15890
  [hep-th]}}.

\bibitem{Iqbal:2020lrt}
N.~Iqbal and N.~Poovuttikul, ``{2-group global symmetries, hydrodynamics and
  holography},'' \href{http://arxiv.org/abs/2010.00320}{{\ttfamily
  arXiv:2010.00320 [hep-th]}}.

\bibitem{Hidaka:2020izy}
Y.~Hidaka, M.~Nitta, and R.~Yokokura, ``{Global 3-group symmetry and 't Hooft
  anomalies in axion electrodynamics},''
  \href{http://dx.doi.org/10.1007/JHEP01(2021)173}{{\em JHEP} {\bfseries 01}
  (2021) 173}, \href{http://arxiv.org/abs/2009.14368}{{\ttfamily
  arXiv:2009.14368 [hep-th]}}.

\bibitem{Brennan:2020ehu}
T.~D. Brennan and C.~Cordova, ``{Axions, higher-groups, and emergent
  symmetry},'' \href{http://dx.doi.org/10.1007/JHEP02(2022)145}{{\em JHEP}
  {\bfseries 02} (2022) 145}, \href{http://arxiv.org/abs/2011.09600}{{\ttfamily
  arXiv:2011.09600 [hep-th]}}.

\bibitem{Closset:2020afy}
C.~Closset, S.~Giacomelli, S.~Schafer-Nameki, and Y.-N. Wang, ``{5d and 4d
  SCFTs: Canonical Singularities, Trinions and S-Dualities},''
  \href{http://dx.doi.org/10.1007/JHEP05(2021)274}{{\em JHEP} {\bfseries 05}
  (2021) 274}, \href{http://arxiv.org/abs/2012.12827}{{\ttfamily
  arXiv:2012.12827 [hep-th]}}.

\bibitem{Thorngren:2020yht}
R.~Thorngren and Y.~Wang, ``{Anomalous symmetries end at the boundary},''
  \href{http://dx.doi.org/10.1007/JHEP09(2021)017}{{\em JHEP} {\bfseries 09}
  (2021) 017}, \href{http://arxiv.org/abs/2012.15861}{{\ttfamily
  arXiv:2012.15861 [hep-th]}}.

\bibitem{Closset:2020scj}
C.~Closset, S.~Schafer-Nameki, and Y.-N. Wang, ``{Coulomb and Higgs Branches
  from Canonical Singularities: Part 0},''
  \href{http://arxiv.org/abs/2007.15600}{{\ttfamily arXiv:2007.15600
  [hep-th]}}.

\bibitem{Bhardwaj:2021pfz}
L.~Bhardwaj, M.~Hubner, and S.~Schafer-Nameki, ``{1-form Symmetries of 4d
  $\mathcal{N}=2$ Class S Theories},''
  \href{http://dx.doi.org/10.21468/SciPostPhys.11.5.096}{{\em SciPost Phys.}
  {\bfseries 11} (2021) 096}, \href{http://arxiv.org/abs/2102.01693}{{\ttfamily
  arXiv:2102.01693 [hep-th]}}.

\bibitem{Nguyen:2021naa}
M.~Nguyen, Y.~Tanizaki, and M.~\"Unsal, ``{Noninvertible 1-form symmetry and
  Casimir scaling in 2D Yang-Mills theory},''
  \href{http://dx.doi.org/10.1103/PhysRevD.104.065003}{{\em Phys. Rev. D}
  {\bfseries 104} no.~6, (2021) 065003},
  \href{http://arxiv.org/abs/2104.01824}{{\ttfamily arXiv:2104.01824
  [hep-th]}}.

\bibitem{Heidenreich:2021xpr}
B.~Heidenreich, J.~McNamara, M.~Montero, M.~Reece, T.~Rudelius, and
  I.~Valenzuela, ``{Non-invertible global symmetries and completeness of the
  spectrum},'' \href{http://dx.doi.org/10.1007/JHEP09(2021)203}{{\em JHEP}
  {\bfseries 09} (2021) 203}, \href{http://arxiv.org/abs/2104.07036}{{\ttfamily
  arXiv:2104.07036 [hep-th]}}.

\bibitem{Apruzzi:2021phx}
F.~Apruzzi, M.~van Beest, D.~S.~W. Gould, and S.~Sch\"afer-Nameki,
  ``{Holography, 1-form symmetries, and confinement},''
  \href{http://dx.doi.org/10.1103/PhysRevD.104.066005}{{\em Phys. Rev. D}
  {\bfseries 104} no.~6, (2021) 066005},
  \href{http://arxiv.org/abs/2104.12764}{{\ttfamily arXiv:2104.12764
  [hep-th]}}.

\bibitem{Apruzzi:2021vcu}
F.~Apruzzi, L.~Bhardwaj, J.~Oh, and S.~Schafer-Nameki, ``{The Global Form of
  Flavor Symmetries and 2-Group Symmetries in 5d SCFTs},''
  \href{http://arxiv.org/abs/2105.08724}{{\ttfamily arXiv:2105.08724
  [hep-th]}}.

\bibitem{Hosseini:2021ged}
S.~S. Hosseini and R.~Moscrop, ``{Maruyoshi-Song flows and defect groups of $
  {\mathrm{D}}_{\mathrm{p}}^{\mathrm{b}} $(G) theories},''
  \href{http://dx.doi.org/10.1007/JHEP10(2021)119}{{\em JHEP} {\bfseries 10}
  (2021) 119}, \href{http://arxiv.org/abs/2106.03878}{{\ttfamily
  arXiv:2106.03878 [hep-th]}}.

\bibitem{Cvetic:2021sxm}
M.~Cvetic, M.~Dierigl, L.~Lin, and H.~Y. Zhang, ``{Higher-form symmetries and
  their anomalies in M-/F-theory duality},''
  \href{http://dx.doi.org/10.1103/PhysRevD.104.126019}{{\em Phys. Rev. D}
  {\bfseries 104} no.~12, (2021) 126019},
  \href{http://arxiv.org/abs/2106.07654}{{\ttfamily arXiv:2106.07654
  [hep-th]}}.

\bibitem{Buican:2021xhs}
M.~Buican and H.~Jiang, ``{1-form symmetry, isolated $ \mathcal{N} $ = 2 SCFTs,
  and Calabi-Yau threefolds},''
  \href{http://dx.doi.org/10.1007/JHEP12(2021)024}{{\em JHEP} {\bfseries 12}
  (2021) 024}, \href{http://arxiv.org/abs/2106.09807}{{\ttfamily
  arXiv:2106.09807 [hep-th]}}.

\bibitem{Bhardwaj:2021zrt}
L.~Bhardwaj, M.~Hubner, and S.~Schafer-Nameki, ``{Liberating confinement from
  Lagrangians: 1-form symmetries and lines in 4d $\mathcal{N}=1$ from 6d
  $\mathcal{N}=(2,0)$},''
  \href{http://dx.doi.org/10.21468/SciPostPhys.12.1.040}{{\em SciPost Phys.}
  {\bfseries 12} no.~1, (2022) 040},
  \href{http://arxiv.org/abs/2106.10265}{{\ttfamily arXiv:2106.10265
  [hep-th]}}.

\bibitem{Iqbal:2021rkn}
N.~Iqbal and J.~McGreevy, ``{Mean string field theory: Landau-Ginzburg theory
  for 1-form symmetries},'' \href{http://arxiv.org/abs/2106.12610}{{\ttfamily
  arXiv:2106.12610 [hep-th]}}.

\bibitem{Braun:2021sex}
A.~P. Braun, M.~Larfors, and P.-K. Oehlmann, ``{Gauged 2-form symmetries in 6D
  SCFTs coupled to gravity},''
  \href{http://dx.doi.org/10.1007/JHEP12(2021)132}{{\em JHEP} {\bfseries 12}
  (2021) 132}, \href{http://arxiv.org/abs/2106.13198}{{\ttfamily
  arXiv:2106.13198 [hep-th]}}.

\bibitem{Cvetic:2021maf}
M.~Cveti\v{c}, J.~J. Heckman, E.~Torres, and G.~Zoccarato, ``{Reflections on
  the Matter of 3D $\mathcal{N} = 1$ Vacua and Local $Spin(7)$
  Compactifications},''
  \href{http://dx.doi.org/10.1103/PhysRevD.105.026008}{{\em Phys. Rev. D}
  {\bfseries 105} no.~2, (2022) 026008},
  \href{http://arxiv.org/abs/2107.00025}{{\ttfamily arXiv:2107.00025
  [hep-th]}}.

\bibitem{Closset:2021lhd}
C.~Closset and H.~Magureanu, ``{The $U$-plane of rank-one 4d $\mathcal{N}=2$ KK
  theories},'' \href{http://dx.doi.org/10.21468/SciPostPhys.12.2.065}{{\em
  SciPost Phys.} {\bfseries 12} no.~2, (2022) 065},
  \href{http://arxiv.org/abs/2107.03509}{{\ttfamily arXiv:2107.03509
  [hep-th]}}.

\bibitem{Sharpe:2021srf}
E.~Sharpe, ``{Topological operators, noninvertible symmetries and
  decomposition},'' \href{http://arxiv.org/abs/2108.13423}{{\ttfamily
  arXiv:2108.13423 [hep-th]}}.

\bibitem{Bhardwaj:2021wif}
L.~Bhardwaj, ``{2-Group symmetries in class S},''
  \href{http://dx.doi.org/10.21468/SciPostPhys.12.5.152}{{\em SciPost Phys.}
  {\bfseries 12} no.~5, (2022) 152},
  \href{http://arxiv.org/abs/2107.06816}{{\ttfamily arXiv:2107.06816
  [hep-th]}}.

\bibitem{Hidaka:2021mml}
Y.~Hidaka, M.~Nitta, and R.~Yokokura, ``{Topological axion electrodynamics and
  4-group symmetry},''
  \href{http://dx.doi.org/10.1016/j.physletb.2021.136762}{{\em Phys. Lett. B}
  {\bfseries 823} (2021) 136762},
  \href{http://arxiv.org/abs/2107.08753}{{\ttfamily arXiv:2107.08753
  [hep-th]}}.

\bibitem{Lee:2021obi}
Y.~Lee and Y.~Zheng, ``{Remarks on compatibility between conformal symmetry and
  continuous higher-form symmetries},''
  \href{http://dx.doi.org/10.1103/PhysRevD.104.085005}{{\em Phys. Rev. D}
  {\bfseries 104} no.~8, (2021) 085005},
  \href{http://arxiv.org/abs/2108.00732}{{\ttfamily arXiv:2108.00732
  [hep-th]}}.

\bibitem{Lee:2021crt}
Y.~Lee, K.~Ohmori, and Y.~Tachikawa, ``{Matching higher symmetries across
  Intriligator-Seiberg duality},''
  \href{http://dx.doi.org/10.1007/JHEP10(2021)114}{{\em JHEP} {\bfseries 10}
  (2021) 114}, \href{http://arxiv.org/abs/2108.05369}{{\ttfamily
  arXiv:2108.05369 [hep-th]}}.

\bibitem{Hidaka:2021kkf}
Y.~Hidaka, M.~Nitta, and R.~Yokokura, ``{Global 4-group symmetry and
  \textquoteright{}t Hooft anomalies in topological axion electrodynamics},''
  \href{http://dx.doi.org/10.1093/ptep/ptab150}{{\em PTEP} {\bfseries 2022}
  no.~4, (2022) 04A109}, \href{http://arxiv.org/abs/2108.12564}{{\ttfamily
  arXiv:2108.12564 [hep-th]}}.

\bibitem{Koide:2021zxj}
M.~Koide, Y.~Nagoya, and S.~Yamaguchi, ``{Non-invertible topological defects in
  4-dimensional $\mathbb {Z}_2$ pure lattice gauge theory},''
  \href{http://dx.doi.org/10.1093/ptep/ptab145}{{\em PTEP} {\bfseries 2022}
  no.~1, (2022) 013B03}, \href{http://arxiv.org/abs/2109.05992}{{\ttfamily
  arXiv:2109.05992 [hep-th]}}.

\bibitem{Apruzzi:2021mlh}
F.~Apruzzi, L.~Bhardwaj, D.~S.~W. Gould, and S.~Schafer-Nameki, ``{2-Group
  symmetries and their classification in 6d},''
  \href{http://dx.doi.org/10.21468/SciPostPhys.12.3.098}{{\em SciPost Phys.}
  {\bfseries 12} no.~3, (2022) 098},
  \href{http://arxiv.org/abs/2110.14647}{{\ttfamily arXiv:2110.14647
  [hep-th]}}.

\bibitem{Kaidi:2021xfk}
J.~Kaidi, K.~Ohmori, and Y.~Zheng, ``{Kramers-Wannier-like duality defects in
  $(3+1)$d gauge theories},''
  \href{http://dx.doi.org/10.1103/PhysRevLett.128.111601}{{\em Phys. Rev.
  Lett.} {\bfseries 128} no.~11, (2022) 111601},
  \href{http://arxiv.org/abs/2111.01141}{{\ttfamily arXiv:2111.01141
  [hep-th]}}.

\bibitem{Bah:2021brs}
I.~Bah, F.~Bonetti, E.~Leung, and P.~Weck, ``{M5-branes Probing Flux
  Backgrounds},'' \href{http://arxiv.org/abs/2111.01790}{{\ttfamily
  arXiv:2111.01790 [hep-th]}}.

\bibitem{Gukov:2021swm}
S.~Gukov, D.~Pei, C.~Reid, and A.~Shehper, ``{Symmetries of 2d TQFTs and
  Equivariant Verlinde Formulae for General Groups},''
  \href{http://arxiv.org/abs/2111.08032}{{\ttfamily arXiv:2111.08032
  [hep-th]}}.

\bibitem{Closset:2021lwy}
C.~Closset, S.~Sch\"afer-Nameki, and Y.-N. Wang, ``{Coulomb and Higgs branches
  from canonical singularities. Part I. Hypersurfaces with smooth Calabi-Yau
  resolutions},'' \href{http://dx.doi.org/10.1007/JHEP04(2022)061}{{\em JHEP}
  {\bfseries 04} (2022) 061}, \href{http://arxiv.org/abs/2111.13564}{{\ttfamily
  arXiv:2111.13564 [hep-th]}}.

\bibitem{Yu:2021zmu}
M.~Yu, ``{Gauging Categorical Symmetries in 3d Topological Orders and Bulk
  Reconstruction},'' \href{http://arxiv.org/abs/2111.13697}{{\ttfamily
  arXiv:2111.13697 [hep-th]}}.

\bibitem{Apruzzi:2021nmk}
F.~Apruzzi, F.~Bonetti, I.~n.~G. Etxebarria, S.~S. Hosseini, and
  S.~Schafer-Nameki, ``{Symmetry TFTs from String Theory},''
  \href{http://arxiv.org/abs/2112.02092}{{\ttfamily arXiv:2112.02092
  [hep-th]}}.

\bibitem{Beratto:2021xmn}
E.~Beratto, N.~Mekareeya, and M.~Sacchi, ``{Zero-form and one-form symmetries
  of the ABJ and related theories},''
  \href{http://dx.doi.org/10.1007/JHEP04(2022)126}{{\em JHEP} {\bfseries 04}
  (2022) 126}, \href{http://arxiv.org/abs/2112.09531}{{\ttfamily
  arXiv:2112.09531 [hep-th]}}.

\bibitem{Bhardwaj:2021mzl}
L.~Bhardwaj, S.~Giacomelli, M.~H\"ubner, and S.~Sch\"afer-Nameki, ``{Relative
  Defects in Relative Theories: Trapped Higher-Form Symmetries and Irregular
  Punctures in Class S},'' \href{http://arxiv.org/abs/2201.00018}{{\ttfamily
  arXiv:2201.00018 [hep-th]}}.

\bibitem{Wang:2021vki}
J.~Wang and Y.-Z. You, ``{Gauge Enhanced Quantum Criticality Between Grand
  Unifications: Categorical Higher Symmetry Retraction},''
  \href{http://arxiv.org/abs/2111.10369}{{\ttfamily arXiv:2111.10369
  [hep-th]}}.

\bibitem{Cvetic:2022uuu}
M.~Cveti\v{c}, M.~Dierigl, L.~Lin, and H.~Y. Zhang, ``{All eight- and
  nine-dimensional string vacua from junctions},''
  \href{http://dx.doi.org/10.1103/PhysRevD.106.026007}{{\em Phys. Rev. D}
  {\bfseries 106} no.~2, (2022) 026007},
  \href{http://arxiv.org/abs/2203.03644}{{\ttfamily arXiv:2203.03644
  [hep-th]}}.

\bibitem{DelZotto:2022fnw}
M.~Del~Zotto, J.~J. Heckman, S.~N. Meynet, R.~Moscrop, and H.~Y. Zhang,
  ``{Higher symmetries of 5D orbifold SCFTs},''
  \href{http://dx.doi.org/10.1103/PhysRevD.106.046010}{{\em Phys. Rev. D}
  {\bfseries 106} no.~4, (2022) 046010},
  \href{http://arxiv.org/abs/2201.08372}{{\ttfamily arXiv:2201.08372
  [hep-th]}}.

\bibitem{Cvetic:2022imb}
M.~Cveti\v{c}, J.~J. Heckman, M.~H\"ubner, and E.~Torres, ``{0-Form, 1-Form and
  2-Group Symmetries via Cutting and Gluing of Orbifolds},''
  \href{http://arxiv.org/abs/2203.10102}{{\ttfamily arXiv:2203.10102
  [hep-th]}}.

\bibitem{DelZotto:2022joo}
M.~Del~Zotto, I.~n.~G. Etxebarria, and S.~Schafer-Nameki, ``{2-Group Symmetries
  and M-Theory},'' \href{http://arxiv.org/abs/2203.10097}{{\ttfamily
  arXiv:2203.10097 [hep-th]}}.

\bibitem{DelZotto:2022ras}
M.~Del~Zotto and I.~Garcia~Etxebarria, ``{Global Structures from the
  Infrared},'' \href{http://arxiv.org/abs/2204.06495}{{\ttfamily
  arXiv:2204.06495 [hep-th]}}.

\bibitem{Bhardwaj:2022yxj}
L.~Bhardwaj, L.~Bottini, S.~Schafer-Nameki, and A.~Tiwari, ``{Non-Invertible
  Higher-Categorical Symmetries},''
  \href{http://arxiv.org/abs/2204.06564}{{\ttfamily arXiv:2204.06564
  [hep-th]}}.

\bibitem{Hayashi:2022fkw}
Y.~Hayashi and Y.~Tanizaki, ``{Non-invertible self-duality defects of
  Cardy-Rabinovici model and mixed gravitational anomaly},''
  \href{http://arxiv.org/abs/2204.07440}{{\ttfamily arXiv:2204.07440
  [hep-th]}}.

\bibitem{Roumpedakis:2022aik}
K.~Roumpedakis, S.~Seifnashri, and S.-H. Shao, ``{Higher Gauging and
  Non-invertible Condensation Defects},''
  \href{http://arxiv.org/abs/2204.02407}{{\ttfamily arXiv:2204.02407
  [hep-th]}}.

\bibitem{Choi:2022jqy}
Y.~Choi, H.~T. Lam, and S.-H. Shao, ``{Non-invertible Global Symmetries in the
  Standard Model},'' \href{http://arxiv.org/abs/2205.05086}{{\ttfamily
  arXiv:2205.05086 [hep-th]}}.

\bibitem{Arias-Tamargo:2022nlf}
G.~Arias-Tamargo and D.~Rodriguez-Gomez, ``{Non-Invertible Symmetries from
  Discrete Gauging and Completeness of the Spectrum},''
  \href{http://arxiv.org/abs/2204.07523}{{\ttfamily arXiv:2204.07523
  [hep-th]}}.

\bibitem{Cordova:2022ieu}
C.~Cordova and K.~Ohmori, ``{Non-Invertible Chiral Symmetry and Exponential
  Hierarchies},'' \href{http://arxiv.org/abs/2205.06243}{{\ttfamily
  arXiv:2205.06243 [hep-th]}}.

\bibitem{Bhardwaj:2022dyt}
L.~Bhardwaj, M.~Bullimore, A.~E.~V. Ferrari, and S.~Schafer-Nameki,
  ``{Anomalies of Generalized Symmetries from Solitonic Defects},''
  \href{http://arxiv.org/abs/2205.15330}{{\ttfamily arXiv:2205.15330
  [hep-th]}}.

\bibitem{Benedetti:2022zbb}
V.~Benedetti, H.~Casini, and J.~M. Magan, ``{Generalized symmetries and
  Noether's theorem in QFT},''
  \href{http://arxiv.org/abs/2205.03412}{{\ttfamily arXiv:2205.03412
  [hep-th]}}.

\bibitem{DelZotto:2022ohj}
M.~Del~Zotto, M.~Liu, and P.-K. Oehlmann, ``{Back to heterotic strings on ALE
  spaces. Part I. Instantons, 2-groups and T-duality},''
  \href{http://dx.doi.org/10.1007/JHEP01(2023)176}{{\em JHEP} {\bfseries 01}
  (2023) 176}, \href{http://arxiv.org/abs/2209.10551}{{\ttfamily
  arXiv:2209.10551 [hep-th]}}.

\bibitem{Bhardwaj:2022scy}
L.~Bhardwaj and D.~S.~W. Gould, ``{Disconnected 0-Form and 2-Group
  Symmetries},'' \href{http://arxiv.org/abs/2206.01287}{{\ttfamily
  arXiv:2206.01287 [hep-th]}}.

\bibitem{Antinucci:2022eat}
A.~Antinucci, G.~Galati, and G.~Rizi, ``{On Continuous 2-Category Symmetries
  and Yang-Mills Theory},'' \href{http://arxiv.org/abs/2206.05646}{{\ttfamily
  arXiv:2206.05646 [hep-th]}}.

\bibitem{Carta:2022spy}
F.~Carta, S.~Giacomelli, N.~Mekareeya, and A.~Mininno, ``{Dynamical
  consequences of 1-form symmetries and the exceptional Argyres-Douglas
  theories},'' \href{http://dx.doi.org/10.1007/JHEP06(2022)059}{{\em JHEP}
  {\bfseries 06} (2022) 059}, \href{http://arxiv.org/abs/2203.16550}{{\ttfamily
  arXiv:2203.16550 [hep-th]}}.

\bibitem{Apruzzi:2022dlm}
F.~Apruzzi, ``{Higher Form Symmetries TFT in 6d},''
  \href{http://arxiv.org/abs/2203.10063}{{\ttfamily arXiv:2203.10063
  [hep-th]}}.

\bibitem{Heckman:2022suy}
J.~J. Heckman, C.~Lawrie, L.~Lin, H.~Y. Zhang, and G.~Zoccarato, ``{6d SCFTs,
  Center-Flavor Symmetries, and Stiefel--Whitney Compactifications},''
  \href{http://arxiv.org/abs/2205.03411}{{\ttfamily arXiv:2205.03411
  [hep-th]}}.

\bibitem{Choi:2022rfe}
Y.~Choi, H.~T. Lam, and S.-H. Shao, ``{Non-invertible Time-reversal
  Symmetry},'' \href{http://arxiv.org/abs/2208.04331}{{\ttfamily
  arXiv:2208.04331 [hep-th]}}.

\bibitem{Bhardwaj:2022lsg}
L.~Bhardwaj, S.~Schafer-Nameki, and J.~Wu, ``{Universal Non-Invertible
  Symmetries},'' \href{http://arxiv.org/abs/2208.05973}{{\ttfamily
  arXiv:2208.05973 [hep-th]}}.

\bibitem{Lin:2022xod}
L.~Lin, D.~Robbins, and E.~Sharpe, ``{Decomposition, condensation defects, and
  fusion},'' \href{http://arxiv.org/abs/2208.05982}{{\ttfamily arXiv:2208.05982
  [hep-th]}}.

\bibitem{Bartsch:2022mpm}
T.~Bartsch, M.~Bullimore, A.~E.~V. Ferrari, and J.~Pearson, ``{Non-invertible
  Symmetries and Higher Representation Theory I},''
  \href{http://arxiv.org/abs/2208.05993}{{\ttfamily arXiv:2208.05993
  [hep-th]}}.

\bibitem{Apruzzi:2022rei}
F.~Apruzzi, I.~Bah, F.~Bonetti, and S.~Schafer-Nameki, ``{Non-Invertible
  Symmetries from Holography and Branes},''
  \href{http://arxiv.org/abs/2208.07373}{{\ttfamily arXiv:2208.07373
  [hep-th]}}.

\bibitem{GarciaEtxebarria:2022vzq}
I.~n. Garc\'\i{}a~Etxebarria, ``{Branes and Non-Invertible Symmetries},''
  \href{http://arxiv.org/abs/2208.07508}{{\ttfamily arXiv:2208.07508
  [hep-th]}}.

\bibitem{Cherman:2022eml}
A.~Cherman, T.~Jacobson, and M.~Neuzil, ``{1-form symmetry versus large N
  QCD},'' \href{http://arxiv.org/abs/2209.00027}{{\ttfamily arXiv:2209.00027
  [hep-th]}}.

\bibitem{Heckman:2022muc}
J.~J. Heckman, M.~H\"ubner, E.~Torres, and H.~Y. Zhang, ``{The Branes Behind
  Generalized Symmetry Operators},''
  \href{http://arxiv.org/abs/2209.03343}{{\ttfamily arXiv:2209.03343
  [hep-th]}}.

\bibitem{Lu:2022ver}
D.-C. Lu and Z.~Sun, ``{On Triality Defects in 2d CFT},''
  \href{http://arxiv.org/abs/2208.06077}{{\ttfamily arXiv:2208.06077
  [hep-th]}}.

\bibitem{Niro:2022ctq}
P.~Niro, K.~Roumpedakis, and O.~Sela, ``{Exploring Non-Invertible Symmetries in
  Free Theories},'' \href{http://arxiv.org/abs/2209.11166}{{\ttfamily
  arXiv:2209.11166 [hep-th]}}.

\bibitem{Kaidi:2022cpf}
J.~Kaidi, K.~Ohmori, and Y.~Zheng, ``{Symmetry TFTs for Non-Invertible
  Defects},'' \href{http://arxiv.org/abs/2209.11062}{{\ttfamily
  arXiv:2209.11062 [hep-th]}}.

\bibitem{Mekareeya:2022spm}
N.~Mekareeya and M.~Sacchi, ``{Mixed Anomalies, Two-groups, Non-Invertible
  Symmetries, and 3d Superconformal Indices},''
  \href{http://arxiv.org/abs/2210.02466}{{\ttfamily arXiv:2210.02466
  [hep-th]}}.

\bibitem{vanBeest:2022fss}
M.~van Beest, D.~S.~W. Gould, S.~Schafer-Nameki, and Y.-N. Wang, ``{Symmetry
  TFTs for 3d QFTs from M-theory},''
  \href{http://arxiv.org/abs/2210.03703}{{\ttfamily arXiv:2210.03703
  [hep-th]}}.

\bibitem{Giaccari:2022xgs}
S.~Giaccari and R.~Volpato, ``{A fresh view on string orbifolds},''
  \href{http://arxiv.org/abs/2210.10034}{{\ttfamily arXiv:2210.10034
  [hep-th]}}.

\bibitem{Bashmakov:2022uek}
V.~Bashmakov, M.~Del~Zotto, A.~Hasan, and J.~Kaidi, ``{Non-invertible
  Symmetries of Class $\mathcal{S}$ Theories},''
  \href{http://arxiv.org/abs/2211.05138}{{\ttfamily arXiv:2211.05138
  [hep-th]}}.

\bibitem{Cordova:2022fhg}
C.~Cordova, S.~Hong, S.~Koren, and K.~Ohmori, ``{Neutrino Masses from
  Generalized Symmetry Breaking},''
  \href{http://arxiv.org/abs/2211.07639}{{\ttfamily arXiv:2211.07639
  [hep-ph]}}.

\bibitem{GarciaEtxebarria:2022jky}
I.~Garcia~Etxebarria and N.~Iqbal, ``{A Goldstone theorem for continuous
  non-invertible symmetries},''
  \href{http://arxiv.org/abs/2211.09570}{{\ttfamily arXiv:2211.09570
  [hep-th]}}.

\bibitem{Choi:2022fgx}
Y.~Choi, H.~T. Lam, and S.-H. Shao, ``{Non-invertible Gauss Law and Axions},''
  \href{http://arxiv.org/abs/2212.04499}{{\ttfamily arXiv:2212.04499
  [hep-th]}}.

\bibitem{Robbins:2022wlr}
D.~Robbins, E.~Sharpe, and T.~Vandermeulen, ``{Decomposition, Trivially-Acting
  Symmetries, and Topological Operators},''
  \href{http://arxiv.org/abs/2211.14332}{{\ttfamily arXiv:2211.14332
  [hep-th]}}.

\bibitem{Bhardwaj:2022kot}
L.~Bhardwaj, S.~Schafer-Nameki, and A.~Tiwari, ``{Unifying Constructions of
  Non-Invertible Symmetries},''
  \href{http://arxiv.org/abs/2212.06159}{{\ttfamily arXiv:2212.06159
  [hep-th]}}.

\bibitem{Bhardwaj:2022maz}
L.~Bhardwaj, L.~E. Bottini, S.~Schafer-Nameki, and A.~Tiwari, ``{Non-Invertible
  Symmetry Webs},'' \href{http://arxiv.org/abs/2212.06842}{{\ttfamily
  arXiv:2212.06842 [hep-th]}}.

\bibitem{Bartsch:2022ytj}
T.~Bartsch, M.~Bullimore, A.~E.~V. Ferrari, and J.~Pearson, ``{Non-invertible
  Symmetries and Higher Representation Theory II},''
  \href{http://arxiv.org/abs/2212.07393}{{\ttfamily arXiv:2212.07393
  [hep-th]}}.

\bibitem{Gaiotto:2020iye}
D.~Gaiotto and J.~Kulp, ``{Orbifold groupoids},''
  \href{http://dx.doi.org/10.1007/JHEP02(2021)132}{{\em JHEP} {\bfseries 02}
  (2021) 132}, \href{http://arxiv.org/abs/2008.05960}{{\ttfamily
  arXiv:2008.05960 [hep-th]}}.

\bibitem{Robbins:2021ibx}
D.~G. Robbins, E.~Sharpe, and T.~Vandermeulen, ``{Quantum symmetries in
  orbifolds and decomposition},''
  \href{http://dx.doi.org/10.1007/JHEP02(2022)108}{{\em JHEP} {\bfseries 02}
  (2022) 108}, \href{http://arxiv.org/abs/2107.12386}{{\ttfamily
  arXiv:2107.12386 [hep-th]}}.

\bibitem{Robbins:2021xce}
D.~G. Robbins, E.~Sharpe, and T.~Vandermeulen, ``{Anomaly resolution via
  decomposition},'' \href{http://dx.doi.org/10.1142/S0217751X21502201}{{\em
  Int. J. Mod. Phys. A} {\bfseries 36} no.~29, (2021) 2150220},
  \href{http://arxiv.org/abs/2107.13552}{{\ttfamily arXiv:2107.13552
  [hep-th]}}.

\bibitem{Huang:2021zvu}
T.-C. Huang, Y.-H. Lin, and S.~Seifnashri, ``{Construction of two-dimensional
  topological field theories with non-invertible symmetries},''
  \href{http://dx.doi.org/10.1007/JHEP12(2021)028}{{\em JHEP} {\bfseries 12}
  (2021) 028}, \href{http://arxiv.org/abs/2110.02958}{{\ttfamily
  arXiv:2110.02958 [hep-th]}}.

\bibitem{Inamura:2021szw}
K.~Inamura, ``{On lattice models of gapped phases with fusion category
  symmetries},'' \href{http://dx.doi.org/10.1007/JHEP03(2022)036}{{\em JHEP}
  {\bfseries 03} (2022) 036}, \href{http://arxiv.org/abs/2110.12882}{{\ttfamily
  arXiv:2110.12882 [cond-mat.str-el]}}.

\bibitem{Cherman:2021nox}
A.~Cherman, T.~Jacobson, and M.~Neuzil, ``{Universal Deformations},''
  \href{http://dx.doi.org/10.21468/SciPostPhys.12.4.116}{{\em SciPost Phys.}
  {\bfseries 12} no.~4, (2022) 116},
  \href{http://arxiv.org/abs/2111.00078}{{\ttfamily arXiv:2111.00078
  [hep-th]}}.

\bibitem{Sharpe:2022ene}
E.~Sharpe, ``{An introduction to decomposition},''
  \href{http://arxiv.org/abs/2204.09117}{{\ttfamily arXiv:2204.09117
  [hep-th]}}.

\bibitem{Bashmakov:2022jtl}
V.~Bashmakov, M.~Del~Zotto, and A.~Hasan, ``{On the 6d Origin of Non-invertible
  Symmetries in 4d},'' \href{http://arxiv.org/abs/2206.07073}{{\ttfamily
  arXiv:2206.07073 [hep-th]}}.

\bibitem{Lee:2022swr}
S.-J. Lee and P.-K. Oehlmann, ``{Geometric Bounds on the 1-Form Gauge
  Sector},'' \href{http://arxiv.org/abs/2212.11915}{{\ttfamily arXiv:2212.11915
  [hep-th]}}.

\bibitem{Inamura:2022lun}
K.~Inamura, ``{Fermionization of fusion category symmetries in 1+1
  dimensions},'' \href{http://arxiv.org/abs/2206.13159}{{\ttfamily
  arXiv:2206.13159 [cond-mat.str-el]}}.

\bibitem{Damia:2022bcd}
J.~A. Damia, R.~Argurio, and E.~Garcia-Valdecasas, ``{Non-Invertible Defects in
  5d, Boundaries and Holography},''
  \href{http://arxiv.org/abs/2207.02831}{{\ttfamily arXiv:2207.02831
  [hep-th]}}.

\bibitem{Lin:2022dhv}
Y.-H. Lin, M.~Okada, S.~Seifnashri, and Y.~Tachikawa, ``{Asymptotic density of
  states in 2d CFTs with non-invertible symmetries},''
  \href{http://arxiv.org/abs/2208.05495}{{\ttfamily arXiv:2208.05495
  [hep-th]}}.

\bibitem{Burbano:2021loy}
I.~M. Burbano, J.~Kulp, and J.~Neuser, ``{Duality Defects in $E_8$},''
  \href{http://arxiv.org/abs/2112.14323}{{\ttfamily arXiv:2112.14323
  [hep-th]}}.

\bibitem{Damia:2022rxw}
J.~A. Damia, R.~Argurio, and L.~Tizzano, ``{Continuous Generalized Symmetries
  in Three Dimensions},'' \href{http://arxiv.org/abs/2206.14093}{{\ttfamily
  arXiv:2206.14093 [hep-th]}}.

\bibitem{Apte:2022xtu}
A.~Apte, C.~Cordova, and H.~T. Lam, ``{Obstructions to Gapped Phases from
  Non-Invertible Symmetries},''
  \href{http://arxiv.org/abs/2212.14605}{{\ttfamily arXiv:2212.14605
  [hep-th]}}.

\bibitem{Chen:2022cyw}
S.~Chen and Y.~Tanizaki, ``{Solitonic symmetry beyond homotopy: invertibility
  from bordism and non-invertibility from TQFT},''
  \href{http://arxiv.org/abs/2210.13780}{{\ttfamily arXiv:2210.13780
  [hep-th]}}.

\bibitem{Nawata:2023rdx}
S.~Nawata, M.~Sperling, H.~E. Wang, and Z.~Zhong, ``{3d $\mathcal{N}=4$ mirror
  symmetry with 1-form symmetry},''
  \href{http://arxiv.org/abs/2301.02409}{{\ttfamily arXiv:2301.02409
  [hep-th]}}.

\bibitem{Bhardwaj:2023zix}
L.~Bhardwaj, M.~Bullimore, A.~E.~V. Ferrari, and S.~Schafer-Nameki,
  ``{Generalized Symmetries and Anomalies of 3d $\mathcal{N}=4$ SCFTs},''
  \href{http://arxiv.org/abs/2301.02249}{{\ttfamily arXiv:2301.02249
  [hep-th]}}.

\bibitem{Kaidi:2023maf}
J.~Kaidi, E.~Nardoni, G.~Zafrir, and Y.~Zheng, ``{Symmetry TFTs and Anomalies
  of Non-Invertible Symmetries},''
  \href{http://arxiv.org/abs/2301.07112}{{\ttfamily arXiv:2301.07112
  [hep-th]}}.

\bibitem{Etheredge:2023ler}
M.~Etheredge, I.~Garcia~Etxebarria, B.~Heidenreich, and S.~Rauch, ``{Branes and
  symmetries for $\mathcal N=3$ S-folds},''
  \href{http://arxiv.org/abs/2302.14068}{{\ttfamily arXiv:2302.14068
  [hep-th]}}.

\bibitem{Lin:2023uvm}
Y.-H. Lin and S.-H. Shao, ``{Bootstrapping Non-invertible Symmetries},''
  \href{http://arxiv.org/abs/2302.13900}{{\ttfamily arXiv:2302.13900
  [hep-th]}}.

\bibitem{Amariti:2023hev}
A.~Amariti, D.~Morgante, A.~Pasternak, S.~Rota, and V.~Tatitscheff, ``{One-form
  symmetries in $\mathcal{N} = 3$$S$-folds},''
  \href{http://arxiv.org/abs/2303.07299}{{\ttfamily arXiv:2303.07299
  [hep-th]}}.

\bibitem{Bhardwaj:2023wzd}
L.~Bhardwaj and S.~Schafer-Nameki, ``{Generalized Charges, Part I: Invertible
  Symmetries and Higher Representations},''
  \href{http://arxiv.org/abs/2304.02660}{{\ttfamily arXiv:2304.02660
  [hep-th]}}.

\bibitem{Bartsch:2023pzl}
T.~Bartsch, M.~Bullimore, and A.~Grigoletto, ``{Higher representations for
  extended operators},'' \href{http://arxiv.org/abs/2304.03789}{{\ttfamily
  arXiv:2304.03789 [hep-th]}}.

\bibitem{Carta:2023bqn}
F.~Carta, S.~Giacomelli, N.~Mekareeya, and A.~Mininno, ``{Comments on
  Non-invertible Symmetries in Argyres-Douglas Theories},''
  \href{http://arxiv.org/abs/2303.16216}{{\ttfamily arXiv:2303.16216
  [hep-th]}}.

\bibitem{Zhang:2023wlu}
C.~Zhang and C.~C\'ordova, ``{Anomalies of $(1+1)D$ categorical symmetries},''
  \href{http://arxiv.org/abs/2304.01262}{{\ttfamily arXiv:2304.01262
  [cond-mat.str-el]}}.

\bibitem{Cao:2023doz}
W.~Cao, L.~Li, M.~Yamazaki, and Y.~Zheng, ``{Subsystem Non-Invertible Symmetry
  Operators and Defects},'' \href{http://arxiv.org/abs/2304.09886}{{\ttfamily
  arXiv:2304.09886 [cond-mat.str-el]}}.

\bibitem{Putrov:2023jqi}
P.~Putrov and J.~Wang, ``{Categorical Symmetry of the Standard Model from
  Gravitational Anomaly},'' \href{http://arxiv.org/abs/2302.14862}{{\ttfamily
  arXiv:2302.14862 [hep-th]}}.

\bibitem{Acharya:2023bth}
B.~S. Acharya, M.~Del~Zotto, J.~J. Heckman, M.~Hubner, and E.~Torres,
  ``{Junctions, Edge Modes, and $G_2$-Holonomy Orbifolds},''
  \href{http://arxiv.org/abs/2304.03300}{{\ttfamily arXiv:2304.03300
  [hep-th]}}.

\bibitem{Inamura:2023qzl}
K.~Inamura and K.~Ohmori, ``{Fusion Surface Models: 2+1d Lattice Models from
  Fusion 2-Categories},'' \href{http://arxiv.org/abs/2305.05774}{{\ttfamily
  arXiv:2305.05774 [cond-mat.str-el]}}.

\bibitem{Dierigl:2023jdp}
M.~Dierigl, J.~J. Heckman, M.~Montero, and E.~Torres, ``{R7-Branes as Charge
  Conjugation Operators},'' \href{http://arxiv.org/abs/2305.05689}{{\ttfamily
  arXiv:2305.05689 [hep-th]}}.

\bibitem{Antinucci:2023uzq}
A.~Antinucci, G.~Galati, G.~Rizi, and M.~Serone, ``{Symmetries and topological
  operators, on average},'' \href{http://arxiv.org/abs/2305.08911}{{\ttfamily
  arXiv:2305.08911 [hep-th]}}.

\bibitem{Cvetic:2023plv}
M.~"Cvetic, J.~J. Heckman, M.~H\"ubner, and E.~Torres, ``{Fluxbranes,
  Generalized Symmetries, and Verlinde's Metastable Monopole},''
  \href{http://arxiv.org/abs/2305.09665}{{\ttfamily arXiv:2305.09665
  [hep-th]}}.

\bibitem{Barkeshli:2014cna}
M.~Barkeshli, P.~Bonderson, M.~Cheng, and Z.~Wang, ``{Symmetry
  Fractionalization, Defects, and Gauging of Topological Phases},''
  \href{http://dx.doi.org/10.1103/PhysRevB.100.115147}{{\em Phys. Rev. B}
  {\bfseries 100} no.~11, (2019) 115147},
  \href{http://arxiv.org/abs/1410.4540}{{\ttfamily arXiv:1410.4540
  [cond-mat.str-el]}}.

\bibitem{Benini:2018reh}
F.~Benini, C.~C\'ordova, and P.-S. Hsin, ``{On 2-Group Global Symmetries and
  their Anomalies},'' \href{http://dx.doi.org/10.1007/JHEP03(2019)118}{{\em
  JHEP} {\bfseries 03} (2019) 118},
  \href{http://arxiv.org/abs/1803.09336}{{\ttfamily arXiv:1803.09336
  [hep-th]}}.

\bibitem{Delmastro:2022pfo}
D.~Delmastro, J.~Gomis, P.-S. Hsin, and Z.~Komargodski, ``{Anomalies and
  Symmetry Fractionalization},''
  \href{http://arxiv.org/abs/2206.15118}{{\ttfamily arXiv:2206.15118
  [hep-th]}}.

\bibitem{Brennan:2022tyl}
T.~D. Brennan, C.~Cordova, and T.~T. Dumitrescu, ``{Line Defect Quantum Numbers
  \& Anomalies},'' \href{http://arxiv.org/abs/2206.15401}{{\ttfamily
  arXiv:2206.15401 [hep-th]}}.

\bibitem{douglas2018fusion}
C.~L. Douglas and D.~J. Reutter, ``Fusion 2-categories and a state-sum
  invariant for 4-manifolds,'' {\em arXiv preprint arXiv:1812.11933} (2018) .

\bibitem{Johnson-Freyd:2020usu}
T.~Johnson-Freyd, ``{On the Classification of Topological Orders},''
  \href{http://dx.doi.org/10.1007/s00220-022-04380-3}{{\em Commun. Math. Phys.}
  {\bfseries 393} no.~2, (2022) 989--1033},
  \href{http://arxiv.org/abs/2003.06663}{{\ttfamily arXiv:2003.06663
  [math.CT]}}.

\bibitem{Moore:1988qv}
G.~W. Moore and N.~Seiberg, ``{Classical and Quantum Conformal Field Theory},''
  \href{http://dx.doi.org/10.1007/BF01238857}{{\em Commun. Math. Phys.}
  {\bfseries 123} (1989) 177}.

\bibitem{Gaiotto:2019xmp}
D.~Gaiotto and T.~Johnson-Freyd, ``{Condensations in higher categories},''
  \href{http://arxiv.org/abs/1905.09566}{{\ttfamily arXiv:1905.09566
  [math.CT]}}.

\bibitem{Antinucci:2022cdi}
A.~Antinucci, C.~Copetti, G.~Galati, and G.~Rizi, ``{''Zoology'' of
  non-invertible duality defects: the view from class $\mathcal{S}$},''
  \href{http://arxiv.org/abs/2212.09549}{{\ttfamily arXiv:2212.09549
  [hep-th]}}.

\bibitem{Putrov:2022pua}
P.~Putrov, ``{$\mathbb{Q}/\mathbb{Z}$ symmetry},''
  \href{http://arxiv.org/abs/2208.12071}{{\ttfamily arXiv:2208.12071
  [hep-th]}}.

\bibitem{Yokokura:2022alv}
R.~Yokokura, ``{Non-invertible symmetries in axion electrodynamics},''
  \href{http://arxiv.org/abs/2212.05001}{{\ttfamily arXiv:2212.05001
  [hep-th]}}.

\bibitem{Freed:2012bs}
D.~S. Freed and C.~Teleman, ``{Relative quantum field theory},''
  \href{http://dx.doi.org/10.1007/s00220-013-1880-1}{{\em Commun. Math. Phys.}
  {\bfseries 326} (2014) 459--476},
  \href{http://arxiv.org/abs/1212.1692}{{\ttfamily arXiv:1212.1692 [hep-th]}}.

\bibitem{Freed:2022qnc}
D.~S. Freed, G.~W. Moore, and C.~Teleman, ``{Topological symmetry in quantum
  field theory},'' \href{http://arxiv.org/abs/2209.07471}{{\ttfamily
  arXiv:2209.07471 [hep-th]}}.

\bibitem{Cordova:2019wpi}
C.~C\'{o}rdova, K.~Ohmori, S.-H. Shao, and F.~Yan, ``{Decorated
  $\mathbb{Z}_{2}$ Symmetry Defects and Their Time-Reversal Anomalies},''
  \href{http://arxiv.org/abs/1910.14046}{{\ttfamily arXiv:1910.14046
  [hep-th]}}.

\bibitem{Carqueville:2018sld}
N.~Carqueville, I.~Runkel, and G.~Schaumann, ``{Orbifolds of Reshetikhin-Turaev
  TQFTs},'' {\em Theor. Appl. Categor.} {\bfseries 35} (2020) 513--561,
  \href{http://arxiv.org/abs/1809.01483}{{\ttfamily arXiv:1809.01483
  [math.QA]}}.

\bibitem{PhysRevB.62.7850}
T.~Senthil and M.~P.~A. Fisher, ``${Z}_{2}$ gauge theory of electron
  fractionalization in strongly correlated systems,''
  \href{https://link.aps.org/doi/10.1103/PhysRevB.62.7850}{{\em Phys. Rev. B}
  {\bfseries 62} (Sep, 2000) 7850--7881}.

\bibitem{Essin:2013rca}
A.~M. Essin and M.~Hermele, ``{Classifying fractionalization: Symmetry
  classification of gapped $\mathbb{Z}_2$ spin liquids in two dimensions},''
  \href{http://dx.doi.org/10.1103/PhysRevB.87.104406}{{\em Phys. Rev. B}
  {\bfseries 87} no.~10, (2013) 104406},
  \href{http://arxiv.org/abs/1212.0593}{{\ttfamily arXiv:1212.0593
  [cond-mat.str-el]}}.

\bibitem{Chen_2015}
X.~Chen, F.~Burnell, A.~Vishwanath, and L.~Fidkowski, ``Anomalous symmetry
  fractionalization and surface topological order,''
  \href{https://doi.org/10.1103%2Fphysrevx.5.041013}{{\em Physical Review X}
  {\bfseries 5} no.~4, (Oct, 2015) }.

\bibitem{Tarantino_2016}
N.~Tarantino, N.~H. Lindner, and L.~Fidkowski, ``Symmetry fractionalization and
  twist defects,''
  \href{https://doi.org/10.1088%2F1367-2630%2F18%2F3%2F035006}{{\em New Journal
  of Physics} {\bfseries 18} no.~3, (Mar, 2016) 035006}.

\bibitem{Bulmash:2021hmb}
D.~Bulmash and M.~Barkeshli, ``{Fermionic symmetry fractionalization in (2+1)
  dimensions},'' \href{http://dx.doi.org/10.1103/PhysRevB.105.125114}{{\em
  Phys. Rev. B} {\bfseries 105} no.~12, (2022) 125114},
  \href{http://arxiv.org/abs/2109.10913}{{\ttfamily arXiv:2109.10913
  [cond-mat.str-el]}}.

\bibitem{davydov2013witt}
A.~Davydov, M.~M{\"u}ger, D.~Nikshych, and V.~Ostrik, ``The witt group of
  non-degenerate braided fusion categories,'' {\em Journal f{\"u}r die reine
  und angewandte Mathematik (Crelles Journal)} {\bfseries 2013} no.~677, (2013)
  135--177.

\bibitem{Kapustin:2010if}
A.~Kapustin and N.~Saulina, ``{Surface operators in 3d Topological Field Theory
  and 2d Rational Conformal Field Theory},''
  \href{http://arxiv.org/abs/1012.0911}{{\ttfamily arXiv:1012.0911 [hep-th]}}.

\bibitem{Kong:2020cie}
L.~Kong, T.~Lan, X.-G. Wen, Z.-H. Zhang, and H.~Zheng, ``{Algebraic higher
  symmetry and categorical symmetry -- a holographic and entanglement view of
  symmetry},'' \href{http://dx.doi.org/10.1103/PhysRevResearch.2.043086}{{\em
  Phys. Rev. Res.} {\bfseries 2} no.~4, (2020) 043086},
  \href{http://arxiv.org/abs/2005.14178}{{\ttfamily arXiv:2005.14178
  [cond-mat.str-el]}}.

\bibitem{Antinucci:2022vyk}
A.~Antinucci, F.~Benini, C.~Copetti, G.~Galati, and G.~Rizi, ``{The holography
  of non-invertible self-duality symmetries},''
  \href{http://arxiv.org/abs/2210.09146}{{\ttfamily arXiv:2210.09146
  [hep-th]}}.

\bibitem{Kitaev:2005hzj}
A.~Kitaev, ``{Anyons in an exactly solved model and beyond},''
  \href{http://dx.doi.org/10.1016/j.aop.2005.10.005}{{\em Annals Phys.}
  {\bfseries 321} no.~1, (2006) 2--111},
  \href{http://arxiv.org/abs/cond-mat/0506438}{{\ttfamily
  arXiv:cond-mat/0506438}}.

\bibitem{Gaiotto:2017zba}
D.~Gaiotto and T.~Johnson-Freyd, ``{Symmetry Protected Topological phases and
  Generalized Cohomology},''
  \href{http://dx.doi.org/10.1007/JHEP05(2019)007}{{\em JHEP} {\bfseries 05}
  (2019) 007}, \href{http://arxiv.org/abs/1712.07950}{{\ttfamily
  arXiv:1712.07950 [hep-th]}}.

\bibitem{Delmastro:2021xox}
D.~Delmastro, D.~Gaiotto, and J.~Gomis, ``{Global anomalies on the Hilbert
  space},'' \href{http://dx.doi.org/10.1007/JHEP11(2021)142}{{\em JHEP}
  {\bfseries 11} (2021) 142}, \href{http://arxiv.org/abs/2101.02218}{{\ttfamily
  arXiv:2101.02218 [hep-th]}}.

\bibitem{Cui:2016bmd}
S.~X. Cui, ``{Four dimensional topological quantum field theories from
  $G$-crossed braided categories},''
  \href{http://dx.doi.org/10.4171/qt/128}{{\em Quantum Topol.} {\bfseries 10}
  no.~4, (2019) 593--676}, \href{http://arxiv.org/abs/1610.07628}{{\ttfamily
  arXiv:1610.07628 [math.QA]}}.

\bibitem{Moore:2006dw}
G.~W. Moore and G.~Segal, ``{D-branes and K-theory in 2D topological field
  theory},'' \href{http://arxiv.org/abs/hep-th/0609042}{{\ttfamily
  arXiv:hep-th/0609042}}.

\bibitem{Cordova:2017vab}
C.~Cordova, P.-S. Hsin, and N.~Seiberg, ``{Global Symmetries, Counterterms, and
  Duality in Chern-Simons Matter Theories with Orthogonal Gauge Groups},''
  \href{http://dx.doi.org/10.21468/SciPostPhys.4.4.021}{{\em SciPost Phys.}
  {\bfseries 4} no.~4, (2018) 021},
  \href{http://arxiv.org/abs/1711.10008}{{\ttfamily arXiv:1711.10008
  [hep-th]}}.

\bibitem{Kapustin:2010hk}
A.~Kapustin and N.~Saulina, ``{Topological boundary conditions in abelian
  Chern-Simons theory},''
  \href{http://dx.doi.org/10.1016/j.nuclphysb.2010.12.017}{{\em Nucl. Phys. B}
  {\bfseries 845} (2011) 393--435},
  \href{http://arxiv.org/abs/1008.0654}{{\ttfamily arXiv:1008.0654 [hep-th]}}.

\bibitem{Kong:2013aya}
L.~Kong, ``{Anyon condensation and tensor categories},''
  \href{http://dx.doi.org/10.1016/j.nuclphysb.2014.07.003}{{\em Nucl. Phys. B}
  {\bfseries 886} (2014) 436--482},
  \href{http://arxiv.org/abs/1307.8244}{{\ttfamily arXiv:1307.8244
  [cond-mat.str-el]}}.

\bibitem{Harlow:2018tng}
D.~Harlow and H.~Ooguri, ``{Symmetries in quantum field theory and quantum
  gravity},'' \href{http://dx.doi.org/10.1007/s00220-021-04040-y}{{\em Commun.
  Math. Phys.} {\bfseries 383} no.~3, (2021) 1669--1804},
  \href{http://arxiv.org/abs/1810.05338}{{\ttfamily arXiv:1810.05338
  [hep-th]}}.

\bibitem{Delmastro:2019vnj}
D.~Delmastro and J.~Gomis, ``{Symmetries of Abelian Chern-Simons Theories and
  Arithmetic},'' \href{http://dx.doi.org/10.1007/JHEP03(2021)006}{{\em JHEP}
  {\bfseries 03} (2021) 006}, \href{http://arxiv.org/abs/1904.12884}{{\ttfamily
  arXiv:1904.12884 [hep-th]}}.

\bibitem{Benini:2022hzx}
F.~Benini, C.~Copetti, and L.~Di~Pietro, ``{Factorization and global symmetries
  in holography},'' \href{http://dx.doi.org/10.21468/SciPostPhys.14.2.019}{{\em
  SciPost Phys.} {\bfseries 14} no.~2, (2023) 019},
  \href{http://arxiv.org/abs/2203.09537}{{\ttfamily arXiv:2203.09537
  [hep-th]}}.

\bibitem{davydov2013structure}
A.~Davydov, D.~Nikshych, and V.~Ostrik, ``On the structure of the witt group of
  braided fusion categories,'' {\em Selecta Mathematica} {\bfseries 19} no.~1,
  (2013) 237--269.

\bibitem{Kapustin:2005py}
A.~Kapustin, ``{Wilson-'t Hooft operators in four-dimensional gauge theories
  and S-duality},'' \href{http://dx.doi.org/10.1103/PhysRevD.74.025005}{{\em
  Phys. Rev. D} {\bfseries 74} (2006) 025005},
  \href{http://arxiv.org/abs/hep-th/0501015}{{\ttfamily arXiv:hep-th/0501015}}.

\bibitem{Choi:2023xjw}
Y.~Choi, B.~C. Rayhaun, Y.~Sanghavi, and S.-H. Shao, ``{Comments on Boundaries,
  Anomalies, and Non-Invertible Symmetries},''
  \href{http://arxiv.org/abs/2305.09713}{{\ttfamily arXiv:2305.09713
  [hep-th]}}.

\end{thebibliography}\endgroup

\end{document}